\documentclass[3p,12pt]{elsarticle}
\usepackage{bm}
\usepackage{amsfonts}
\usepackage{amssymb}
\newcommand{\fD}{\frak D}
\newcommand{\fP}{\frak P}
\newcommand{\fT}{\frak T}
\journal{Annals of Physics}

\newcommand{\rT}{^{\rm T}}
\newcommand{\rL}{^{\rm L}}
\newcommand{\rM}{_{\rm M}}
\newcommand{\rc}{_{\rm c}}
\newcommand{\rd}{{\rm d}}
\newcommand{\re}{{\rm e}}
\newcommand{\rf}{{\rm f}}
\newcommand{\ri}{{\rm i}}
\newcommand{\rp}{_{\rm p}}
\newcommand{\rs}{_{\rm s}}
\newcommand{\iO}{{\it \Omega}}
\newcommand{\iP}{{\it \Psi}}
\newcommand{\iX}{{\it \Xi}}
\newcommand{\iPi}{{\it \Pi}}

\newcommand{\cC}{{\cal C}}
\newcommand{\cD}{{\cal D}}

\newcommand{\cH}{{\cal H}}
\newcommand{\cI}{{\cal I}}
\newcommand{\cJ}{{\cal J}}
\newcommand{\cK}{{\cal K}}
\newcommand{\cP}{{\cal P}}
\newcommand{\cR}{{\cal R}}
\newcommand{\cS}{{\cal S}}
\newcommand{\cT}{{\cal T}}
\newcommand{\cU}{{\cal U}}
\newcommand{\cV}{{\cal V}}

\newcommand{\0}{{\bm 0}}
\def\fr#1#2{{\textstyle{\frac{#1}{#2}}}}
\def\ft#1#2{{\frac{\textstyle #1}{\textstyle #2}}}
\def\elabel#1{{\label{#1}}}
\def\msp{\vbox to 15 pt {}}
\begin{document}

\title{Solutions of the Maxwell equations and photon wave
functions}

\newcommand{\addrGaithersburg}{National Institute of Standards and Technology,
Gaithersburg, MD 20899-8420, USA}

\author{Peter J.~Mohr}
\address{\addrGaithersburg}

\date{\today}

%\keywords{photons; wave functions; Maxwell equations; Dirac
%equation; Green function}

\begin{abstract}

Properties of six-component electromagnetic field solutions of a
matrix form of the Maxwell equations, analogous to the
four-component solutions of the Dirac equation, are described.
It is shown that the six-component equation, including sources,
is invariant under Lorentz transformations.  Complete sets of
eigenfunctions of the Hamiltonian for the electromagnetic
fields, which may be interpreted as photon wave functions, are
given both for plane waves and for angular-momentum eigenstates.
Rotationally invariant projection operators are used to identify
transverse or longitudinal electric and magnetic fields.  For
plane waves, the velocity transformed transverse wave functions
are also transverse, and the velocity transformed longitudinal
wave functions include both longitudinal and transverse
components.  A suitable sum over these eigenfunctions provides a
Green function for the matrix Maxwell equation, which can be
expressed in the same covariant form as the Green function for
the Dirac equation.  Radiation from a dipole source and from a
Dirac atomic transition current are calculated to illustrate
applications of the Maxwell Green function.

\end{abstract}
%
%\pacs{03.50.De, 92.60.Ta, 11.30.Cp}
%
\maketitle
%
%%%%%%%%%%%%%%%%%%%%%%%%%%%%%%%%%%%%%%%%%%%%%%%%%%%%%%%%%%%%%%%%%%%%%

%\tableofcontents

\section{Introduction}
\elabel{sec:intro}

For quantum mechanics to provide a complete description of
nature, it is necessary to have a wave function for something as
important as electromagnetic radiation or photons.  This has
become increasingly relevant as the number of experiments on
single photon production and detection, motivated by interest in
the fields of quantum computation and quantum cryptography, has
grown rapidly over the past two decades \cite{migdal}.  The
history of theoretical efforts to define photon wave functions
dates back to the early days of quantum mechanics and is still
unfolding.  Overviews have been given in \cite{1996213, 1997197,
2005358}.  However, there is not yet a consensus on the form a
photon wave function should take or the properties it should
have.  Further investigation of these questions is warranted,
and possible answers are given in this paper.

Quantum electrodynamics (QED) accurately describes the
interaction of radiation with free electrons and electrons bound
in atoms, but as it is formulated in terms of an $S$ matrix,
asymptotic states, and Feynman diagrams, it does not readily
lend itself to the description of the time evolution of
radiation.  In particular, interference effects or the
space-time behavior of a photon wave packet would be more
naturally described in the framework of wave mechanics with a
wave function for a single photon.

There are a number of requirements that need to be be imposed on
a formalism for a quantum mechanical description of photons.
First, the predicted behavior of radiation should be consistent
with the Maxwell equations.  Time-dependent solutions of the
Maxwell equations provide the basis for both classical
electromagnetic theory and QED, and it can be expected that a
photon wave function should also be based on solutions of the
Maxwell equations.  This means that the wave function is
simultaneously the solution of both of the first-order Maxwell
equations with time derivatives and not just a solution of a
second-order scalar wave equation.

A second requirement is that the wave functions obey the quantum
mechanical principle of linear superposition.  Simply stated,
this means that if two wave functions describe possible states
of radiation, then a linear combination of these wave functions
also describes a possible state.  For example, a wave function
for circularly polarized radiation can be written as a linear
combination of two wave functions for linearly polarized
radiation, all of which must be solutions of the same wave
equation.

Another requirement is that the formalism be Lorentz invariant
in order to properly describe the space-time behavior of
radiation.  An approximation scheme like the reduction of the
Dirac equation to obtain the Schr\"odinger equation for electron
velocities that are small compared to the speed of light is not
an option for radiative photons.

Finally, it is necessary for the formalism to provide the tools
for methods associated with quantum mechanics.  This includes a
wave equation with a Hamiltonian that describes the time
development of states, wave functions that comprise a complete
set of eigenfunctions of the Hamiltonian, normalizable states
with a probability distribution that corresponds to the location
of the photon, a law of conservation of probability, operators
with expectation values for observables, and wave packets that
realistically describe the propagation of photons in space and
time.

To arrive at a wave equation that addresses these requirements,
we examine an approach in which the four-component matrix Dirac
equation for a spin one-half electron is adapted to a
six-component form of the Maxwell equations for a spin-one
photon.  This version of the Maxwell equations is a direct
extension of the Dirac equation for the electron in which
two-by-two Pauli matrices are replaced by analogous
three-by-three matrices.  Since the quantum mechanical
properties of the Dirac equation, Hamiltonian, and wave
functions are well understood and tested experimentally, it is
natural to consider the analogous Maxwell equation, Hamiltonian,
and wave functions as a quantum mechanical description of
photons.  

There are fundamental differences between the Dirac equation and
the matrix Maxwell equation, so the extension requires a
detailed analysis.  The most prominent difference is the fact
that there is a possible source term in the Maxwell equation
which has no analog for the Dirac equation \cite{1931004}.
Also, some properties of the three-by-three spin matrices differ
from those of the Pauli matrices, even though they have the same
commutation relations.  

Linear operators that are representations of the inhomogeneous
Lorentz group can replace the wave equation of a system for free
electrons or transverse photons with no sources
\cite{1948004,1964012,1964011}, but the source terms and
longitudinal solutions of the Maxwell equation fall outside this
framework.  Taking the view that the Maxwell equation with a
source is the most direct contact with experiment, our approach
is to start from the matrix Maxwell equation with a source term
and explicitly work out the Lorentz transformations of the
solutions.  It is shown that the six-component equation is
invariant under Lorentz transformations, as it should be, but
this is not self-evident, since the source term is essentially
the three-vector current density.

Next, six-component solutions are constructed and shown to be
complete sets of orthogonal coordinate-space eigenfunctions of
the Maxwell Hamiltonian, parameterized by physical properties,
such as linear momentum, angular momentum, and parity. These
properties are associated with operators that commute with the
Hamiltonian.  Complete sets of both plane-wave solutions and
angular-momentum eigenfunctions are given.  Bilinear products of
normalizable linear combinations of these functions provide
expressions for the probability density and flux.  The
eigenfunctions are further classified according to whether they
represent transverse or longitudinal states.  These properties
are associated with the electric and magnetic fields, with the
result that under a velocity boost, the transformed transverse
solutions are also transverse, unlike solutions corresponding to
a transverse vector potential.  Moreover, by summing over both
transverse and longitudinal solutions, we obtain a covariant
Green function for the Maxwell equation, which is of the same
form as the Green function for the Dirac equation.

Solutions are obtained directly from the Maxwell equation, with
no recourse to a vector potential.  This avoids problems such as
extra polarization components and ambiguities associated with
gauge transformations \cite{1960007, 1962012, 1962010}.
Although integrals over a closed path of the potential may be
observables, as discussed in \cite{1959009}, such integrals can
be expressed in terms of the magnetic flux through the loop
\cite{2001380}, so it is expected that the fields alone provide
a complete description of electrodynamics.  It is also clear
that photon wave functions are closely aligned with electric and
magnetic fields, and an approach that starts with classical
electrodynamics expressed in terms of fields only provides a
natural framework for the transition to the wave mechanics of
photons.  A possible advantage of using a vector potential is
that it is the solution of a scalar wave equation, which has a
well-known Green function.  However, this advantage is offset by
the fact that we provide a covariant Green function for the
Maxwell equation.

This paper is organized as follows.  In Sec.~\ref{sec:maxdirac}
the vector Maxwell equations and the Dirac equation are stated
to define notation.  The algebra of three-component spin
matrices is reviewed in Sec.~\ref{sec:spinmat}, where both a
spherical basis, which is the direct extension of the Pauli
matrices to three components, and a Cartesian basis with real
components, are defined.  The Maxwell equations are written in
terms of the spherical spin matrices and combined into the Dirac
equation form in Sec.~\ref{sec:mmaxeq}.  In
Sec.~\ref{sec:tlfields}, transverse and longitudinal projection
operators are defined and used to separate the Maxwell equations
and solutions into the corresponding disjoint sectors.  Lorentz
invariance is addressed in Sec.~\ref{sec:lorentz}, where
transformations of the coordinates and derivatives,
transformations of the Maxwell equation, and transformations of
the solutions are explicitly written.  In Sec.~\ref{sec:pwf},
plane-wave solutions, which are eigenfunctions of the momentum
operator as well as the Hamiltonian, are given for both
transverse and longitudinal states, and the set of solutions is
shown to be complete.  The explicit action of Lorentz
transformations on the plane-wave solutions is described.
Properties of normalizable wave packets formed from the
plane-wave solutions are illustrated.  The angular-momentum
operator and the corresponding eigenfunctions are given and
shown to be complete in Sec.~\ref{sec:pswf}.  In
Sec.~\ref{sec:prop}, the Maxwell Green function is written as an
integral over the plane-wave solutions in a form analogous to
the Dirac Green function.  As examples of applications of the
Maxwell Green function, formulas for radiation from a point
dipole source and from a Dirac current source are derived in
Sec.~\ref{sec:appmpf}.  A summary of the main points of the
paper is in Sec.~\ref{sec:summary} and brief concluding remarks
are made in Sec.~\ref{sec:conclusion}.

The relation of the present study to earlier work is indicated
in the sections where the particular topics are discussed.  

\section{Three-vector Maxwell equations and the Dirac equation}
\elabel{sec:maxdirac}

The Maxwell equations in vacuum, in the
International System of Units (SI), are
\begin{eqnarray}
\bm{ \nabla \cdot E } &=& \frac{\rho}{\epsilon_0}, \elabel{eq:maxeq1}\\
\bm{ \nabla \times B } - \frac{1}{c^2} \frac{\partial \bm E}{ \partial
t}&=& \mu_0\bm J, \elabel{eq:maxeq2}\\
\bm{ \nabla \times E } + \frac{\partial \bm B}{ \partial t} &=& 0,
\elabel{eq:maxeq3}\\
\bm{ \nabla \cdot B } &=& 0, \elabel{eq:maxeq4}
\end{eqnarray}
where $\bm E$ and $\bm B$ are the electric and magnetic fields,
$\rho$ and $\bm J$ are the charge and current densities,
$\epsilon_0$ and $\mu_0$ are the electric and magnetic
constants, and $c = (\epsilon_0\mu_0)^{-1/2}$ is the speed of
light.  The continuity equation 
\begin{eqnarray}
\frac{\partial \rho }{ \partial t} + \bm{\nabla\cdot J} &=& 0
\elabel{eq:ce}
\end{eqnarray}
follows from Eqs.~(\ref{eq:maxeq1}) and (\ref{eq:maxeq2}).

The form of the Maxwell equations considered here is analogous
to the Dirac equation for the electron.  The Dirac wave function
$\phi(x)$ is a four-component column matrix that is a function
of the four-vector $x$.  For a free electron, the Dirac equation
is
\begin{eqnarray}
\left(\,\ri \, \hbar \gamma^\mu \partial_\mu -m_{\rm e}c\right)\phi(x) =
0,
\elabel{eq:dirac}
\end{eqnarray}
where $\hbar$ is the Planck constant divided by $2\pi$, $m_{\rm e}$ is
the mass of the electron, $\gamma^\mu$, $\mu=0,1,2,3$, are the
$4\times4$ Dirac
gamma matrices, given by
\begin{eqnarray}
\gamma^0 = \left(\begin{array}{rrr} I && 0 \\ 
0 && -I \end{array}\right)\! ; \quad
\gamma^i = \left(\begin{array}{rrr} 0 && \sigma^i \\
-\sigma^i && 0 \end{array}\right)\!, \ i=1,2,3,
\elabel{eq:gammas}
\\ \nonumber
\end{eqnarray}
$I$ and $0$ are the $2\times2$ identity and zero matrices,
$\sigma^i$, $i=1,2,3$, are the
Pauli spin matrices
\begin{eqnarray}
\sigma^1 = \left(\begin{array}{rrr} 0 && 1 
\\ 1 && 0 \end{array}\right)\!, \
\sigma^2 = \left(\begin{array}{rrr} 0 && -\ri 
\\ \ri && 0 \end{array}\right)\!, \
\sigma^3 = \left(\begin{array}{rrr} 1 && 0 
\\ 0 && -1 \end{array}\right)\!, \
\elabel{eq:pauli}
\\ \nonumber
\end{eqnarray}
and the derivatives $\partial_\mu$ are
\begin{eqnarray}
\partial_0 = \frac{\partial}{\partial ct}; \quad
\partial_i = \frac{\partial}{\partial x^i}, \ i = 1,2,3.
\end{eqnarray}
We take the metric tensor $g^{\mu\nu}$ to be
\begin{eqnarray}
g^{00} = 1; \quad g^{ii} = -1, \ i = 1,2,3; 
\quad g^{\mu\nu} = 0, \ \mu\ne\nu. 
\quad
\end{eqnarray}
In terms of the spin matrices, the derivative term in
Eq.~(\ref{eq:dirac}) can be written as
\begin{eqnarray}
\gamma^\mu\partial_\mu &=& \left(\begin{array}{ccc}
I\,\ft{\partial}{\partial ct} && \bm\sigma\cdot\bm\nabla
\\
-\bm\sigma\cdot\bm\nabla && -I\,\ft{\partial}{\partial ct}
\end{array}\right) \!.
\end{eqnarray}

The Pauli spin matrices act on two-component spin matrices in
the electron wave function.  Oppenheimer has suggested that
since the Maxwell equations involve three-vectors,
three-component matrices should be considered for constructing a
photon wave function  \cite{1931004}.  Here we implement such an
extension by replacing the Pauli spin matrices in the Dirac
equation by the analogous $3\times3$ matrices described in the
next section.

\section{Three-component spin matrices}
\elabel{sec:spinmat}

As is well known, three-vectors and operations among them are
interchangeable with three-component matrices and matrix
operations.  In this section, formulas for these matrices
relevant to subsequent work are given.  Some of these formulas
have been given in \cite{1994171}.  It is useful to define
both Cartesian and spherical matrices to represent
three-vectors.

The Cartesian matrix representing a vector $\bm a$ may be written as
\begin{eqnarray}
\bm a\rc &=& \left(\begin{array}{c} a^1 \\ a^2 \\ a^3
\end{array}\right)
\end{eqnarray}
where $a^1,~a^2,~a^3$ are the rectangular components of the
vector $\bm a$, and a spherical representation is denoted by
\begin{eqnarray}
\bm a\rs &=& 
 \bm M  \bm a\rc ,
\elabel{eq:as}
\end{eqnarray}
where $\bm M$ is a $3\times3$ unitary matrix specified in the
following.  The dot product of two vectors is
\begin{eqnarray}
\bm a\cdot\bm b &=& \bm a\rc^\dagger\bm b\rc
= \bm a\rs^\dagger\bm b\rs,
\elabel{eq:dot}
\end{eqnarray}
where $\dagger$ denotes the combined operations of matrix
transposition and complex conjugation.

Explicit Hermitian $\bm \tau$ matrices ($\bm\tau^\dagger = \bm
\tau$), which are $3\times3$ versions of the Pauli matrices, are
obtained by taking $\tau^3$ to be diagonal 
\begin{eqnarray}
\tau^3 &=& \left(\begin{array}{rrrrr} 1    &&    0     &&     0
\\ 0    &&    0     &&     0   \\ 0    &&    0     &&    -1
\end{array}\right) \elabel{eq:t3} 
\end{eqnarray} 
and applying
appropriate rotation matrices to obtain $\tau^1$ and $\tau^2$:
\begin{eqnarray} 
\elabel{eq:t1} 
\tau^1 &=& \bm
\fD^{(1)}(\{0,\fr{\pi}{2},0\})\,\tau^3\,\bm
\fD^{(1)}(\{0,-\fr{\pi}{2},0\}) =
\frac{1}{\sqrt{2}} \left(\begin{array}{rrrrr} 0   && 1  &&     0
\\ 1 &&   0     &&  1 \\ 0   && 1  &&     0
\end{array}\right)\!, \\[12 pt] \elabel{eq:t2} 
\tau^2 &=& \bm \fD^{(1)}(\{0,0,\fr{\pi}{2}\})\,\tau^1\,\bm
\fD^{(1)}(\{0,0,-\fr{\pi}{2}\})
 =
\frac{\ri}{\sqrt{2}} \left(\begin{array}{rrrrr} 0    &&  -1  &&
0   \\ 1 &&    0     &&  -1 \\ 0    && 1  &&     0
\end{array}\right)\!, 
\end{eqnarray} 
where $\bm \fD^{(1)}(\{\alpha,\beta,\gamma\})$ is the $j=1$
representation of the rotation group, parameterized by the Euler
angles $\alpha,\beta,\gamma$ \citep{1959010}.  In particular,
$\bm \fD^{(1)}(\{0,\fr{\pi}{2},0\})$ represents the rotation
about the 2 axis by the angle $\pi/2$ and $\bm
\fD^{(1)}(\{0,0,\fr{\pi}{2}\})$ represents the rotation about
the 3 axis by the angle $\pi/2$.  The same rotations starting
from $\sigma^3$, with the $j=\fr{1}{2}$ representation,
reproduce $\sigma^1$ and $\sigma^2$.  The $\bm \tau$ matrices
are related by 
\begin{eqnarray} 
\left[\tau^i,\tau^j\right] = \ri \,
\epsilon_{ijk} \, \tau^k, 
\end{eqnarray} 
where $\epsilon_{ijk}$ is the Levi-Civita symbol.\footnote{The
tau matrices defined by Oppenheimer \cite{1931004} are $\bm
N^\dagger \tau^i \bm N$, where $\bm N = \left(\begin{array}{ccc}
1 & 0 & 0 \\ 0 & \ri & 0 \\ 0 & 0 & -1 \\ \end{array}\right)$.
[The minus sign in $\tau^2$ in Eq.~(10) of that paper apparently
is a typographical error, as indicated by inspection of
Eq.~(11).] The matrices defined by Majorana \cite{1974037} are
$\bm M^\dagger \tau^i \bm M$, where $\bm M$ is given in
Eq.~(\ref{eq:meq}).}

The cross product of two vectors $\bm a$ and $\bm b$
can be written in terms of the scalar product of the tau
matrices with the vector $\bm a$
\begin{eqnarray}
\bm \tau\cdot\bm a &=& \tau^i\,a^i
\end{eqnarray}
acting on the spherical matrix for the vector $\bm b$ as
\begin{eqnarray}
\bm \tau\cdot\bm a \ \bm b\rs &=& 
\ri\left(\bm a\times\bm b\right)\rs,
\elabel{eq:cross}
\end{eqnarray}
provided the matrix $\bm M$ in Eq.~(\ref{eq:as}) is suitably
chosen.  To determine $\bm M$, we take the Cartesian
definition
\begin{eqnarray}
(\bm a\times\bm b)^i &=& \epsilon_{ijk}a^jb^k
\end{eqnarray}
and write Eq.~(\ref{eq:cross}) as
\begin{eqnarray}
\bm \tau\cdot\bm a \, \bm M \, \bm b\rc &=& 
\ri \bm M(\bm a\times\bm b)\rc.
\\ \nonumber
\end{eqnarray}
Imposing the requirement that this equation be valid for any vectors
$\bm a$ and $\bm b$ fixes $\bm M$, up to a phase factor, to be
\begin{eqnarray}
\bm M &=& \frac{1}{\sqrt{2}}\left(\begin{array}{ccc}
 -1 & \ri & 0 \\
 0 &   0  & \sqrt{2} \\
 1 &  \ri & 0 \end{array}\right),
 \elabel{eq:meq}
\end{eqnarray}
which yields
\begin{eqnarray}
\bm a\rs
&=& \left(\begin{array}{c}-{1\over\sqrt{2}}(a^1 - {\rm i}\,a^2) \\
a^3 \\
{1\over\sqrt{2}}(a^1 + {\rm i}\,a^2)\end{array}\right).
\elabel{eq:asph}
\end{eqnarray}

Consequences of Eq.~(\ref{eq:cross}) are
\begin{eqnarray}
\bm \tau\cdot \bm a \ \bm a\rs &=& 0,
\elabel{eq:tda}
\\
\bm \tau\cdot \bm a \ \bm b\rs  + \bm \tau\cdot \bm b \ \bm a\rs &=& 0,
\elabel{eq:asum}
\\
\bm a\rs^\dagger \, \bm \tau\cdot \bm b \ \bm c\rs &=& \ri \,
\bm a \cdot(\bm b \times \bm c),
\\
(\bm\tau\cdot\bm a\,\bm c\rs)\cdot
(\bm\tau\cdot\bm b\,\bm d\rs) &=&
(\bm a\times \bm c)\cdot(\bm b \times\bm d)
\nonumber \\ &=&
\bm a\cdot\bm b \ \bm c\cdot\bm d
-\bm a\cdot\bm d \ \bm c\cdot\bm b .
\elabel{eq:crcr}\quad
\end{eqnarray}
Equation~(\ref{eq:crcr}) can be written as
\begin{eqnarray}
\bm c\rs^\dagger(\bm \tau\cdot\bm a)^\dagger
\bm \tau\cdot\bm b\,\bm d\rs &=&
\bm c\rs^\dagger(\bm a\cdot\bm b- \bm b\rs\,\bm a\rs^\dagger)\bm
d\rs \quad
\end{eqnarray}
for any vectors $\bm c$ and $\bm d$, which yields the relation
\begin{eqnarray}
(\bm\tau\cdot\bm a)^\dagger  \bm \tau\cdot \bm b &=& 
\bm a\cdot\bm b - \bm b\rs\bm a\rs^\dagger ,
\elabel{eq:bmeq}
\end{eqnarray}
where it is understood that the first term on the right includes
the $3\times3$ identity matrix as a factor and the second term
is also a $3\times3$ matrix.  If $\bm a\rc$ has real components,
then
\begin{eqnarray}
(\bm \tau\cdot\bm a)^\dagger &=&  \bm \tau\cdot\bm a,
\\
(\bm \tau\cdot\bm a)^3 &=& \bm a^2 \ \bm \tau\cdot\bm a,
\end{eqnarray}
where $\bm a^2 = \bm a\cdot \bm a$ is the ordinary real vector
scalar product.  

Real Cartesian tau matrices $\bm{\tilde \tau}$ may be defined so
that
\begin{eqnarray} \bm{\tilde\tau}\cdot\bm a \ \bm b\rc
&=& \left(\bm a\times\bm b\right)\rc.  \elabel{eq:ccross}
\end{eqnarray}
This relation follows from Eqs.~(\ref{eq:cross}) and
(\ref{eq:as}) with the definition
\begin{eqnarray} \tilde\tau^i &=& -\ri
\bm M^\dagger\,\tau^i \,\bm M , \quad i = 1,2,3.  
\end{eqnarray}
These matrices are antisymmetric $\bm{\tilde \tau}^\top = -
\bm{\tilde \tau}$, where $\top$ denotes matrix transposition, in
contrast to $\bm \tau^\dagger = \bm \tau$.
For vectors $\bm a$ and $\bm b$
with real Cartesian components, we have
\begin{eqnarray}
\bm{\tilde \tau}\cdot\bm a \ \bm{\tilde \tau}\cdot \bm b &=& 
\bm b\rc\bm a\rc^\top -
\bm a\cdot \bm b,
\\
(\bm{\tilde \tau}\cdot\bm a)^3 &=& -\bm a^2\,(\bm{\tilde \tau}\cdot\bm a),
\\
(\bm{\tilde \tau}\cdot\bm a)^{ij} &=& -\epsilon_{ijk}a^k.
\end{eqnarray}
The matrix
\begin{eqnarray}
\bm{\tilde\tau}\cdot \bm a &=& \left(\begin{array}{ccc}
 0   & -a^3 &  a^2 \\
 a^3 &   0  & -a^1 \\
-a^2 & \ a^1 &   0 
\end{array}\right)
\elabel{eq:lrp}
\end{eqnarray}
has the form of the lower right portion of the electromagnetic
field-strength tensor $F^{\mu\nu}$, as given in
\cite{1998165} for example.  

\section{Matrix Maxwell equation}
\elabel{sec:mmaxeq}

In terms of the notation of the previous section, the matrix
forms of the Maxwell equations in (\ref{eq:maxeq2}) and
(\ref{eq:maxeq3}), for the source-free case ($\bm J = 0$), are
\begin{eqnarray}
\ri\,\bm \tau\cdot\bm\nabla  \bm B\rs +
\frac{1}{c} \frac{\partial \bm E\rs}{ \partial
ct}&=& 0, \elabel{eq:mateq2}\\
\ri\,\bm \tau\cdot\bm\nabla  \bm E\rs -
c \frac{\partial \bm B\rs}{ \partial ct} &=& 0.
\elabel{eq:mateq3}
\end{eqnarray}
These equations may be written as two uncoupled equations
\begin{eqnarray}
\left(\bm I\frac{\partial}{ \partial ct} + \bm \tau\cdot \bm\nabla\right)
\left(\bm E\rs + \ri \, c \bm B\rs\right) = 0,
\elabel{eq:old1}
\\
\left(\bm I\frac{\partial}{ \partial ct} - \bm \tau\cdot \bm\nabla\right)
\left(\bm E\rs - \ri \, c \bm B\rs\right) = 0,
\elabel{eq:old2}
\end{eqnarray}
where $\bm I$ is the $3\times3$ identity matrix.  In the
Cartesian basis, Eqs.~(\ref{eq:old1}) and (\ref{eq:old2}) are
\begin{eqnarray}
\left(\bm I\frac{\partial}{ \partial ct} 
+ \ri \bm{\tilde\tau}\cdot \bm\nabla\right)
\left(\bm E\rc + \ri \, c \bm B\rc\right) = 0,
\elabel{eq:cold1}
\\
\left(\bm I\frac{\partial}{ \partial ct} 
- \ri \bm{\tilde\tau}\cdot \bm\nabla\right)
\left(\bm E\rc - \ri \, c \bm B\rc\right) = 0.
\elabel{eq:cold2}
\end{eqnarray}
In these expressions, it is evident that for real electric and
magnetic fields, Eqs.~(\ref{eq:cold1}) and (\ref{eq:cold2}) are
complex conjugates of each other and reduce to a single complex
equation.  It was recognized in lectures by Riemann in the
nineteenth century that this complex combination of $\bm E$ and
$\bm B$ is a solution of a single equation \cite{1901001}.  This
fact was also discussed in \cite{1907002, 1907003} and is
included in many works up to the present.
Equations~(\ref{eq:cold1}) and (\ref{eq:cold2}) may be
interpreted as Maxwell equations for right- and left-circularly
polarized radiation, analogous to the Weyl equations for right-
and left-handed neutrino fields \cite{1964011,1994171}.

However, in this paper, we consider the more restrictive case of
complex electric and magnetic fields that are simultaneously
solutions of both Eqs~(\ref{eq:cold1}) and (\ref{eq:cold2}), or
equivalently both Eqs.~(\ref{eq:mateq2}) and (\ref{eq:mateq3}),
for any polarization of radiation.  The question of whether such
solutions can be found is answered by their explicit
construction in subsequent sections of the paper.  To formulate
this approach, we follow the Dirac equation and write
\begin{eqnarray}
\left(\begin{array}{ccc}
\bm I\,\ft{\partial}{\partial ct} && \bm\tau\cdot\bm\nabla
\\
-\bm\tau\cdot\bm\nabla && -\bm I\,\ft{\partial}{\partial ct}
\end{array}\right) \left(\begin{array}{c}
\bm E\rs \\ \ri \, c \bm B\rs \msp \end{array}\right) = 0,
\elabel{eq:block6}
\end{eqnarray}
which is a restatement of Eqs.~(\ref{eq:mateq2}) and
(\ref{eq:mateq3}) in the form of the Dirac equation for an
electron wave function.  It is a matrix equation with six
components that may be viewed as a single equation equivalent to
Eqs.~(\ref{eq:mateq2}) and (\ref{eq:mateq3}) for any
polarization of the fields.  Any complex solution of
Eq.~(\ref{eq:block6}) is a solution of both Eqs.~(\ref{eq:old1})
and (\ref{eq:old2}).  Similar wave functions have been discussed
in \cite{1994172, 2007351, wang}.  It should be noted that this
formulation is different from the six-component form considered
by Oppenheimer in which the upper-three components and
lower-three components represent opposite helicity states
\cite{1931004}.

If we define $6\times6$ gamma matrices by
\begin{eqnarray}
\gamma^0 = \left(\begin{array}{rrr} \bm I && \0 \\ 
\0 && -\bm I \end{array}\right)\! ; \quad
\gamma^i = \left(\begin{array}{rrr} \0 && \tau^i \\
-\tau^i && \0 \end{array}\right)\!, \ i=1,2,3, \quad
\elabel{eq:ggammas}
\\ \nonumber
\end{eqnarray}
where $\0$ is the $3\times3$ zero matrix, and
write
\begin{eqnarray}
\iP(x) &=& \left(\begin{array}{c}
\bm E\rs(x) \\ \ri \, c \bm B\rs(x) \msp \end{array}\right),
\elabel{eq:dwf}
\end{eqnarray}
then Eq.~(\ref{eq:block6}) takes the covariant Dirac equation form
\begin{eqnarray}
\gamma^\mu\partial_\mu \iP(x) &=& 0,
\elabel{eq:dehom}
\end{eqnarray}
which provides a concise expression for two of the Maxwell equations.
We can also write this as
\begin{eqnarray}
\overline{\iP}(x) 
\overleftarrow{\partial}_\mu 
{\gamma^\mu}
&=& 0,
\elabel{eq:dehomc}
\end{eqnarray}
where $\overline{\iP}(x) = \iP^\dagger(x)\,\gamma^0$ and
$\overleftarrow{\partial}_\mu$ denotes differentiation of the
function to the left.  Although these equations are simply
algebraic rearrangements of the two Maxwell equations, the
resemblance to the Dirac equation and wave function is
suggestive of a form that photon wave functions might take.

It is of interest to note that for solutions of the Dirac
equation for the hydrogen atom, the lower two components are
small and approach zero in the nonrelativistic limit, {\it
i.e.,} as the velocity of the bound electron approaches zero.
Similarly, for local electromagnetic fields generated by moving
charges, the magnetic field, given by the lower three components
of $\iP$, also approaches zero in the limit as the velocity of
the charges approaches zero.

To take source currents into account,
Eq.~(\ref{eq:maxeq2}) is written as
\begin{eqnarray}
\ri\,\bm \tau\cdot\bm\nabla  \bm B\rs +
\frac{1}{c} \frac{\partial \bm E\rs}{ \partial
ct}&=& -\mu_0\bm J\rs, \elabel{eq:smateq2}
\end{eqnarray}
and a source term $\iX$ is defined to be
\begin{eqnarray}
\iX(x) &=& \left(\begin{array}{c}
-\mu_0c\bm J\rs(x) \\ \0 \msp \end{array}\right),
\elabel{eq:des}
\end{eqnarray}
where $\0$ is a $3\times1$ matrix of zeros.
This yields the expressions
\begin{eqnarray}
\gamma^\mu\partial_\mu \iP(x) &=& \iX(x)
\elabel{eq:dms}
\end{eqnarray}
and
\begin{eqnarray}
\overline{\iP}(x) 
\overleftarrow{\partial}_\mu 
{\gamma^\mu}
&=& \overline{\iX}(x) ,
\elabel{eq:dmsc}
\end{eqnarray}
either of which is referred to as the Maxwell equation here.
The source term in Eq.~(\ref{eq:dms}) or (\ref{eq:dmsc})
represents a fundamental difference between the Dirac equation
and the Maxwell equation, as mentioned in Sec.~\ref{sec:intro}
\cite{1931004}.

In this framework, an energy-momentum density operator is
\begin{eqnarray}
p^\mu &=& \frac{\epsilon_0}{2c} \, \gamma^\mu,
\elabel{eq:emdo}
\end{eqnarray}
which gives
\begin{eqnarray}
\overline{\iP} \,c p^0 \iP &=& 
\frac{1}{2}\left(\epsilon_0 |\bm E|^2 + 
\frac{1}{\mu_0}|\bm B|^2\right)  = u,
\\
\overline{\iP} \, \bm p \iP &=&
\frac{\ri\,\epsilon_0}{2}
\left(\bm E\rs^\dagger \bm \tau \bm B\rs -
\bm B\rs^\dagger \bm \tau \bm E\rs \right)
\nonumber\\&=&
\frac{1}{c^2\mu_0}\,{\rm Re}\,\bm E \!\times \bm B^*
= \bm g.
\elabel{eq:pmd}
\end{eqnarray}
Eqs.~(\ref{eq:dms}) and (\ref{eq:dmsc}) imply that
\begin{eqnarray}
\partial_\mu \overline{\iP}(x){\gamma^\mu}\iP(x) &=& 
  \overline{\iX}(x)\iP(x) 
  + \overline{\iP}(x)\,\iX(x) , \quad
\elabel{eq:derivpsis}
\end{eqnarray}
which is a complex form of the Poynting theorem [see
Eq.~(\ref{eq:dot})]
\begin{eqnarray}
\frac{\partial u}{\partial t} + \bm\nabla\cdot\bm S = 
-{\rm Re}\, \bm E\cdot\bm J ,
\end{eqnarray}
where
\begin{eqnarray}
\bm S = c^2\bm g,
\elabel{eq:srelg}
\end{eqnarray}
which gives the conventional result if the fields and current are
real \cite{1998165}.

\section{Transverse and longitudinal fields}
\elabel{sec:tlfields}

To make a Helmholtz decomposition of electromagnetic fields
expressed in matrix form into transverse and longitudinal
components, we define $3\times3$ matrix transverse and
longitudinal Hermitian projection operators $\bm{\iPi}\rs\rT(\bm
a)$ and $\bm{\iPi}\rs\rL(\bm a)$ to be
\begin{eqnarray}
\bm{\iPi}\rs\rT(\bm a) &=& 
\frac{(\bm\tau\cdot\bm a)^\dagger(\bm\tau\cdot\bm a)}
{\bm a\cdot\bm a} ,
\\
\bm{\iPi}\rs\rL(\bm a) &=& \frac{\bm a\rs\bm a\rs^\dagger}
{\bm a\cdot\bm a} .
\end{eqnarray}
Based on identities in Sec.~\ref{sec:spinmat}, these operators
have the following properties:
\begin{eqnarray}
\left[\bm{\iPi}\rs\rT(\bm a)\right]^2 &=& 
\bm{\iPi}\rs\rT(\bm a) ,
\\
\left[\bm{\iPi}\rs\rL(\bm a)\right]^2 &=& 
\bm{\iPi}\rs\rL(\bm a) ,
\\
\bm{\iPi}\rs\rT(\bm a) + 
\bm{\iPi}\rs\rL(\bm a) &=& \bm I ,
\\
\bm{\iPi}\rs\rT(\bm a) \, 
\bm{\iPi}\rs\rL(\bm a) &=& 0 ,
\\
\bm{\iPi}\rs\rT(\bm a)\, \bm a\rs &=& 0 ,
\\
\bm{\iPi}\rs\rL(\bm a)\, \bm a\rs &=& \bm a\rs .
\end{eqnarray}
Acting on the matrix of an arbitrary vector $\bm b$, the
operators project the components perpendicular to and parallel
to the argument $\bm a$
\begin{eqnarray}
\bm{\iPi}\rs\rT(\bm a)\,\bm b\rs &=& \bm b\rs - 
\frac{\bm a\cdot\bm b}{\bm a\cdot\bm a} \, \bm a\rs,
\\
\bm{\iPi}\rs\rL(\bm a)\,\bm b\rs &=& 
\frac{\bm a\cdot\bm b}{\bm a\cdot\bm a} \, \bm a\rs.
\end{eqnarray}

In addition to these algebraic relations, usefulness of the
projection operators arises from an extension to include
differential and integral operations acting on coordinate-space
functions.  Formally, we write 
\begin{eqnarray}
\bm{\iPi}\rs\rT(\bm\nabla) &=& \frac{(\bm\tau\cdot\bm\nabla)^2}{\bm\nabla^2} \, ,
\elabel{eq:tpjop}
\\
\bm{\iPi}\rs\rL(\bm\nabla) &=&
\frac{\bm\nabla\rs\bm\nabla\rs^\dagger}{\bm\nabla^2} \, ,
\elabel{eq:lpjop}
\end{eqnarray}
which takes into account that fact that $\bm \nabla$ has real
Cartesian components (in the sense that they give real values
when acting on a real function).
The inverse Laplacian is defined by the relation
\begin{eqnarray}
\frac{1}{\bm\nabla^2}\,f(\bm x) &=& -\frac{1}{4\pi} \int{\rm d}\,\bm
x^\prime \, \frac{1}{|\bm x - \bm x^\prime|} \,f(\bm x^\prime),
\elabel{eq:invlap}
\end{eqnarray}
which yields
\begin{eqnarray}
\bm\nabla^2\,\frac{1}{\bm\nabla^2}\,f(\bm x) &=& -\frac{1}{4\pi} \int{\rm d}\,\bm
x^\prime \, \bm\nabla^2 \frac{1}{|\bm x - \bm x^\prime|} \,f(\bm x^\prime)
 = f(\bm x),
\elabel{eq:nabsnab}
\end{eqnarray}
based on
\begin{eqnarray}
\bm\nabla^2\, \frac{1}{|\bm x - \bm x^\prime|} &=& 
-4\pi\,\delta(\bm x - \bm x^\prime),
\end{eqnarray}
where
\begin{eqnarray}
\delta(\bm x - \bm x^\prime) &=& 
\delta(x^1 - x^{\prime 1}) \,
\delta(x^2 - x^{\prime 2}) \,
\delta(x^3 - x^{\prime 3}). \qquad
\end{eqnarray}
Equation~(\ref{eq:nabsnab}) indicates that the Laplacian operator follows
analogs of the rules of algebra in this context.  For example, we have
\begin{eqnarray}
\left[\bm{\iPi}\rs\rT(\bm \nabla)\right]^2 &=&
\left[\frac{(\bm\tau\cdot\bm\nabla)^2}{\bm\nabla^2}\right]^2
=
\frac{\bm\nabla^2\,(\bm\tau\cdot\bm\nabla)^2}{\bm\nabla^2\bm\nabla^2}
\nonumber\\ &=&
\frac{(\bm\tau\cdot\bm\nabla)^2}{\bm\nabla^2}
=\bm{\iPi}\rs\rT(\bm \nabla)\,,
\end{eqnarray}
where either simply canceling the $\bm \nabla^2$ factors in the
numerator and denominator or applying the definition in
Eq.~(\ref{eq:invlap}) for the operators acting on a suitable
function gives the same result.  Transverse and longitudinal
components of the electric and magnetic fields and the current
density are identified by writing
\begin{eqnarray}
\bm F\rs &=& \bm F\rs\rT + \bm F\rs\rL ,
\end{eqnarray}
where
\begin{eqnarray}
\bm F\rs\rT &=& \bm{\iPi}\rs\rT(\bm \nabla)\bm F\rs ,
\\
\bm F\rs\rL &=& \bm{\iPi}\rs\rL(\bm \nabla)\bm F\rs ,
\end{eqnarray}
and $\bm F\rs$ may be any of $\bm E\rs$, $\bm
B\rs$, or $\bm J\rs$.

The separation of the Maxwell equations into two independent
sets of equations which involve either transverse components or
longitudinal components takes the following form.  In terms of
the spherical matrices, Eq.~(\ref{eq:maxeq1}) is
\begin{eqnarray}
\bm\nabla\rs^\dagger \bm E\rs\rL  &=& 
\frac{\rho}{\epsilon_0}, \elabel{eq:smaxeq1}
\end{eqnarray}
where the transverse component of the electric field is absent,
because $ \bm\nabla\rs^\dagger \bm{\iPi}\rs\rT(\bm \nabla) = 0$.
Equations~(\ref{eq:maxeq1}) and (\ref{eq:smaxeq1}) are
equivalent, in the sense that each can be derived from the
other; Eq.~(\ref{eq:maxeq1}) follows from Eq.~(\ref{eq:smaxeq1})
if the vanishing transverse component is added to the latter
equation.  In the separated form, it is evident that the
equation neither contains information about or places any
constraint on the transverse component $\bm E\rs\rT$.  The
transverse and longitudinal projection operators acting on
Eq.~(\ref{eq:smateq2}), the matrix form of
Eq.~(\ref{eq:maxeq2}), yield
\begin{eqnarray}
\ri\,\bm \tau\cdot\bm\nabla  \bm B\rs\rT +
\frac{1}{c} \frac{\partial \bm E\rs\rT}{ \partial
ct}&=& -\mu_0\bm J\rs\rT, \elabel{eq:smateq2t}
\\
\frac{1}{c} \frac{\partial \bm E\rs\rL}{ \partial
ct} &=& -\mu_0\bm J\rs\rL
\elabel{eq:smateq2l}
\end{eqnarray}
respectively, which take into account the commutation relation
$[\bm{\iPi}\rs\rT(\bm \nabla), \bm \tau\cdot\bm\nabla]=0$ and
the fact that $\bm{\iPi}\rs\rL(\bm
\nabla)\,\bm\tau\cdot\bm\nabla = 0$.  Together, these equations
are equivalent to Eq.~(\ref{eq:smateq2}) which can be restored
by writing the sum of Eq.~(\ref{eq:smateq2t}) and Eq.~(\ref{eq:smateq2l})
and adding the term that vanishes.  Evidently, this pair of
equations is independent of $\bm B\rs\rL$.  Similarly,
Eq.~(\ref{eq:maxeq3}), or equivalently Eq.~(\ref{eq:mateq3}),
can be written as the pair
\begin{eqnarray}
\ri\,\bm \tau\cdot\bm\nabla  \bm E\rs\rT -
c \frac{\partial \bm B\rs\rT}{ \partial ct} &=& 0,
\elabel{eq:smateq3t}
\\
\frac{\partial \bm B\rs\rL}{ \partial ct} &=& 0,
\elabel{eq:smateq3l}
\end{eqnarray}
which are independent of $\bm E\rs\rL$.
Equation~(\ref{eq:maxeq4}) takes the form
\begin{eqnarray}
\bm\nabla\rs^\dagger \bm B\rs\rL  &=& 0,
\elabel{eq:smaxeq4}
\end{eqnarray}
independent of $\bm B\rs\rT$.  The transverse and longitudinal
equations comprise two independent sets.  

Six-dimensional transverse and longitudinal projection operators
are defined by
\begin{eqnarray}
\iPi\rT(\bm \nabla) &=& \left(\begin{array}{ccc}
\bm{\iPi}\rs\rT(\bm\nabla) && \0 \\
\0 && \bm{\iPi}\rs\rT(\bm\nabla) \msp \end{array}\right),
\elabel{eq:ptop}
\\
\iPi\rL(\bm \nabla) &=& \left(\begin{array}{ccc}
\bm{\iPi}\rs\rL(\bm\nabla) && \0 \\
\0 && \bm{\iPi}\rs\rL(\bm\nabla) \msp \end{array}\right),
\elabel{eq:plop}
\end{eqnarray}
where $\0$ is the $3\times3$ matrix of zeros, $\iPi\rT(\bm
\nabla)+\iPi\rL(\bm \nabla) = \cI$, and $\cI$ is the $6\times6$
identity matrix.  The transverse equations are summarized by
writing 
\begin{eqnarray}
\gamma^\mu\partial_\mu \iP\rT(x) &=& \iX\rT(x),
\elabel{eq:dmst}
\end{eqnarray}
where
\begin{eqnarray}
\iP\rT(x) &=& \iPi\rT(\bm \nabla)\iP(x),
\\
\iX\rT(x) &=& \iPi \rT(\bm \nabla)\,\iX(x).
\end{eqnarray}
Equation~(\ref{eq:dmst}) also follows directly from
Eq.~(\ref{eq:dms}) and the fact that $\left[\iPi\rT(\bm \nabla)
, \gamma^\mu\partial_\mu\right] = 0$.  The longitudinal
equations are Eqs.~(\ref{eq:smaxeq1}), (\ref{eq:smateq2l}),
(\ref{eq:smateq3l}), and (\ref{eq:smaxeq4}), together with the
continuity equation, Eq.~(\ref{eq:ce}), which can be expressed
as 
\begin{eqnarray}
\frac{\partial \rho }{ \partial t} + \bm\nabla\rs^\dagger \bm J\rs\rL
&=& 0 .
\elabel{eq:sceq}
\end{eqnarray}
Since the continuity equation follows from Eqs.~(\ref{eq:smaxeq1}) and
(\ref{eq:smateq2l}), it is not necessary to include it in an independent
set of equations; it is listed here only to show that it provides no
restriction on $\bm J\rs\rT$.  Equations (\ref{eq:smateq3l}) and
(\ref{eq:smaxeq4}) are eliminated from consideration by taking 
\begin{eqnarray}
\bm B\rs\rL = 0.  
\elabel{eq:bl}
\end{eqnarray}
Constant fields are eliminated by the requirement that static
fields vanish at infinite distances for finite source
distributions.  However, a constant magnetic field may be
approximated by the field at the center of a current loop with a
radius that is large compared to the extent of the region of
interest.  Such a steady-state current density is transverse, as
shown by Eq.~(\ref{eq:smateq2l}), and so the magnetic field,
given by Eq.~(\ref{eq:smateq2t}), is also transverse, which is
consistent with Eq.~(\ref{eq:bl}).  A complete set of equations,
equivalent to the set of Maxwell equations, is provided by
Eqs.~(\ref{eq:dms}), (\ref{eq:smaxeq1}), and (\ref{eq:bl}),
and the transverse fields are completely described by
Eq.~(\ref{eq:dmst}).

\section{Lorentz transformations}
\elabel{sec:lorentz}

Lorentz transformations of the matrix Maxwell equation are
examined here in order to confirm that this form of the Maxwell
equations is Lorentz invariant.  We adopt the convention that
transformations apply to the physical system rather than to the
observer's coordinates.

To represent four-vector coordinates, the Cartesian matrices are
extended to include a time component $x^0 = ct$, so coordinate
vectors take the form
\begin{eqnarray}
x = \left(\begin{array}{c} x^0 \\ x^1\\ x^2\\ x^3 \end{array}\right)
=\left(\begin{array}{c} ct \\ \bm x\rc \end{array}\right).
\end{eqnarray}
We employ the Cartesian basis for coordinate and momentum
vectors and the spherical basis for fields and currents, with a
few exceptions that will be apparent.  The notation $x$
represents either the four-coordinate argument of a function or
a column matrix, depending on the context.  It is sufficient for
our purpose to consider only homogeneous Lorentz transformations
and to consider rotations and velocity transformations
separately.  These transformations acting on four-vectors leave
the scalar product
\begin{eqnarray}
x\cdot x &=&x^\top g \, x = (ct)^2 - \bm x^2 
\elabel{eq:sprod}
\end{eqnarray}
invariant, where $g$ is the metric tensor given by
\begin{eqnarray}
g = \left(\begin{array}{ccc}
1 && \0 \\
\0 && -\bm I \msp \end{array}\right).
\elabel{eq:gc}
\end{eqnarray}

A remark on notation is that a boldface $\0$ means either a
$3\times3$, a $1\times3$, or a $3\times1$ rectangular array of
zeros, as appropriate.  We take the liberty of using an ordinary
zero on the right-hand side of equations to mean whatever sort
of zero matches the left-hand side.

\subsection{Rotation of coordinates}

Rotations are parameterized by a vector $\bm u = \theta \bm{\hat u}$,
where $\bm{\hat u}$ is a unit vector in the direction of the axis of the
rotation and $\theta$ is the angle of rotation.  An infinitesimal
rotation $\delta\theta\,\bm{\hat u}$ changes the point at position $\bm
x$ to the point at position $\bm x^\prime$, where 
\begin{eqnarray}
\bm x^\prime &=& \bm x + \delta\theta\,\bm {\hat u}\times\bm x 
+ \dots \ ,
\end{eqnarray}
or
\begin{eqnarray}
\bm x\rc^\prime&=& \left(\bm I+\delta\theta\,\bm{\tilde
\tau}\cdot\bm{\hat u} \right)\bm x\rc
+ \dots \ .
\end{eqnarray}
For a finite rotation, the operation is exponentiated to give
\begin{eqnarray}
\bm x\rc^\prime &=& \re^{\bm{\tilde \tau}\cdot \bm u}\bm x\rc
=\bm R\rc(\bm u)\,\bm x\rc .
\elabel{eq:finrot}
\end{eqnarray}
Expansion of the exponential function in powers of $\theta$, taking into
account the fact that $(\bm{\tilde \tau}\cdot\bm{\hat u})^3 = 
-\bm{\tilde \tau}\cdot\bm{\hat u}$, yields
\begin{eqnarray}
\bm R\rc(\bm u) &=& \bm I + 
(\bm{\tilde \tau}\cdot\bm{\hat u})^2 \left(1-\cos{\theta}\right) +
\bm{\tilde \tau}\cdot\bm{\hat u} \,\sin{\theta}
\quad
\nonumber \\ &=&
\bm{\hat u}\rc\bm{\hat u}\rc^\top
-(\bm{\tilde \tau}\cdot\bm{\hat u})^2 \,\cos{\theta} +
\bm{\tilde \tau}\cdot\bm{\hat u} \,\sin{\theta}.
\end{eqnarray}
Evidently, $\bm R\rc^{-1}(\bm u) = \bm R\rc(-\bm u) = \bm
R\rc^\top(\bm u)$.  It is confirmed that this operator has the
appropriate action on a vector by calculating
\begin{eqnarray}
\bm x^\prime &=& \bm{\hat u} \, \bm{\hat u}\cdot \bm x
-\bm{\hat u}\times(\bm{\hat u}\times \bm x)\cos{\theta}
+ \bm{\hat u}\times \bm x \sin{\theta}.
\qquad
\elabel{eq:xtrans}
\end{eqnarray}
We use the notation 
\begin{eqnarray}
\bm x^\prime = \bm R(\bm u)\bm x
\elabel{eq:xrot}
\end{eqnarray}
to represent the transformation in Eq.~(\ref{eq:xtrans}).

Rotations of a four-vector only change the spatial coordinates
and are written as
\begin{eqnarray}
x^\prime &=& R(\bm u)\, x
=\left(\begin{array}{c} ct \\ \bm R\rc(\bm u)\,\bm x\rc \msp
\end{array}\right) ,
\end{eqnarray}
where
\begin{eqnarray}
R(\bm u) &=& \left(\begin{array}{ccc}
1 && \0 \\
\0 && \bm R\rc(\bm u) \msp \end{array}\right).
\elabel{eq:4rotmat}
\end{eqnarray}
The scalar product $x\cdot x$ is invariant under
rotations, since $\bm x^2$ is invariant.

The spatial coordinate rotation operator in the spherical basis, 
which follows from
\begin{eqnarray}
\bm R\rs(\bm u) &=& \bm{M}\bm R\rc(\bm u)\bm{M}^\dagger,
\elabel{eq:srotcrot}
\end{eqnarray}
is
\begin{eqnarray}
 \bm R\rs(\bm u) &=& \re^{-\ri\bm\tau\cdot\bm{u}}
=
\bm{\hat u}\rs\bm{\hat u}\rs^\dagger
+(\bm{\tau}\cdot\bm{\hat u})^2 \,\cos{\theta} 
- \ri \, \bm{\tau}\cdot\bm{\hat u} \,\sin{\theta},
\qquad
\elabel{eq:rots}
\end{eqnarray}
and
$\bm R\rs^{-1}(\bm u) = \bm R\rs(-\bm u) = \bm R\rs^\dagger(\bm u)$.
Starting from the geometrical constraint that the rotated cross
product of two vectors is the cross product of the rotated
vectors, written as
\begin{eqnarray}
\bm R\rs(\bm u) (\bm a\times\bm b)\rs &=&
(\bm a^\prime \times \bm b^\prime)\rs ,
\end{eqnarray}
we have 
\begin{eqnarray}
\bm R\rs(\bm u)\,\bm{\tau}\cdot\bm a\,\bm b\rs &=&
\bm{\tau}\cdot\bm a^\prime\,\bm b\rs^\prime =
\bm{\tau}\cdot\bm a^\prime \bm R\rs(\bm u)\,\bm b\rs.
\end{eqnarray}
Since this relation holds for any vector $\bm b$, it yields
\begin{eqnarray}
\bm R\rs(\bm u)\,\bm{\tau}\cdot\bm a &=&
\bm{\tau}\cdot\bm a^\prime \bm R\rs(\bm u)
\elabel{eq:tautrans}
\end{eqnarray}
and
\begin{eqnarray}
\bm R\rs(\bm u)\,\bm{\tau}\cdot\bm a \, \bm R^{-1}\rs(\bm u)&=&
\bm{\tau}\cdot\bm a^\prime .
\elabel{eq:sigvecrot}
\end{eqnarray}
A direct calculation provides the same result.

The relation between the rotated and the unrotated gradient operators is
given by
\begin{eqnarray}
{\nabla^\prime}^i &=& \frac{\partial}{\partial {x^\prime}^i}
=\frac{\partial x^j}{\partial{x^\prime}^i}\,\frac{\partial}{\partial x^j}
=\frac{\partial x^j}{\partial{x^\prime}^i}\,\nabla^j,
\end{eqnarray}
where, from Eq.~(\ref{eq:finrot}), we have
\begin{eqnarray}
\frac{\partial x^j}{\partial{x^\prime}^i} &=&
\bm R_{{\rm c}\,ji}^{-1}(\bm u) =
\bm R_{{\rm c}\,ij}(\bm u),
\end{eqnarray}
so that
\begin{eqnarray}
\bm \nabla\rc^\prime &=& \bm R\rc(\bm u)\,\bm\nabla\rc,
\end{eqnarray}
and from Eq.~(\ref{eq:srotcrot}), 
\begin{eqnarray}
\bm \nabla\rs^\prime &=& \bm R\rs(\bm u)\,\bm\nabla\rs.
\elabel{eq:nabtrans}
\end{eqnarray}
Since the spherical gradient operator transforms as a spherical vector,
we also have
\begin{eqnarray}
\bm R\rs(\bm u)\,\bm{\tau}\cdot\bm \nabla \bm R^{-1}\rs(\bm u)&=&
\bm{\tau}\cdot\bm \nabla^\prime 
\elabel{eq:tdtrans}
\end{eqnarray}
from Eq.~(\ref{eq:sigvecrot}).  Equations~(\ref{eq:nabtrans}) and
(\ref{eq:tdtrans}) imply that transverse and longitudinal projection
operators transform according to
\begin{eqnarray}
\bm{\iPi}\rs\rT(\bm \nabla^\prime) &=& \bm R\rs(\bm
u)\,\bm{\iPi}\rs\rT(\bm
\nabla)\bm R^{-1}\rs(\bm u),
\elabel{eq:ptt}
\\
\bm{\iPi}\rs\rL(\bm \nabla^\prime) &=& \bm R\rs(\bm
u)\,\bm{\iPi}\rs\rL(\bm
\nabla)\bm R^{-1}\rs(\bm u).
\elabel{eq:plt}
\end{eqnarray}
The action of the inverse Laplacian in terms of the rotated
coordinates is the same as it is for unrotated coordinates,
which follows either because $\nabla^{\prime 2}= \nabla^2$ from
Eq.~(\ref{eq:nabtrans}) or by the definition in
Eq.~(\ref{eq:invlap}), taking into account the fact that the
Jacobian for a rotation is unity.

\subsection{Velocity transformation of coordinates}

Velocity transformations are parameterized by a velocity vector
$\bm v = c\,\tanh{\zeta} \,\bm{\hat v}$.  If a space-time point
is given an infinitesimal velocity boost of $\delta\zeta\,
c\,\bm{\hat v}$, its spatial coordinate will change to
\begin{eqnarray}
\bm x^\prime &=& \bm x + \delta\zeta\,ct \,\bm{\hat v} +
\dots \ ,
\end{eqnarray}
and its time coordinate must transform in such a way that the
scalar product is invariant.  In particular, we require
$x^\prime \cdot x^\prime = x\cdot x$, which yields
\begin{eqnarray}
ct^\prime &=& ct + \delta\zeta\,\bm{\hat v}\cdot \bm x
+ \dots \ .
\end{eqnarray}
The complete infinitesimal transformation is
\begin{eqnarray}
\left(\begin{array}{c} ct^\prime \\ \bm x\rc^\prime \end{array}\right)
&=&
\left[I + \delta\zeta\left(\begin{array}{ccc}
0 && \bm{\hat v}\rc^\top \\
\bm{\hat v}\rc && \0 \end{array}\right) \right]
\left(\begin{array}{c} ct \\ \bm x\rc \end{array}\right)
+ \dots \ . \qquad
\end{eqnarray}
This may be written in terms of a $4\times4$ matrix valued
function of the velocity direction:
\begin{eqnarray}
K(\bm{\hat v}) &=& \left(\begin{array}{cc}
0 & \bm{\hat v}\rc^\top \\
\bm{\hat v}\rc & \0 \end{array}\right),
\end{eqnarray}
for which
\begin{eqnarray}
K^2(\bm{\hat v}) &=& \left(\begin{array}{cc}
1 & \0 \\
\0 & \bm{\hat v}\rc\bm{\hat v}\rc^\top \end{array}\right) 
\end{eqnarray}
and $K^3(\bm{\hat v}) = K(\bm{\hat v})$.
For a finite velocity, the transformation is exponentiated to give
\begin{eqnarray}
x^\prime &=& \re^{\zeta K(\bm{\hat v})}\, x = V(\bm v)\, x .
\elabel{eq:xvtr}
\end{eqnarray}
Expansion in powers of $\zeta$ yields
\begin{eqnarray}
V(\bm v) &=& I
+ K^2(\bm{\hat v})(\cosh{\zeta}-1)
+ K(\bm{\hat v})\sinh{\zeta}
\qquad
\nonumber\\[8 pt] &=& \left(\begin{array}{ccc}
\cosh{\zeta} && \bm{\hat v}\rc^\top\sinh{\zeta} \\
\bm{\hat v}\rc\sinh{\zeta} && \bm I
+\bm{\hat v}\rc\bm{\hat v}\rc^\top\left(\cosh{\zeta}-1\right)
\msp\end{array}\right).
\elabel{eq:cvt}
\end{eqnarray}
\msp The relations $V^\top\!(\bm v) = V(\bm v)$ and $gV(\bm v)=V^{-1}(\bm v)g$
confirm the invariance of the scalar product:
\begin{eqnarray}
x^\prime\cdot x^\prime &=& x^\top V^\top\!(\bm v)\,g\,V(\bm v)\,x =
x\cdot x.
\end{eqnarray}
The transformation yields
\begin{eqnarray}
ct^\prime &=& ct\,\cosh{\zeta} + \bm{\hat v}\cdot\bm x \sinh{\zeta},
\\
\bm x^\prime &=& \bm x + \bm{\hat v}\,\bm{\hat v}\cdot\bm x
(\cosh{\zeta}-1) + ct \, \bm{\hat v} \,\sinh{\zeta}.
\end{eqnarray}
A point with $\bm x = 0$ has the boosted velocity
\begin{eqnarray}
\frac{\bm x^\prime}{t^\prime} &=& c
\tanh{\zeta}\,\bm{\hat v} = \bm v.
\end{eqnarray}

The spherical counterpart of the operator $V(\bm v)$, in the
velocity transformation
\begin{eqnarray}
\left(\begin{array}{c} ct^\prime \\ \bm x\rs^\prime \end{array}\right)
&=& V\rs(\bm v)
\left(\begin{array}{c} ct \\ \bm x\rs \end{array}\right),
\end{eqnarray}
is
\begin{eqnarray}
V\rs(\bm v) &=& 
\left(\begin{array}{cc} 1 & \0 \\ \0 & \bm M \end{array}\right) 
V(\bm v)
\left(\begin{array}{cc} 1 & \0 \\ \0 & \bm M^\dagger \end{array}\right) 
\nonumber\\ &=& \left(\begin{array}{ccc}
\cosh{\zeta} && \bm{\hat v}\rs^\dagger\sinh{\zeta} \\
\bm{\hat v}\rs\sinh{\zeta} && I
+\bm{\hat v}\rs\bm{\hat
v}\rs^\dagger\left(\cosh{\zeta}-1\right)\msp\end{array}\right) .
\qquad
\elabel{eq:sav}
\end{eqnarray}

For the four-gradient operator, we have
\begin{eqnarray}
\partial_\mu^\prime &=& \frac{\partial}{\partial {x^\prime}^\mu}
=\frac{\partial x^\nu}{\partial{x^\prime}^\mu}\,
\frac{\partial}{\partial x^\nu}
=\frac{\partial x^\nu}{\partial{x^\prime}^\mu}\,\partial_\nu,
\elabel{eq:chain}
\end{eqnarray}
and from Eq.~(\ref{eq:xvtr}), which can be written as
\begin{eqnarray}
x &=& V^{-1}(\bm v) \, x^\prime = V(-\bm v) \, x^\prime
\end{eqnarray}
or
\begin{eqnarray}
{x^\nu} &=& V_{\nu\mu}(-\bm v)\,{x^\prime}^\mu,
\end{eqnarray}
we also have
\begin{eqnarray}
\frac{\partial x^\nu}{\partial{x^\prime}^\mu} &=&
V_{\nu\mu}(-\bm v) = V_{\mu\nu}(-\bm v),
\end{eqnarray}
which yields
\begin{eqnarray}
\partial_\mu^\prime &=& V_{\mu\nu}(-\bm v)\,\partial_\nu.
\elabel{eq:fgt}
\end{eqnarray}
If a Cartesian gradient operator is defined as
\begin{eqnarray}
\partial\rc &=& 
\left(\begin{array}{c} \ft{\partial}{\partial ct} \\ 
\msp -\bm \nabla\rc \end{array}\right),
\elabel{eq:cgo}
\end{eqnarray}
then Eq.~(\ref{eq:fgt}) gives
\begin{eqnarray}
g\,\partial\rc^\prime &=& 
V(-\bm v) \,
g\,\partial\rc
\end{eqnarray}
or
\begin{eqnarray}
\partial\rc^\prime &=& 
V(\bm v) \,
\partial\rc,
\elabel{eq:cgt}
\end{eqnarray}
since $g\,V(-\bm v)\,g = V(\bm v)$.

\subsection{Parity and time reversal of coordinates}
\elabel{ssec:patcoord}

Lorentz transformations that leave the scalar product in
Eq.~(\ref{eq:sprod}) invariant include the parity transformation
$P = g$, time reversal $T = -g$, and total inversion $PT = -I$
operations.  These transformations have the following defining
effects on the coordinate vectors:
\begin{eqnarray}
Px &=&
\left(\begin{array}{c} ct \\ -\bm x\rc \end{array}\right),
\\
Tx &=&
\left(\begin{array}{c} - ct \\ \bm x\rc \end{array}\right),
\\
PTx &=& -x.  
\end{eqnarray}
It is sufficient for the present purpose to consider only $P$
and $T$.  The coordinate derivatives transform as
\begin{eqnarray}
P\partial\rc &=& 
\left(\begin{array}{c} \ft{\partial}{\partial ct} \\ 
\msp \bm \nabla\rc \end{array}\right),
\elabel{eq:derivparity}
\\
T\partial\rc &=& 
\left(\begin{array}{c} -\ft{\partial}{\partial ct} \\ 
\msp -\bm \nabla\rc \end{array}\right).
\elabel{eq:derivtime}
\end{eqnarray}

Comparison of Eqs.~(\ref{eq:gc}) and (\ref{eq:4rotmat}) shows
that parity transformations commute with rotations.  On the
other hand, for velocity transformations, the relation
\begin{eqnarray}
PV(\bm v) &=& V(-\bm v)P
\elabel{eq:pvrel}
\end{eqnarray}
applies as it should, because the space reflection of a point
moving with a velocity $\bm v$ is a point at the reflected
position moving with a velocity $-\bm v$.  Similar conclusions
follow for time-reversal transformations.

\subsection{Rotation of $\iP(x)$}
\elabel{ssec:psirots}

The result of a rotation, parameterized by the vector $\bm u$, applied
to the field $\iP(x)$ in Eq.~(\ref{eq:dms}) is the field
$\iP^\prime(x)$ given by
\begin{eqnarray}
\iP^\prime(x) &=& \cR(\bm u)\iP\!\big(R^{-1}(\bm u)\,x\big),
\elabel{eq:solrot}
\end{eqnarray}
where $\cR(\bm u)$ is a $6\times6$ matrix that gives the local
transformation of the field $\iP(x)$ at any point $x$.  The
inverse transformation of the argument on the right-hand-side
takes into account the fact that the transformed field at the
point $x$ originated from the field at the point that is mapped
into $x$ by the transformation.  Lorentz invariance is confirmed
by showing that the transformed field satisfies the same
equation as the original field.  We expect the current to
transform in the same way as $\iP$ and write  
\begin{eqnarray}
\iX^\prime(x) &=& \cR(\bm u)\,\iX\!\big(R^{-1}(\bm u)\,x\big).
\elabel{eq:sourot}
\end{eqnarray}
The objective is to show that
\begin{eqnarray}
\gamma^\mu\partial_\mu \iP^\prime(x) &=& \iX^\prime(x),
\elabel{eq:transeq}
\end{eqnarray}
for a suitable transformation $\cR(\bm u)$.
In terms of the original field and source, Eq.~(\ref{eq:transeq}) is
given by
\begin{eqnarray}
\gamma^\mu\partial_\mu \cR(\bm u)\iP\big(R^{-1}(\bm u)\,x\big) &=&
\cR(\bm u)\,\iX\big(R^{-1}(\bm u)\,x\big) \qquad
\end{eqnarray}
or
\begin{eqnarray}
\gamma^\mu\partial_\mu^\prime \cR(\bm u)\iP(x) &=&
\cR(\bm u)\,\iX(x), \qquad
\end{eqnarray}
where the variable $x$ has been replaced by $x^\prime = R(\bm u)\,x$.
Thus Eq.~(\ref{eq:transeq}) will follow if
\begin{eqnarray}
\cR^{-1}(\bm u)\gamma^\mu\partial_\mu^\prime \cR(\bm u) &=&
\gamma^\mu\partial_\mu.
\elabel{eq:opeq}
\end{eqnarray}
We expect $\cR(\bm u)$ to be of the form
\begin{eqnarray}
\cR(\bm u) &=& \left(\begin{array}{cc} \bm R\rs(\bm u) & \0  \\
\0 & \bm R\rs(\bm u) \end{array}\right),
\elabel{eq:crot}
\end{eqnarray}
which yields
\begin{eqnarray}
\cR^{-1}(\bm u)\gamma^\mu\partial_\mu^\prime \cR(\bm u) &=&
\left(\begin{array}{cc}
\bm I\,\ft{\partial}{\partial ct} & 
\bm R^{-1}\rs(\bm u)\, \bm\tau\cdot\bm\nabla^\prime \bm R\rs(\bm u)
\\
- \bm R^{-1}\rs(\bm u)\, \bm\tau\cdot\bm\nabla^\prime \bm R\rs(\bm u)
&
-\bm I\,\ft{\partial}{\partial ct}
\msp\end{array}\right) ,
\end{eqnarray}
so Eq.~(\ref{eq:opeq}) follows from
\begin{eqnarray}
\bm R^{-1}\rs(\bm u)\, \bm\tau\cdot\bm\nabla^\prime \bm R\rs(\bm u)
&=& 
\bm\tau\cdot\bm\nabla,
\end{eqnarray}
which, in turn, follows from Eq.~(\ref{eq:tdtrans}).  We
conclude that, as expected, the solution and source terms,
transformed according to Eqs.~(\ref{eq:solrot}) and
(\ref{eq:sourot}), where $\cR(\bm u)$ is given in
Eq.~(\ref{eq:crot}), satisfy the same equation as the original
solution and source terms.  The six-dimensional rotation
operator $\cR(\bm u)$ may be written as
\begin{eqnarray}
\cR(\bm u) &=& \re^{-\ri\bm \cS\cdot\bm u},
\elabel{eq:srotop}
\end{eqnarray}
where
\begin{eqnarray}
\bm \cS &=& \left(\begin{array}{cc} \bm \tau & \0 \\
\0 & \bm \tau \end{array}\right).
\elabel{eq:spinop}
\end{eqnarray}

Equations~(\ref{eq:solrot}), (\ref{eq:sourot}), and
(\ref{eq:crot}) correspond to the separate equations 
\begin{eqnarray}
\bm E\rs^\prime(x) &=& \bm R\rs(\bm u)\,\bm E\rs\big(R^{-1}(\bm u)\,x\big),
\\
\bm B\rs^\prime(x) &=& \bm R\rs(\bm u)\,\bm B\rs\big(R^{-1}(\bm u)\,x\big),
\\
\bm J\rs^\prime(x) &=& \bm R\rs(\bm u)\,\bm J\rs\big(R^{-1}(\bm u)\,x\big).
\end{eqnarray}
It can be confirmed that Eqs.~(\ref{eq:maxeq1}) and
(\ref{eq:maxeq4}) in spherical form are invariant under rotations.  In
particular, Eq.~(\ref{eq:maxeq1}) for the rotated electric field and
charge density $\rho^\prime(x) = \rho\big(R^{-1}(\bm u)x\big)$ is
\begin{eqnarray}
\bm\nabla\rs^\dagger\,\bm E^\prime\rs(x) &=& 
\frac{\rho^\prime(x)}{\epsilon_0}
\end{eqnarray}
or
\begin{eqnarray}
\bm\nabla\rs^\dagger\,\bm R\rs(\bm u)\bm E\rs\big(R^{-1}(\bm u)\,x\big) &=& 
\frac{\rho\big(R^{-1}(\bm u)\,x\big)}{\epsilon_0}.
\end{eqnarray}
The substitution $x\rightarrow R(\bm u)\,x$ gives
\begin{eqnarray}
\bm\nabla\rs^{\prime\dagger}\,\bm R\rs(\bm u)\bm E\rs(x) &=& 
\frac{\rho(x)}{\epsilon_0},
\end{eqnarray}
and Eq.~(\ref{eq:nabtrans}) yields
\begin{eqnarray}
\bm\nabla\rs^\dagger\,\bm E\rs(x) &=& 
\frac{\rho(x)}{\epsilon_0}.
\end{eqnarray}
Hence, the transformed field and charge density satisfy
Eq.~(\ref{eq:maxeq1}) if the original field and charge density do.
Similarly,
\begin{eqnarray}
\bm\nabla\rs^\dagger\,\bm B^\prime\rs(x) &=& 
\bm\nabla\rs^\dagger\,\bm B\rs(x) = 0.
\end{eqnarray}
Thus, all of the Maxwell equations in matrix form are invariant under
rotations.  

The separation into transverse and longitudinal components
of the electric and magnetic fields is also invariant under rotations.
This can be seen by considering the expression
$\bm{\iPi}\rs(\bm \nabla)\,\bm F\rs(x)$, where $\bm{\iPi}\rs(\bm \nabla)$ is
either $\bm{\iPi}\rT\rs(\bm \nabla)$ or $\bm{\iPi}\rL\rs(\bm \nabla)$ and $\bm
F\rs(x)$ is any of $\bm E\rs(x)$, $\bm B\rs(x)$, or $\bm J\rs(x)$.
We have
\begin{eqnarray}
\bm{\iPi}\rs(\bm \nabla)\,\bm F^\prime\rs(x) &=&
\bm{\iPi}\rs(\bm \nabla)\,
\bm R\rs(\bm u)\,\bm F\rs\big(R^{-1}(\bm u)\,x\big)
 \qquad
\end{eqnarray}
or
\begin{eqnarray}
\bm{\iPi}\rs(\bm \nabla^\prime)\,\bm F^\prime\rs(x^\prime) &=&
\bm{\iPi}\rs(\bm \nabla^\prime)\, \bm R\rs(\bm u)\,\bm F\rs(x)
\nonumber\\ &=& 
\bm R\rs(\bm u)\,\bm{\iPi}\rs(\bm \nabla)\, \bm F\rs(x),
\elabel{eq:rotproj}
\end{eqnarray}
where the last line follows from either Eq.~(\ref{eq:ptt}) or
Eq.~(\ref{eq:plt}).  This means that if the original field is transverse
or longitudinal, then the rotated field has the same character.
These results extend directly to the six-dimensional projection
operators $\iPi(\bm \nabla)$, solution $\iP(x)$, and source
$\iX(x)$.

\subsection{Velocity transformation of $\iP(x)$}
\elabel{ssec:vtpsi}

The result of the velocity transformation, by a velocity $\bm v$,
applied to the field $\iP(x)$ in Eq.~(\ref{eq:dms}) is the
function $\iP^\prime(x)$ given by
\begin{eqnarray}
\iP^\prime(x) &=& \cV(\bm v)\iP\!\big(V^{-1}(\bm v)\,x\big),
\elabel{eq:solvel}
\end{eqnarray}
where $\cV(\bm u)$ is a $6\times6$ matrix that gives the local
transformation of the field $\iP(x)$ at any point.  The inverse
transformation of the argument on the right-hand-side plays the
same role as for rotations.  Our objective is to establish the
covariance of Eq.~(\ref{eq:dms}) by showing that if $\iP(x)$ is
a solution of that equation with a source $\iX(x)$, then
\begin{eqnarray}
\gamma^\mu\partial_\mu \iP^\prime(x) &=& \iX^\prime(x) ,
\elabel{eq:transheq}
\end{eqnarray}
where $\iX^\prime(x)$ is a suitably transformed source term.
Equation~(\ref{eq:transheq}) can be written as
\begin{eqnarray}
\gamma^\mu\partial_\mu \cV(\bm v)\iP\big(V^{-1}(\bm v)\,x\big) &=&
\iX^\prime(x) \qquad
\end{eqnarray}
or
\begin{eqnarray}
\gamma^\mu\partial_\mu^\prime \cV(\bm v)\iP(x) &=& 
\iX^\prime\left(V(\bm v)\,x\right),
\elabel{eq:vteq}
\end{eqnarray}
where the variable $x$ has been replaced by $x^\prime = V(\bm
v)\,x$.  The $6\times6$ matrix $\cV(\bm v)$ is based on the
conventional local velocity transformation of the electric and
magnetic fields as discussed in \ref{app:vtft}.  Here
we write it as
\begin{eqnarray}
\cV(\bm v) &=& \re^{\zeta\bm \cK \cdot \bm{\hat v}},
\elabel{eq:vdef}
\end{eqnarray}
where
\begin{eqnarray}
\bm \cK = \left(\begin{array}{cc} \0 & \bm \tau \\ \bm \tau & \0
\end{array}\right).
\elabel{eq:kdef}
\end{eqnarray}
Expansion of the exponential function in Eq.~(\ref{eq:vdef}) in powers of
$\zeta$ yields
\begin{eqnarray}
\cV(\bm v) &=& 
\cI + \left(\bm \cK \cdot \bm{\hat v}\right)^2(\cosh{\zeta}-1)
+\left(\bm{\bm \cK \cdot \hat v}\right)\,\sinh{\zeta} 
\nonumber\\[10 pt]&=&
 \left(\begin{array}{cc}
\bm I + (\bm \tau \cdot \bm{\hat v})^2 \,(\cosh{\zeta}-1) 
& \bm \tau \cdot \bm{\hat v} \,\sinh{\zeta} \\
\bm \tau \cdot \bm{\hat v}\,\sinh{\zeta} &
\bm I + \left(\bm \tau \cdot \bm{\hat v}\right)^2(\cosh{\zeta}-1)
\end{array}\right) ,
\elabel{eq:vtrans}
\end{eqnarray}
where $(\bm \cK \cdot \bm{\hat v})^3 = \bm \cK \cdot \bm{\hat v}$ is
taken into account, so that
\begin{eqnarray}
\cV(\bm v)\iP &=& 
\left(\begin{array}{c}
\bm E\rs^\prime \\ \ri \, c \bm B\rs^\prime \msp \end{array}\right)
=
\left(\begin{array}{c}
\bm E\rs + (\bm \tau \cdot \bm{\hat v})^2 \bm E\rs
(\cosh{\zeta}-1) + \ri \, \bm \tau \cdot \bm{\hat v} 
\, c \bm B\rs \sinh{\zeta}
\\ \ri \left[c\bm B\rs + (\bm \tau \cdot \bm{\hat v})^2 
c\bm B\rs (\cosh{\zeta}-1) 
- \ri \, \bm \tau \cdot \bm{\hat v} \, \bm E\rs
\sinh{\zeta}\,\right] 
\msp \end{array}\right). \qquad
\end{eqnarray}

In Eq.~(\ref{eq:vteq}), we have 
\begin{eqnarray}
\gamma^\mu \partial^{\,\prime}_\mu
&=& \left(\begin{array}{ccc}
\bm I\,\frac{\textstyle \partial}{\textstyle \partial ct^\prime} 
&& \bm\tau\cdot\bm\nabla^\prime
\\
-\bm\tau\cdot\bm\nabla^\prime && -\bm I\,\frac{\textstyle
\partial}{\textstyle \partial ct^\prime}
\end{array}\right) ,
\elabel{eq:tdiffop}
\end{eqnarray}
where, from Eqs.~(\ref{eq:cgo}) and (\ref{eq:cgt}),
\begin{eqnarray}
\frac{\partial}{\partial ct^\prime} &=&
\cosh{\zeta}\,\frac{\partial}{\partial ct}
- \sinh{\zeta} \,\bm{\hat v}\cdot\bm\nabla,
\\
\bm \nabla^\prime &=& 
\bm \nabla + (\cosh{\zeta} - 1)
\bm{\hat v}\,\bm{\hat v}\cdot\bm\nabla
-\sinh{\zeta}\,\bm{\hat v}\,\frac{\partial}{\partial
ct}. \qquad
\end{eqnarray}
Multiplication of Eq.~(\ref{eq:vtrans}) by
Eq.~(\ref{eq:tdiffop}) yields the identity
\begin{eqnarray}
\gamma^\mu \partial^{\,\prime}_\mu \, {\cal V}(\bm v)
&=&
\left(\begin{array}{ccc}
\bm I+\bm{\hat v}_{\rm s}\bm{\hat v}_{\rm s}^\dagger
(\cosh{\zeta}-1) && \0 \\
\0 &&
\bm I+\bm{\hat v}_{\rm s}\bm{\hat v}_{\rm s}^\dagger
(\cosh{\zeta}-1)
 \end{array} \right)
\gamma^\mu \, \partial_\mu 
 \nonumber\\[6 pt]&&
 +
 \left(\begin{array}{ccc}
 -\sinh{\zeta}\,\bm{\hat v}_{\rm s}\bm\nabla_{\rm s}^\dagger &&
 \0 \\
 \0 && \sinh{\zeta}\,\bm{\hat v}_{\rm s}\bm\nabla_{\rm s}^\dagger
 \end{array}\right)
\elabel{eq:ident}
\end{eqnarray}
and hence
\begin{eqnarray}
\gamma^\mu\partial_\mu^\prime \cV(\bm v)\iP(x) &=& 
\left(\begin{array}{ccc}
\bm I+\bm{\hat v}_{\rm s}\bm{\hat v}_{\rm s}^\dagger
(\cosh{\zeta}-1) && \0 \\
\0 &&
\bm I+\bm{\hat v}_{\rm s}\bm{\hat v}_{\rm s}^\dagger
(\cosh{\zeta}-1)
 \end{array} \right)
\iX(x)
\nonumber\\[4 pt]&&
 +
 \left(\begin{array}{c}
 -\sinh{\zeta}\,\bm{\hat v}_{\rm s}\,\bm\nabla\cdot \bm E(x)
 \\
 \ri\sinh{\zeta}\,\bm{\hat v}_{\rm s}\,c\,\bm\nabla\cdot \bm B(x)
 \end{array}\right)
 \nonumber\\[10 pt] &=& 
\left(\begin{array}{c}
-\mu_0c\left[\bm J_{\rm s}(x)+\bm{\hat v}_{\rm s}
\bm{\hat v} \cdot \bm J(x)
(\cosh{\zeta}-1)
+ \sinh{\zeta} \, \bm{\hat v}_{\rm s} c \rho(x)\right] \\
\0 
 \end{array} \right)
 \nonumber\\[10 pt]&=&
\left(\begin{array}{c}
-\mu_0c\bm J_{\rm s}^\prime(x) \\
\0 
 \end{array} \right),
 \elabel{eq:ltrans}
\end{eqnarray}
where $\bm J_{\rm s}^\prime(x)$ is the velocity transformed
three-vector source current, and  
\begin{eqnarray}
\iX^\prime(x) &=& \left(\begin{array}{c}
-\mu_0c\bm J\rs^\prime\big(V^{-1}(\bm v)\,x\big) \\ 
\0 \msp \end{array}\right).
\end{eqnarray}
The result in Eq.~(\ref{eq:ltrans}) takes into account the
additional two Maxwell equations in (\ref{eq:maxeq1}) and
(\ref{eq:maxeq4}), besides Eqs.~(\ref{eq:maxeq2}) and
(\ref{eq:maxeq3}) used to construct Eq.~(\ref{eq:dms}).  It also
requires the conventional result that the current
\begin{eqnarray}
\left(\begin{array}{c} c\rho(x) \\ \bm J\rs(x) \end{array}\right)
\end{eqnarray}
transforms as a four-vector under the velocity transformation
given by Eq.~(\ref{eq:sav}).  Equation~(\ref{eq:ltrans})
establishes the validity of Eq.~(\ref{eq:transheq}), provided
the transformed three-vector source current is the three-vector
component of the transformed four-vector current.

The covariance of Eq.~(\ref{eq:dms}), even though the charge
density does not appear in the source term, is linked to the
fact that the current density satisfies the continuity equation.
Since the continuity equation follows from the Maxwell
equations, it cannot be expected that consistent solutions may
be found for an arbitrary four-vector current density.  However,
for valid sources, information about the charge density may be
obtained from the three-vector current density and the
continuity equation.  For example, if the electric field is
specified at a particular time, then the charge density at that
time is known from Eq.~(\ref{eq:maxeq1}) and may be determined
at any other time from knowledge of the three-vector current
density by use of the continuity equation.  Thus the time
evolution of the electromagnetic fields can be described
relativistically with no reference to the charge density.  The
Maxwell Green function, which provides the solutions of
Eq.~(\ref{eq:dms}), is discussed in Sec.~\ref{sec:prop}.

Since $\cV^\dagger \gamma^0 \cV = \gamma^0$ and $\cV^\dagger
\gamma^0 \eta \cV = \gamma^0 \eta$, where
\begin{eqnarray}
\eta = \left(\begin{array}{ccc} \0 && \bm I \\
\bm I && \0 \end{array}\right),
\elabel{eq:eta}
\end{eqnarray}
the invariance of the quantities
\begin{eqnarray}
\overline{\iP}\iP = |\bm E|^2 - c^2|\bm B|^2,
\\
\overline{\iP}\eta\iP = 2\,\ri\, c \,{\rm Re} \, \bm E\cdot \bm B
\end{eqnarray}
is evident.

\subsection{Parity and time-reversal transformations
of~$\iP(x)$}
\elabel{ssec:ptpsi}

We expect the fields $\iP(x)$ to transform under a parity change
according to
\begin{eqnarray}
\iP^\prime(x) &=& \cP\iP\!\big(\,P^{-1}\,x\big),
\elabel{eq:solparity}
\end{eqnarray}
where $\cP$ is a $6\times6$ matrix,  
and we assume that the current transforms in the same way, so that
\begin{eqnarray}
\iX^\prime(x) &=& \cP\iX\!\big(\,P^{-1}\,x\big).
\end{eqnarray}
We can obtain
\begin{eqnarray}
\gamma^\mu\partial_\mu \iP^\prime(x) &=& \iX^\prime(x),
\elabel{eq:partr}
\end{eqnarray}
by finding  a suitable matrix $\cP$.
In terms of the original field and source, Eq.~(\ref{eq:partr}) is
given by
\begin{eqnarray}
\gamma^\mu\partial_\mu \cP\iP\big(P^{-1}x\big) &=&
\cP\iX\big(P^{-1}x\big) \qquad
\end{eqnarray}
or
\begin{eqnarray}
\gamma^\mu\partial_\mu^\prime \cP\iP(x) &=&
\cP\iX(x), \qquad
\end{eqnarray}
where $\partial_\mu^\prime$ is the parity transformed derivative
given by Eq.~(\ref{eq:derivparity}).
Equation~(\ref{eq:partr}) will follow if
\begin{eqnarray}
\cP^{-1}\gamma^\mu\partial_\mu^\prime \cP &=&
\gamma^\mu\partial_\mu,
\end{eqnarray}
which corresponds to
\begin{eqnarray}
\cP^{-1}\gamma^0 \cP &=& \gamma^0,
\\
\cP^{-1}\gamma^i \,\cP &=& -\gamma^i, \quad i = 1,2,3.
\end{eqnarray}
Solutions of these equations are provided by
\begin{eqnarray}
\cP &=& \pm\left(\begin{array}{ccc} \bm I && \0 \\
\0 && -\bm I \end{array}\right).
\elabel{eq:twoparity}
\end{eqnarray}

The minus sign corresponds to the conventional choice of how
classical electric and magnetic fields transform under a parity
change, that is, the current and electric fields change sign and
the magnetic fields do not \cite{1998165}.  

To examine time-reversal invariance, we first consider $\iP(x)$
as a field which is real in the Cartesian basis.  In this case,
the conventional use of an anti-unitary operator for time
reversal is unnecessary, and the same can be expected to be true
for fields expressed in the spherical basis.  We write
\begin{eqnarray}
\iP^\prime(x) &=& \cT\iP\!\big(\,T^{-1}\,x\big)
\elabel{eq:soltime}
\end{eqnarray}
and
\begin{eqnarray}
\iX^\prime(x) &=& -\cT\,\iX\!\big(\,T^{-1}\,x\big),
\end{eqnarray}
where $\cT$ is a suitable $6\times6$ matrix.  The minus sign
for the current provides the result that the electric field does
not change sign under time reversal and the current does.  The
objective is to find a matrix $\cT$ such that
\begin{eqnarray}
\gamma^\mu\partial_\mu \iP^\prime(x) &=& \iX^\prime(x)
\end{eqnarray}
or
\begin{eqnarray}
\gamma^\mu\partial_\mu \cT\iP\big(T^{-1}x\big) &=&
-\cT\iX\big(T^{-1}x\big), \qquad
\end{eqnarray}
and so
\begin{eqnarray}
\gamma^\mu\partial_\mu^\prime \cT\iP(x) &=&
-\cT\iX(x), \qquad
\end{eqnarray}
where $\partial_\mu^\prime$ is given by
Eq.~(\ref{eq:derivtime}).  Such a matrix satisfies
\begin{eqnarray}
\cT^{-1}\gamma^\mu\partial_\mu^\prime \cT &=&
-\gamma^\mu\partial_\mu
\end{eqnarray}
or
\begin{eqnarray}
\cT^{-1}\gamma^0 \cT &=& \gamma^0,
\\
\cT^{-1}\gamma^i \cT &=& - \gamma^i, \quad i = 1,2,3,
\end{eqnarray}
with a solution provided by
\begin{eqnarray}
\cT &=& \left(\begin{array}{ccc} \bm I && \0 \\
\0 && - \bm I \end{array}\right).
\end{eqnarray}
The matrices $\cP$ and $\cT$ commute with the rotation matrix, 
as they should, and we have the result that the matrix form of 
the Maxwell equations is invariant under parity and time-reversal
transformations.

For quantum-mechanical time reversal, the time-reversal operator
is anti-unitary and includes complex conjugation, or Hermitian
conjugation in the case of a matrix solution, which has the
effect of interchanging initial and final states.  For an
example where such an interchange corresponds to observable
consequences in QED, see \cite{1978036,dunford}.  We thus write
\begin{eqnarray}
\iP^\prime(x) &=& \fT\iP\!\big(\,T^{-1}\,x\big) =
\iP^\dagger\big(\,T^{-1}\,x\big)\,\cU^{-1},
\\
\iX^\prime(x) &=& -\fT\iX\!\big(\,T^{-1}\,x\big) =
-\iX^\dagger\big(\,T^{-1}\,x\big)\,\cU^{-1},
\end{eqnarray}
where $\fT = \cC\,\cU$ is the product of the Hermitian
conjugation operator $\cC$, which has the action $\cC\iP(x) =
\iP^\dagger(x)$ and a unitary matrix $\cU$.  The objective is to
find a $\cU$ such that
\begin{eqnarray}
\iP^{\,\prime}(x) \gamma^0
\overleftarrow{\partial}_{\!\mu} 
{\gamma^\mu}
&=& \iX^{\,\prime}(x) \gamma^0
\elabel{eq:tconjeq}
\end{eqnarray}
if $\iP(x)$ is a solution of Eq.~(\ref{eq:dmsc}).
Equation~(\ref{eq:tconjeq}) can be written as
\begin{eqnarray}
\iP^\dagger(x)\,\cU^{-1}\gamma^0
\overleftarrow{\partial^\prime}_{\!\mu}
{\gamma^\mu} &=& -\iX^\dagger(x)\,\cU^{-1} \gamma^0,
\elabel{eq:qtr}
\end{eqnarray}
where $\partial^\prime_\mu$ is given by
Eq.~(\ref{eq:derivtime}).  Equation~(\ref{eq:qtr})
follows from Eq.~(\ref{eq:dmsc}) provided
\begin{eqnarray}
\gamma^0\,\cU^{-1}\gamma^0
\overleftarrow{\partial^\prime}_{\!\mu}
{\gamma^\mu} \gamma^0\,\cU\gamma^0&=&
-\overleftarrow{\partial}_{\!\mu}
{\gamma^\mu},
\end{eqnarray}
which has as a solution
\begin{eqnarray}
\cU &=& \cT
\end{eqnarray}
and
\begin{eqnarray}
\fT &=& \cC\,\cT.
\end{eqnarray}

\section{Plane-wave eigenfunctions}
\elabel{sec:pwf}

Following the analogy with the Dirac equation, the Hamiltonian
for the Maxwell equation is 
\begin{eqnarray}
\cH &=& c \, \bm \alpha \cdot \bm p 
=  -\ri\,\hbar  c \, \bm \alpha \cdot \bm \nabla ,
\elabel{eq:phham}
\end{eqnarray}
where $\alpha^i = \gamma^0\,\gamma^i$, and the wave functions
for the photon may be identified as the complete set of
eigenfunctions of $\cH$.  The solutions considered here are
coordinate-space plane waves characterized by a wave vector $\bm
k$ and a polarization vector $\bm{\hat \epsilon}_\lambda$; both
positive- and negative-energy solutions, as well as
zero-energy solutions are included to form a complete set.
These solutions are also eigenfunctions of the momentum operator
\begin{eqnarray}
\bm \cP &=& \cI \bm p = -\ri \, \hbar \, \cI \, \bm \nabla,
\elabel{eq:phmom}
\end{eqnarray}
which commutes with $\cH$.

The plane-wave solutions are not normalizable, because their
modulus squared is independent of $\bm x$ and the integral over
all space does not exist.  As a result, the solutions include an
arbitrary multiplicative factor, that could be a function of
$\bm k$.  Here a factor is chosen to provide the simple result
in Eq.~(\ref{eq:orth1}).

\subsection{Transverse plane-wave photons}
\elabel{ssec:tph}

We first consider transverse photons, {\it i.,e.,} photons for
which the electric and magnetic fields are perpendicular to the
wave vector.  The polarization vector is a unit vector
proportional to the electric or magnetic fields, represented by
a three component, possibly complex, vector in the spherical
basis.  As such, the polarization vector does not transform as
the spatial component of a four-vector under velocity
transformations.

Two polarization vectors, both in the plane perpendicular to
$\bm{\hat k}$, are denoted by
\begin{eqnarray}
\bm{\hat \epsilon}_\lambda(\bm{\hat k}) \, ; \qquad \lambda =
1,2.
\elabel{eq:epst}
\end{eqnarray}
They have the orthonormality properties
\begin{eqnarray}
\bm{\hat \epsilon}_{\lambda_2}^\dagger(\bm{\hat k}) \,
\bm{\hat \epsilon}_{\lambda_1}(\bm{\hat k}) &=&
\delta_{{\lambda_2},{\lambda_1}},
\elabel{eq:epsorth}
\\
\bm{\hat k}_{\rm s}^\dagger\bm{\hat \epsilon}_\lambda(\bm{\hat
k})
&=& 0, 
\elabel{eq:epskorth}
\end{eqnarray}
and the completeness property
\begin{eqnarray}
\sum_{\lambda=1}^2
\bm{\hat \epsilon}_\lambda(\bm{\hat k}) \,
\bm{\hat \epsilon}_\lambda^\dagger(\bm{\hat k}) &=& \bm I
- \bm{\hat k}_{\rm s} \, \bm{\hat k}_{\rm s}^\dagger
= (\bm \tau \cdot \bm{\hat k})^2
 =
 \bm \iPi\rs\rT(\bm{\hat k}).  
\label{eq:bewf}
\end{eqnarray}
From Eq.~(\ref{eq:epskorth}), we also have
\begin{eqnarray}
(\bm \tau \cdot \bm{\hat k})^2 \,
\bm{\hat \epsilon}_\lambda(\bm{\hat k})
=
\bm{\hat \epsilon}_\lambda(\bm{\hat k}).
\end{eqnarray}
The polarization vectors can represent linear polarization,
circular polarization, or any combination by a suitable choice
of $\bm{\hat \epsilon}_\lambda(\bm{\hat k})$.  For example, for
$\bm{k}$ in the $\bm{\hat e}^3$ direction, linear
polarization vectors in the $\bm{\hat e}^1$ and $\bm{\hat e}^2$
directions are
\begin{eqnarray}
\bm{\hat \epsilon}_1(\bm{\hat e}^3) &=& 
\left(\begin{array}{c}-{1\over\sqrt{2}} \\ 
0 \\ 
{1\over\sqrt{2}}\end{array}\right); \quad
\bm{\hat \epsilon}_2(\bm{\hat e}^3) =
\left(\begin{array}{c}{\ri\over\sqrt{2}} \\ 
0 \\ 
{\ri\over\sqrt{2}}\end{array}\right), \quad
\elabel{eq:lin}
\end{eqnarray}
according to Eq.~(\ref{eq:asph}). 
Similarly, circular polarization vectors
are (see Sec.~\ref{ssec:he})
\begin{eqnarray}
\bm{\hat \epsilon}_1(\bm{\hat e}^3) &=& 
\left(\begin{array}{c}
1 \\ 
0 \\ 
0 \end{array}\right) ; \quad
\bm{\hat \epsilon}_2(\bm{\hat e}^3) = 
\left(\begin{array}{c}
0 \\ 
0 \\ 
1 \end{array}\right).
\elabel{eq:circ}
\end{eqnarray}
These polarization vectors can be transformed to the vectors
corresponding to any direction of $\bm k$ with the rotation
operator in Eq.~(\ref{eq:rots}).  (See also
Sec.~\ref{ssec:rtwf}.)

Positive $(+)$ and negative $(-)$ energy transverse photon wave
functions are given by
\begin{eqnarray}
\psi_{\bm{k},\lambda}^{(\pm)}(\bm x) &=& 
\frac{1}{\sqrt{2(2\pi)^3}}
\left(\begin{array}{c}\bm{\hat \epsilon}_\lambda(\bm{\hat k}) \\
                      \bm \tau \cdot \bm{\hat k} \
                      \bm{\hat \epsilon}_\lambda(\bm{\hat k})
		      \end{array}\right)
\re^{\pm\ri\bm k \cdot \bm x}. \quad
\elabel{eq:trwf}
\end{eqnarray}
Although constructed geometrically to be transverse by the
choice of polarization vectors, these wave functions are also
transverse in the sense defined in Sec.~\ref{sec:tlfields}.
We have (see \ref{app:invlap} for more detail)
\begin{eqnarray}
\iPi\rT(\bm\nabla)\, \psi_{\bm{k},\lambda}^{(\pm)}(\bm x)
&=&
\iPi\rT(\bm{\hat k})\, \psi_{\bm{k},\lambda}^{(\pm)}(\bm x)
=
\psi_{\bm{k},\lambda}^{(\pm)}(\bm x) , \qquad 
\\
\iPi\rL(\bm\nabla)\, \psi_{\bm{k},\lambda}^{(\pm)}(\bm x)
&=&
\iPi\rL(\bm{\hat k})\, \psi_{\bm{k},\lambda}^{(\pm)}(\bm x)
= 0 ,
\end{eqnarray}
where $\iPi\rT$ and $\iPi\rL$ are defined in
Eqs.~(\ref{eq:ptop}) and (\ref{eq:plop}).  The wave functions in
Eq.~(\ref{eq:trwf}) are eigenfunctions of the Hamiltonian in
Eq.~(\ref{eq:phham}) with eigenvalues $\pm\hbar c \,|\bm k|$.
In particular,
\begin{eqnarray}
\cH \, \psi_{\bm{k},\lambda}^{(\pm)}(\bm x) 
&=&
\pm \hbar c \, \bm \alpha\cdot\bm k \, 
\psi_{\bm{k},\lambda}^{(\pm)}(\bm x) ,
\end{eqnarray}
and
\begin{eqnarray}
\left(\begin{array}{cc}
\0 & \bm \tau \cdot \bm{k} \\
\bm \tau \cdot \bm{k} & \0 \end{array}\right)
\left(\begin{array}{c}\bm{\hat \epsilon}_\lambda(\bm{\hat k}) \\
                      \bm \tau \cdot \bm{\hat k} \
                      \bm{\hat \epsilon}_\lambda(\bm{\hat k})
		      \end{array}\right)
&=& |\bm k|
\left(\begin{array}{c}\bm{\hat \epsilon}_\lambda(\bm{\hat k}) \\
                      \bm \tau \cdot \bm{\hat k} \
                      \bm{\hat \epsilon}_\lambda(\bm{\hat k})
		      \end{array}\right) ,
\end{eqnarray}
so that
\begin{eqnarray}
\cH \, \psi_{\bm{k},\lambda}^{(\pm)}(\bm x) 
&=&
\pm \hbar c \, |\bm k| \,
\psi_{\bm{k},\lambda}^{(\pm)}(\bm x) .
\end{eqnarray}
Also,
\begin{eqnarray}
\bm \cP \, \psi_{\bm{k},\lambda}^{(\pm)}(\bm x) 
&=&
\pm \hbar \, \bm k \,
\psi_{\bm{k},\lambda}^{(\pm)}(\bm x) .
\end{eqnarray}
The wave functions have the expected property
\begin{eqnarray}
\overline{\psi}_{\bm{k},\lambda}^{(\pm)}(\bm x)
\,\psi_{\bm{k},\lambda}^{(\pm)}(\bm x)
&=& 0,
\end{eqnarray}
since the electric and magnetic field strengths are equal for a
transverse photon.
Normalization and orthogonality relations are
\begin{eqnarray}
\int \rd \bm x \ 
\psi_{\bm{k}_2,\lambda_2}^{(\pm)\dagger}(\bm x) \,
\psi_{\bm{k}_1,\lambda_1}^{(\pm)}(\bm x)
&=& \delta_{\lambda_2\lambda_1}
\delta(\bm k_2 - \bm k_1),\qquad
\elabel{eq:orth1}
\\
\int \rd \bm x \ 
\psi_{\bm{k}_2,\lambda_2}^{(\pm)\dagger}(\bm x) \,
\psi_{\bm{k}_1,\lambda_1}^{(\mp)}(\bm x)
&=& 0 .
\elabel{eq:orth2}
\end{eqnarray}
The latter relation follows from a cancellation of terms between
the upper-three and lower-three components of the wave function:
\begin{eqnarray}
\int \rd \bm x \ 
\psi_{\bm{k}_2,\lambda_2}^{(\pm)\dagger}(\bm x) \,
\psi_{\bm{k}_1,\lambda_1}^{(\mp)}(\bm x)
&=&
\frac{1}{2(2\pi)^3}
\int \rd \bm x \ 
\bm{\hat \epsilon}_{\lambda_2}^\dagger(\bm{\hat k}_2)
\left[
\bm I +\bm \tau\cdot\bm{\hat k}_2
\,\bm \tau\cdot\bm{\hat k}_1
\right]\bm{\hat \epsilon}_{\lambda_1}(\bm{\hat k}_1)
\,\re^{\mp\ri (\bm k_2 + \bm k_1)\cdot\bm x} 
\quad
\nonumber\\&=&
\frac{1}{2} \,
\bm{\hat \epsilon}_{\lambda_2}^\dagger(\bm{\hat k}_2)
\left[
\bm I +\bm \tau\cdot\bm{\hat k}_2
\,\bm \tau\cdot\bm{\hat k}_1
\right]\bm{\hat \epsilon}_{\lambda_1}(\bm{\hat k}_1)
\delta(\bm k_2 + \bm k_1) = 0.
\elabel{eq:orth2l}
\end{eqnarray}

The transverse wave functions constitute a complete set of such
functions.  The completeness is established by writing
\begin{eqnarray}
&&\sum_{\lambda=1}^2 \int \rd \bm k \ 
\psi_{\bm{k},\lambda}^{(\pm)}(\bm x_2)
\psi_{\bm{k},\lambda}^{(\pm)\dagger}(\bm x_1)
\nonumber\\&&= 
\sum_{\lambda=1}^2 \int \rd \bm k \ 
\frac{1}{2(2\pi)^3}\, 
\left(\begin{array}{ccc}
\bm{\hat \epsilon}_\lambda(\bm{\hat k})\,
\bm{\hat \epsilon}_\lambda^\dagger(\bm{\hat k})
&&
\bm{\hat \epsilon}_\lambda(\bm{\hat k})\,
\bm{\hat \epsilon}_\lambda^\dagger(\bm{\hat k})
\bm \tau \cdot \bm{\hat k} \\
\bm \tau \cdot \bm{\hat k} \,
\bm{\hat \epsilon}_\lambda(\bm{\hat k})\,
\bm{\hat \epsilon}_\lambda^\dagger(\bm{\hat k})
&&
\bm \tau \cdot \bm{\hat k} \,
\bm{\hat \epsilon}_\lambda(\bm{\hat k})\,
\bm{\hat \epsilon}_\lambda^\dagger(\bm{\hat k})
\bm \tau \cdot \bm{\hat k} 
\end{array}\right)
\re^{\pm\ri\bm k \cdot (\bm x_2 - \bm x_1)}
\qquad
\nonumber\\[6 pt]&&=
\int \rd \bm k \ 
\frac{1}{2(2\pi)^3}\, 
\left(\begin{array}{ccc}
(\bm \tau \cdot \bm{\hat k})^2
&&
\bm \tau \cdot \bm{\hat k} \\
\bm \tau \cdot \bm{\hat k} \,
&&
(\bm \tau \cdot \bm{\hat k})^2
\end{array}\right)
\re^{\pm\ri\bm k \cdot (\bm x_2 - \bm x_1)} .
\elabel{eq:trcomp}
\end{eqnarray}
To evaluate the integrals for the sum over positive and
negative energy solutions, we use $\kappa$ to represent
either a plus sign or a minus sign and write
\begin{eqnarray}
\sum_{\kappa\rightarrow\pm} \int \rd \bm k \ 
(\bm \tau \cdot \bm{\hat k})^2 \,
\re^{\kappa\ri\bm k \cdot (\bm x_2 - \bm x_1)}
= 2(2\pi)^3  \, \bm \iPi\rs\rT(\bm\nabla_2)
\,\delta(\bm x_2 - \bm x_1)
\end{eqnarray}
and
\begin{eqnarray}
\sum_{\kappa\rightarrow\pm} \int \rd \bm k \ 
\bm \tau \cdot \bm{\hat k} \,
\re^{\kappa\ri\bm k \cdot (\bm x_2 - \bm x_1)}
&=& 0 ,
\end{eqnarray}
which yields the transverse completeness relation
\begin{eqnarray}
&&\sum_{\kappa\rightarrow\pm}\sum_{\lambda=1}^2 \int \rd \bm k \ 
\psi_{\bm{k},\lambda}^{(\kappa)}(\bm x_2)
\psi_{\bm{k},\lambda}^{(\kappa)\dagger}(\bm x_1)
=
\iPi\rT(\bm\nabla_2)\, \delta(\bm x_2 - \bm x_1).
\qquad
\elabel{eq:tcr}
\end{eqnarray}

\subsection{Longitudinal plane-wave photons}
\elabel{ssec:lph}

The transverse wave functions alone do not provide a complete
description of electromagnetic fields.  For example, the field
of a point charge $q$ at rest at the origin, given by
\begin{eqnarray}
\bm E\rs(\bm x) &=& \frac{q}{4\pi\epsilon_0}\,
\frac{\bm x\rs}{|\bm x|^3} 
= -\frac{q}{4\pi\epsilon_0}\bm \nabla\rs\,
\frac{1}{|\bm x|},
\end{eqnarray}
is purely longitudinal, because
\begin{eqnarray}
\bm{\iPi}\rs\rL(\bm\nabla)\,\bm E\rs(\bm x) &=& \bm E\rs(\bm x),
\\
\bm{\iPi}\rs\rT(\bm\nabla)\,\bm E\rs(\bm x) &=&0.
\end{eqnarray}
The longitudinal photons are represented by a third polarization
state, labeled $\lambda = 0$, with the polarization vector taken
to be
\begin{eqnarray}
\bm{\hat \epsilon}_0(\bm{\hat k}) &=& \bm{\hat k}_{\rm s}.
\elabel{eq:epsl}
\end{eqnarray}
If $\bm{k}$ is in the $\bm{\hat e}^3$
direction, the longitudinal polarization vector is
(up to a phase factor)
\begin{eqnarray}
\bm{\hat \epsilon}_0(\bm{\hat e}^3) &=& 
\left(\begin{array}{c} 0 \\ 
1 \\ 0 \end{array}\right) .
\elabel{eq:lpv}
\end{eqnarray}
Longitudinal wave functions are
\begin{eqnarray}
\psi_{\bm{k},0}^{(+)}(\bm x) &=&
\frac{1}{\sqrt{(2\pi)^3}}
\left(\begin{array}{c}\bm{\hat \epsilon}_0(\bm{\hat k}) \\
\0
\end{array}\right)
\re^{\ri\bm k \cdot \bm x} \quad
\elabel{eq:plwf}
\end{eqnarray}
or
\begin{eqnarray}
\psi_{\bm{k},0}^{(-)}(\bm x) &=&
\frac{1}{\sqrt{(2\pi)^3}}
\left(\begin{array}{c}\0\\
\bm{\hat \epsilon}_0(\bm{\hat k})
\end{array}\right)
\re^{-\ri\bm k \cdot \bm x}. \quad
\elabel{eq:nlwf}
\end{eqnarray}
This polarization state has the property that
\begin{eqnarray}
\iPi\rL(\bm\nabla)\, \psi_{\bm{k},0}^{(\pm)}(\bm x)
&=&
\iPi\rL(\bm{\hat k})\, \psi_{\bm{k},0}^{(\pm)}(\bm x)
=
\psi_{\bm{k},0}^{(\pm)}(\bm x) , \qquad 
\\
\iPi\rT(\bm\nabla)\, \psi_{\bm{k},0}^{(\pm)}(\bm x)
&=&
\iPi\rT(\bm{\hat k})\, \psi_{\bm{k},0}^{(\pm)}(\bm x)
= 0 .
\end{eqnarray}
The wave function has an energy eigenvalue of zero,
\begin{eqnarray}
\cH \, \psi_{\bm{k},0}^{(\pm)}(\bm x) 
&=&
\pm \hbar c \, \bm \alpha\cdot\bm k \, 
\psi_{\bm{k},0}^{(\pm)}(\bm x) = 0 ,
\end{eqnarray}
because $\bm \tau \cdot\bm k \, \bm{\hat \epsilon}_0(\bm{\hat
k}) =0$.  
However,
\begin{eqnarray}
\bm \cP \, \psi_{\bm{k},0}^{(\pm)}(\bm x) 
&=&
\pm \hbar \, \bm k \,
\psi_{\bm{k},0}^{(\pm)}(\bm x) .
\end{eqnarray}

Normalization and orthogonality of the $\lambda=0$
wave functions, as well as the transverse wave functions, are
given by Eqs.~(\ref{eq:orth1}) and (\ref{eq:orth2}), where
$\lambda_1$ and $\lambda_2$ may take on any of the values 0, 1,
or 2.  

Completeness relations are given by
\begin{eqnarray}
\int \rd \bm k \ 
\psi_{\bm{k},0}^{(+)}(\bm x_2)
\psi_{\bm{k},0}^{(+)\dagger}(\bm x_1)
&=&
\int \rd \bm k \ 
\frac{1}{(2\pi)^3}\, 
\left(\begin{array}{ccc}
\bm{\hat \epsilon}_0(\bm{\hat k})\,
\bm{\hat \epsilon}_0^\dagger(\bm{\hat k})
&&
\0
\\
\0
&&
\0
\end{array}\right)
\re^{\ri\bm k \cdot (\bm x_2 - \bm x_1)}
\elabel{eq:plongcomp}
\end{eqnarray}
and
\begin{eqnarray}
\int \rd \bm k \ 
\psi_{\bm{k},0}^{(-)}(\bm x_2)
\psi_{\bm{k},0}^{(-)\dagger}(\bm x_1)
&=&
\int \rd \bm k \ 
\frac{1}{(2\pi)^3}\, 
\left(\begin{array}{ccc}
\0
&&
\0
\\
\0
&&
\bm{\hat \epsilon}_0(\bm{\hat k})\,
\bm{\hat \epsilon}_0^\dagger(\bm{\hat k})
\end{array}\right)
\re^{-\ri\bm k \cdot (\bm x_2 - \bm x_1)},
\elabel{eq:nlongcomp}
\end{eqnarray}
where
\begin{eqnarray}
\int \rd \bm k \ 
\bm{\hat \epsilon}_0(\bm{\hat k})\,
\bm{\hat \epsilon}_0^\dagger(\bm{\hat k})
\re^{\pm\ri\bm k \cdot (\bm x_2 - \bm x_1)}
= (2\pi)^3  \,
\bm \iPi\rs\rL(\bm \nabla_2)\,
\delta(\bm x_2 - \bm x_1),\qquad
\end{eqnarray}
which yields
\begin{eqnarray}
\sum_{\kappa\rightarrow\pm}\int \rd \bm k \ 
\psi_{\bm{k},0}^{(\kappa)}(\bm x_2)
\psi_{\bm{k},0}^{(\kappa)\dagger}(\bm x_1)
=\iPi\rL(\bm \nabla_2)\,
\delta(\bm x_2 - \bm x_1).
\qquad
\elabel{eq:lcr}
\end{eqnarray}

For the example of a point charge at the origin, we have
\begin{eqnarray}
\iP\rp(\bm x) &=& \frac{q}{4 \pi\epsilon_0 |\bm x|^3}
\left(\begin{array}{c} \bm x\rs \\ \0 \end{array}\right),
\end{eqnarray}
which may be written as (see \ref{app:coultr} for some
detail)
\begin{eqnarray}
\iP\rp(\bm x) &=& 
\iPi\rL(\bm\nabla)\iP\rp(\bm x)
=
\int\rd\bm x_1 \iPi\rL(\bm\nabla)\,
\delta(\bm x - \bm x_1)\iP\rp(\bm x_1)
\nonumber\\ &=&
\sum_{\kappa\rightarrow\pm}\int \rd \bm k \ 
\psi_{\bm{k},0}^{(\kappa)}(\bm x)
\int\rd\bm x_1 \,
\psi_{\bm{k},0}^{(\kappa)\dagger}(\bm x_1)
\iP\rp(\bm x_1)
\nonumber\\&=&
- \frac{\ri\,q}{\sqrt{(2\pi)^3}\,\epsilon_0}
\int \rd \bm k \, 
\frac{1}{|\bm k|} \,
\psi_{\bm{k},0}^{(+)}(\bm x).
\elabel{eq:coultr}
\end{eqnarray}

\subsection{Full orthogonality and completeness of the plane
wave solutions}
\elabel{ssec:focpls}

The full orthogonality relations are
\begin{eqnarray}
\int \rd \bm x \ 
\psi_{\bm{k}_2,\lambda_2}^{(\kappa_2)\dagger}(\bm x) \,
\psi_{\bm{k}_1,\lambda_1}^{(\kappa_1)}(\bm x)
&=& \delta_{\kappa_2\kappa_1}\delta_{\lambda_2\lambda_1}
\delta(\bm k_2 - \bm k_1),\qquad
\elabel{eq:forth}
\end{eqnarray}
where the factor $\delta_{\kappa_2 \kappa_1}$ is 1 if $\kappa_2$
and $\kappa_1$ represent the same sign and is 0 for opposite
signs, and $\lambda_2$ and $\lambda_1$ may be any of 0,1,2.  The
combined result of the transverse and longitudinal completeness
relations, Eqs.~(\ref{eq:tcr}) and (\ref{eq:lcr}), is 
\begin{eqnarray}
\sum_{\kappa\rightarrow\pm}\sum_{\lambda=0}^2 \int \rd \bm k \ 
\psi_{\bm{k},\lambda}^{(\kappa)}(\bm x_2)
\psi_{\bm{k},\lambda}^{(\kappa)\dagger}(\bm x_1)
= \cI \,
\delta(\bm x_2 - \bm x_1), 
\qquad
\end{eqnarray}
where $\iPi\rT(\bm \nabla) + \iPi\rL(\bm \nabla) = \cI$.

\subsection{Time dependence of the wave functions}
\elabel{ssec:tdwf}

The time dependence of the photon wave functions is given
by\footnote{The notation $f(\bm x) = f(x)\big|_{t=0}$ is
employed throughout this paper.}
\begin{eqnarray}
\psi_{\bm{k},\lambda}^{(\kappa)}(x) &=& 
\psi_{\bm{k},\lambda}^{(\kappa)}(\bm x) \,
\re^{-\kappa\ri\omega t},
\elabel{eq:tdwf}
\end{eqnarray}
where $\omega$ is determined by the equation
\begin{eqnarray}
\gamma^\mu\partial_\mu\,
\psi_{\bm{k},\lambda}^{(\kappa)}(x) &=& 
\gamma^0\left(\cI\,\frac{\partial}{\partial ct} +
\bm\alpha\cdot\bm \nabla\right)
\psi_{\bm{k},\lambda}^{(\kappa)}(x) = 0.
\elabel{eq:freqeq}
\end{eqnarray}
For transverse photons
\begin{eqnarray}
\omega &=& c|\bm k|,
\elabel{eq:transfreq}
\end{eqnarray}
and for longitudinal photons
\begin{eqnarray}
\omega &=& 0 .
\elabel{eq:longfreq}
\end{eqnarray}
The complete exponential factor is thus
\begin{eqnarray}
\re^{-\kappa \ri k\cdot x},
\elabel{eq:phasefactor}
\end{eqnarray}
where
\begin{eqnarray}
k &=& \left(\begin{array}{c} |\bm k| \\
\bm k\rc \end{array}\right) \quad \mbox{or}\quad
 \left(\begin{array}{c} 0 \\
\bm k\rc \end{array}\right),
\elabel{eq:kvector}
\end{eqnarray}
depending on whether the photon is transverse or longitudinal.
This corresponds to the eigenvalue equation
\begin{eqnarray}
\cH \, \psi_{\bm{k},\lambda}^{(\kappa)}(\bm x)
&=& \kappa \hbar\omega
 \, \psi_{\bm{k},\lambda}^{(\kappa)}(\bm x)
\end{eqnarray}
and a time dependence given by
\begin{eqnarray}
\psi_{\bm{k},\lambda}^{(\kappa)}(x)
&=&
\re^{-\ri \cH t/\hbar}
\psi_{\bm{k},\lambda}^{(\kappa)}(\bm x).
\elabel{eq:tdep}
\end{eqnarray}

It is also of interest to consider the effect of a hypothetical
photon mass $m_\gamma$ on the longitudinal photon wave function.
Such a modification, with an infinitesimal mass, resolves an
ambiguity in the construction of the Green function as discussed
in Sec.~\ref{sec:prop}.  Following the form of the Dirac
equation in Eq.~(\ref{eq:dirac}), we have
\begin{eqnarray}
\left(
\ri\,\hbar\gamma^\mu\partial_\mu - m_\gamma c
\right)
\psi_{\bm{k},0}^{(\kappa)}(x) = 0
\elabel{eq:dmeq}
\end{eqnarray}
or
\begin{eqnarray}
\left(\frac{\hbar\kappa\omega}{c}\gamma^0
- m_\gamma c \, \cI\right)
\psi_{\bm{k},0}^{(\kappa)}(x) = 0,
\end{eqnarray}
which yields
\begin{eqnarray}
\hbar\omega &=& m_\gamma c^2,
\end{eqnarray}
since
\begin{eqnarray}
\gamma^0
\psi_{\bm{k},0}^{(\kappa)}(x) &=&
\kappa \,
\psi_{\bm{k},0}^{(\kappa)}(x).
\end{eqnarray}

\subsection{Rotation of the wave functions}
\elabel{ssec:rtwf}

The rotations of the wave functions follow from the discussion
of Sec.~\ref{ssec:psirots}, with an additional consideration of
the vector $\bm k$.  On physical grounds, a rotation
parameterized by the vector $\bm u$ of the state of a photon
means rotation of the vector $\bm k$ into the vector $\bm
k^\prime$, according to
\begin{eqnarray}
\bm k^\prime &=& \bm R(\bm u) \bm k,
\end{eqnarray}
where $\bm R(\bm u)$ is defined by Eq.~(\ref{eq:xrot}).
Similarly, the polarization vector is transformed by the
spherical rotation operator
\begin{eqnarray}
\bm{\hat\epsilon}_\lambda(\bm{\hat k}^\prime) &=&
\bm R\rs(\bm u) \, \bm{\hat\epsilon}_\lambda
(\bm{\hat k})
\elabel{eq:polrot}
\end{eqnarray}
and
\begin{eqnarray}
\bm \tau\cdot \bm{\hat k}^\prime \,
\bm{\hat\epsilon}_\lambda(\bm{\hat k}^\prime) &=&
\bm R\rs(\bm u) \, 
\bm \tau\cdot\bm{\hat k} \, \bm R\rs^{-1}(\bm u)
\bm R\rs(\bm u) \, \bm{\hat\epsilon}_\lambda
(\bm{\hat k}),
\qquad
\end{eqnarray}
\\[2 pt]
where the rotation angle $\theta$, axis direction $\bm{\hat u}$,
$\bm{\hat k}$, and $\bm {\hat k}^\prime$ are related by
\begin{eqnarray}
\bm{\hat u} \sin{\theta} &=& \bm{\hat k}\times\bm{\hat
k}^\prime. 
\elabel{ukkp}\\[-5 pt] \nonumber
\end{eqnarray}
For applications, the rotation operator can be expressed as a
function of $\bm{\hat k}$ and $\bm{\hat k}^\prime$ rather than
$\bm u$.  From
\begin{eqnarray}
 \bm R\rs(\bm u) &=&
\bm I
-(\bm{\tau}\cdot\bm{\hat u})^2 \left(1-\cos{\theta}\right) 
- \ri \, \bm{\tau}\cdot\bm{\hat u} \,\sin{\theta},
\qquad
\end{eqnarray}
one has for transverse polarization, $\lambda =
1,2$,\footnote{The identity $\bm \tau\cdot\bm k\times \bm
k^\prime
 = \ri(\bm k^\prime\rs\,\bm k\rs^\dagger - \bm k\rs \,\bm
k\rs^{\prime\dagger})$ is useful here.}
\begin{eqnarray}
\bm{\hat\epsilon}_\lambda(\bm{\hat k}^\prime) &=&
\frac{(\bm\tau\cdot\bm{\hat
k}^\prime)^2 + \bm\tau\cdot\bm{\hat
 k}^\prime\,\bm\tau\cdot\bm{\hat k}}{1+\bm {\hat k}^\prime
 \cdot\bm{\hat k}} \,
\bm{\hat\epsilon}_\lambda
(\bm{\hat k}),
\\
\bm\tau\cdot\bm{\hat k}^\prime\,
\bm{\hat\epsilon}_\lambda(\bm{\hat k}^\prime) &=&
\frac{(\bm\tau\cdot\bm{\hat
k}^\prime)^2 + \bm\tau\cdot\bm{\hat
 k}^\prime\,\bm\tau\cdot\bm{\hat k}}{1+\bm {\hat k}^\prime
 \cdot\bm{\hat k}} \, \bm\tau\cdot\bm{\hat k} \,
\bm{\hat\epsilon}_\lambda
(\bm{\hat k}), \qquad
\end{eqnarray}
and for longitudinal polarization, $\lambda = 0$,
\begin{eqnarray}
\bm{\hat\epsilon}_0(\bm{\hat k}^\prime) &=&
\bm{\hat k}\rs^\prime \bm{\hat k}\rs^\dagger\,
  \bm{\hat\epsilon}_0
  (\bm{\hat k}).
  \end{eqnarray}

We thus have for a rotation of the wave functions characterized
by the vector $\bm u$
\begin{eqnarray}
\psi_{\bm{k}^\prime\!,\,\lambda}^{(\kappa)}(x) &=&
\cR(\bm u) \,
\psi_{\bm{k},\lambda}^{\,(\kappa)}\big(R^{-1}(\bm
u)\, x\big),
\elabel{eq:rotform}
\end{eqnarray}
where $\bm k^\prime\cdot\bm x = \bm k\cdot\bm R^{-1}(\bm u)\,\bm
x$ in the exponent of the wave function.  This yields the result 
\begin{eqnarray} 
\gamma^\mu\partial_\mu
\psi_{\bm{k}^\prime\!,\,\lambda}^{(\kappa)}(x) &=& 0,
\elabel{eq:roteq}
\end{eqnarray}
according to the discussion in Sec.~\ref{ssec:psirots}.

The expected transformation in Eq.~(\ref{eq:rotform}) can be
confirmed by an explicit calculation based on the completeness of
the wave functions.  For a rotation of a transverse wave
function, we write
\begin{eqnarray}
&&\cR(\bm u)\,
\psi_{\bm{k},\lambda}^{(\kappa)}
\big(R^{-1}(\bm u)\, x\big) =
\int \rd \bm x_1 \, \delta(\bm x - \bm x_1) \,
\cR(\bm u)\,
\psi_{\bm{k},\lambda}^{(\kappa)}
\big(\bm R^{-1}(\bm u)\, \bm x_1\big)\,\re^{-\kappa\ri\omega t}
\nonumber\\ &&\qquad=
\sum_{\kappa_1\rightarrow\pm}\sum_{\lambda_1=1}^2 \int \rd \bm k_1
\int\rd \bm x_1 \
\psi_{\bm{k}_1,\lambda_1}^{(\kappa_1)}(\bm x)
\psi_{\bm{k}_1,\lambda_1}^{(\kappa_1)\dagger}(\bm x_1)
\cR(\bm u)\,
\psi_{\bm{k},\lambda}^{(\kappa)} 
\big(\bm R^{-1}(\bm u)\, \bm x_1\big) \,\re^{-\kappa\ri\omega t}  ,
\qquad
\elabel{eq:rotcalc} 
\end{eqnarray}
where $\lambda = 1$ or $2$, and from Eq.~(\ref{eq:rotproj}) and
the subsequent remarks, it follows that the rotated wave
function is also transverse, so $\lambda_1$ is
restricted to 1 or 2.  The evaluation requires the matrix
element
\begin{eqnarray}
&&\int\rd \bm x_1 \
\psi_{\bm{k}_1,\lambda_1}^{(\kappa_1)\dagger}(\bm x_1)
\cR(\bm u)\,
\psi_{\bm{k},\lambda}^{(\kappa)}
\big(\bm R^{-1}(\bm u)\, \bm x_1\big)
\nonumber\\&&\qquad
= \frac{1}{2(2\pi)^3}\int \rd \bm x_1
\left[
\bm{\hat \epsilon}_{\lambda_1}^\dagger(\bm{\hat k}_1) \,
\bm R\rs(\bm u) \,
\bm{\hat \epsilon}_\lambda(\bm{\hat k}) 
+
\bm{\hat \epsilon}_{\lambda_1}^\dagger(\bm{\hat k}_1) \,
\bm \tau\cdot\bm{\hat k}_1 \,
\bm R\rs(\bm u) \, \bm \tau\cdot\bm{\hat k} \,
\bm{\hat \epsilon}_\lambda(\bm{\hat k}) 
\right]
\nonumber\\ &&\qquad\qquad\qquad\qquad\times
\re^{-\kappa_1\ri \bm k_1\cdot \bm x_1}
\re^{\kappa\ri \bm k\cdot \bm R^{-1}(\bm u) \bm x_1}
\nonumber\\&& \qquad
= \frac{1}{2}\,\delta_{\kappa_1 \kappa}
\left[
\bm{\hat \epsilon}_{\lambda_1}^\dagger(\bm{\hat k^\prime}) \,
\bm R\rs(\bm u) \,
\bm{\hat \epsilon}_\lambda(\bm{\hat k}) 
+
\bm{\hat \epsilon}_{\lambda_1}^\dagger(\bm{\hat k^\prime}) \,
\bm \tau\cdot\bm{\hat k^\prime} \,
\bm R\rs(\bm u) \, \bm \tau\cdot\bm{\hat k} \,
\bm{\hat \epsilon}_\lambda(\bm{\hat k}) 
\right]
\delta(\bm k_1 - \bm k^\prime) \,
\qquad
\nonumber\\&& \qquad
= \delta_{\kappa_1 \kappa} \,
\delta_{\lambda_1 \lambda} \,
\delta(\bm k_1 - \bm k^\prime).
\elabel{eq:matel}
\end{eqnarray}
In the exponent in Eq.~(\ref{eq:matel}), we have $\bm k\cdot \bm
R^{-1}(\bm u) \bm x_1 = \bm R(\bm u)\bm k \cdot \bm x_1 = \bm
k^\prime \cdot \bm x_1$.  The factor $\delta_{\kappa_1 \kappa}$
results from the requirement that $\bm k^\prime \rightarrow \bm
k$ as the rotation angle $\theta \rightarrow 0$, {\it i.e.,}
that $\bm k$ does not change sign for an infinitesimal rotation.
The factor $\delta_{\lambda_1 \lambda}$ follows from
Eq.~(\ref{eq:polrot}) and the discussion that follows it,
together with the orthonormality of the polarization vectors.
Substitution of Eq.~(\ref{eq:matel}) into (\ref{eq:rotcalc})
yields
\begin{eqnarray}
\cR(\bm u)\,
\psi_{\bm{k},\lambda}^{(\kappa)}
\big(R^{-1}(\bm u)\, x\big) &=&
\psi_{\bm{k}^\prime,\lambda}^{(\kappa)}(x), 
\end{eqnarray}
in accord with the general arguments leading to
Eq.~(\ref{eq:rotform}).

For a rotation of a longitudinal wave function, we have
\begin{eqnarray}
&&\cR(\bm u)\,
\psi_{\bm{k},0}^{(\kappa)}
\big(R^{-1}(\bm u)\, x\big) 
=
\int \rd \bm x_1 \, \delta(\bm x - \bm x_1) \,
\cR(\bm u)\,
\psi_{\bm{k},0}^{(\kappa)}
\big(\bm R^{-1}(\bm u)\, \bm x_1\big)
\nonumber\\&&\qquad\qquad =
\sum_{\kappa_1\rightarrow\pm} \int \rd \bm k_1
\int\rd \bm x_1 \
\psi_{\bm{k}_1,0}^{(\kappa_1)}(\bm x)
\psi_{\bm{k}_1,0}^{(\kappa_1)\dagger}(\bm x_1)
\cR(\bm u)\,
\psi_{\bm{k},0}^{(\kappa)} 
\big(\bm R^{-1}(\bm u)\, \bm x_1\big)
\qquad
\elabel{eq:rotcalc0} 
\end{eqnarray}
where only longitudinal wave functions contribute, and
the evaluation requires the matrix element
\begin{eqnarray}
&&\int\rd \bm x_1 \
\psi_{\bm{k}_1,0}^{(\kappa_1)\dagger}(\bm x_1)
\cR(\bm u)\,
\psi_{\bm{k},0}^{(\kappa)}
\big(\bm R^{-1}(\bm u)\, \bm x_1\big)
\nonumber\\&&\qquad\qquad=
 \frac{1}{(2\pi)^3}\int \rd \bm x_1
\delta_{\kappa_1 \kappa} \,
\bm{\hat \epsilon}_{0}^\dagger(\bm{\hat k}_1) \,
\bm R\rs(\bm u) \,
\bm{\hat \epsilon}_0(\bm{\hat k}) \,
\re^{-\kappa_1\ri \bm k_1\cdot \bm x_1}
\re^{\kappa\ri \bm k\cdot \bm R^{-1}(\bm u) \bm x_1}
\qquad
\nonumber\\&&\qquad\qquad=
\delta_{\kappa_1 \kappa} \,
\bm{\hat \epsilon}_0^\dagger(\bm{\hat k^\prime}) \,
\bm R\rs(\bm u) \,
\bm{\hat \epsilon}_0(\bm{\hat k})  \,
\delta(\bm k_1 - \bm k^\prime) \,
\nonumber\\&&\qquad\qquad=
 \delta_{\kappa_1 \kappa} \,
\delta(\bm k_1 - \bm k^\prime).
\elabel{eq:matel0}
\end{eqnarray}
Substitution of Eq.~(\ref{eq:matel0}) into (\ref{eq:rotcalc0})
yields
\begin{eqnarray}
\cR(\bm u)\,
\psi_{\bm{k},0}^{(\kappa)}
\big(R^{-1}(\bm u)\, x\big) &=&
\psi_{\bm{k}^\prime,0}^{(\kappa)}(x), 
\end{eqnarray}
in agreement with Eq.~(\ref{eq:rotform}).

\subsection{Velocity transformation of the wave functions}
\elabel{ssec:vtwf}

\subsubsection{Transverse wave functions}
\elabel{sssec:twf}

Under the velocity transformation of a transverse photon by a
velocity $\bm v$, the four-vector
\begin{eqnarray}
k &=& \left(\begin{array}{c} |\bm k| \\ \bm k\rc
\end{array}\right)
\end{eqnarray}
is transformed to
\begin{eqnarray}
k^\prime &=& V(\bm v) \, k
=\left(\begin{array}{c} |\bm k|
\left(\cosh{\zeta} + \bm{\hat v}\cdot\bm{\hat
k}\sinh{\zeta}\right)
\\
\bm k\rc + |\bm k|\bm{\hat v}\rc \!
\left[ \sinh{\zeta} +
\bm{\hat v}\cdot\bm{\hat k}(\cosh{\zeta}-1)
\right]
\end{array}\right), \qquad
\end{eqnarray}
and the wave function is expected to transform according to
\begin{eqnarray}
\xi\,\psi_{\bm{k}^\prime\!,\,\lambda}^{\prime\,(\kappa)}(x) &=&
\cV(\bm v)\,
\psi_{\bm{k},\lambda}^{\,(\kappa)}
\big(V^{-1}(\bm v)\, x\big).
\elabel{eq:trtr}
\end{eqnarray}
The prime on the transformed function indicates that it is a
function of modified polarization vectors, which in general are
not simply rotated vectors corresponding to $\bm{\hat k}
\rightarrow \bm{\hat k}^\prime$.  The transformed transverse
wave function is proportional to 
\begin{eqnarray}
\cV(\bm v) 
\left(\begin{array}{c}\bm{\hat \epsilon}_\lambda(\bm{\hat k}) \\
       \bm \tau \cdot \bm{\hat k} \
       \bm{\hat \epsilon}_\lambda(\bm{\hat k})
       \end{array}\right) &=&
\left(\begin{array}{c}
\left\{ I +(\bm \tau\cdot\bm{\hat v})^2 \, (\cosh{\zeta}-1)
+\bm \tau\cdot\bm{\hat v}\,\bm \tau \cdot \bm{\hat k} \, \sinh{\zeta}
\right\} \bm{\hat \epsilon}_\lambda(\bm{\hat k}) 
		      \\
\left\{\bm \tau\cdot\bm{\hat v} \sinh{\zeta} +
\left[I +(\bm \tau\cdot\bm{\hat v})^2 \, (\cosh{\zeta}-1)
\right] \bm \tau \cdot \bm{\hat k} \
	\right\}
        \bm{\hat \epsilon}_\lambda(\bm{\hat k})
    \end{array}\right) 
    \qquad
    \nonumber\\&=& \xi
\left(\begin{array}{c}\bm{\hat
\epsilon}_\lambda^\prime(\bm{\hat k}^\prime) \\
       \bm \tau \cdot \bm{\hat k}^\prime \
       \bm{\hat \epsilon}_\lambda^\prime(\bm{\hat k}^\prime)
       \end{array}\right),
\elabel{eq:eptr}
\end{eqnarray}
where the fact that transformed wave function can be written in
the form given at the right-hand end is based on the three
identities:
\begin{eqnarray}
\left|\left\{ I +(\bm \tau\cdot\bm{\hat v})^2 \, (\cosh{\zeta}-1)
+\bm \tau\cdot\bm{\hat v}\,\bm \tau \cdot \bm{\hat k} \, \sinh{\zeta}
\right\} \bm{\hat \epsilon}_\lambda(\bm{\hat k})\right| &=& 
\cosh{\zeta}+\bm{\hat v}\cdot\bm{\hat k}\sinh{\zeta},
\qquad
\elabel{eq:key0}
\end{eqnarray}
\begin{eqnarray}
{\bm k\rs^\prime}^\dagger 
\left\{ I +(\bm \tau\cdot\bm{\hat v})^2 \, (\cosh{\zeta}-1)
+\bm \tau\cdot\bm{\hat v}\,\bm \tau \cdot \bm{\hat k} \, \sinh{\zeta}
\right\} \bm{\hat \epsilon}_\lambda(\bm{\hat k}) &=& 0,
\elabel{eq:key1}
\end{eqnarray}
\begin{eqnarray}
&&\bm \tau\cdot\bm{\hat v} \sinh{\zeta} +
\left[I +(\bm \tau\cdot\bm{\hat v})^2 \, (\cosh{\zeta}-1)
\right] \bm \tau \cdot \bm{\hat k} \
\nonumber\\&&\qquad\qquad
 = \bm\tau\cdot\bm{\hat k}^\prime
\left\{ I +(\bm \tau\cdot\bm{\hat v})^2 \, (\cosh{\zeta}-1)
+\bm \tau\cdot\bm{\hat v}\,\bm \tau \cdot \bm{\hat k} \, \sinh{\zeta}
\right\}, \quad
\elabel{eq:key2}
\end{eqnarray}
where
\begin{eqnarray}
\bm{\hat k}^\prime &=& \frac{\bm{\hat k} + \bm{\hat v} \!
\left[ \sinh{\zeta} +
\bm{\hat v}\cdot\bm{\hat k}(\cosh{\zeta}-1) \right]}
{\cosh{\zeta}+\bm{\hat v}\cdot\bm{\hat k}\sinh{\zeta}}.
\end{eqnarray}
Equation~(\ref{eq:key0}) determines the scalar multiplicative
factor in Eq.~(\ref{eq:trtr}) to be
\begin{eqnarray}
\xi &=& \cosh{\zeta}+\bm{\hat v}\cdot\bm{\hat k}\sinh{\zeta}.
\end{eqnarray}
According to Eq.~(\ref{eq:key1}), the transformed polarization
vector is in the plane perpendicular to $\bm k^\prime$, that is,
the transformed transverse polarization vector is also
transverse.  This is in contrast to the vector potential, which
does not maintain transversality under a velocity
transformation, so the Coulomb, or radiation, gauge condition is
not preserved.  The difference is just the fact that the vector
potential transforms as a vector, so that the angle between the
space component of the vector potential and the space component
of the wave vector is not necessarily preserved under a velocity
transformation, whereas the polarization vector transforms as
the electric field, {\it i.e.,} as a component of a second-rank
tensor, and Eq.~(\ref{eq:key1}) shows that for this case the
transversality is preserved.  Equation~(\ref{eq:key2}) shows
that the lower components of the transformed wave function, as
given in Eq.~(\ref{eq:eptr}), can be written as
$\bm\tau\cdot\bm{\hat k}^\prime$ times the upper components in
the same expression.  This together with the relation
$k^\prime\cdot x = k\cdot V^{-1}(\bm v)\,x$ in the exponent of
the wave function insures that the transformed wave function is
a solution of the source-free Maxwell equation.  It also follows
from Eq.~(\ref{eq:eptr}) that if $\lambda_2 \ne \lambda_1$, then
\begin{eqnarray}
\bm{\hat\epsilon}_{\lambda_2}^{\prime\,\dagger}(\bm{\hat k}^\prime)
\,
\bm{\hat\epsilon}_{\lambda_1}^\prime(\bm{\hat k}^\prime) = 0,
\elabel{eq:tror}
\end{eqnarray}
so the orthogonality of the polarization vectors is preserved by
the velocity transformation.  

Equation~(\ref{eq:trtr}) can be be obtained by an
explicit calculation based on the completeness of the wave
functions, as for rotations.
A velocity transformation of a transverse wave function is given by
\begin{eqnarray}
&&\cV(\bm v)\,
\psi_{\bm{k},\lambda}^{(\kappa)}
\big(V^{-1}(\bm v)\, x\big) 
\nonumber\\&&\qquad\qquad=
\int \rd \bm x_1 \, \delta(\bm x - \bm x_1) \,
\cV(\bm v)\,
\psi_{\bm{k},\lambda}^{(\kappa)}
\big(V^{-1}(\bm v)\, x_1\big)
\nonumber\\&&\qquad\qquad=
\sum_{\kappa_1\rightarrow\pm}\sum_{\lambda_1=0}^2 \int \rd \bm k_1
\int\rd \bm x_1 \
\psi_{\bm{k}_1,\lambda_1}^{\prime\,(\kappa_1)}(\bm x)
\psi_{\bm{k}_1,\lambda_1}^{\prime\,(\kappa_1)\dagger}(\bm x_1)
\cV(\bm v)\,
\psi_{\bm{k},\lambda}^{(\kappa)} 
\big(V^{-1}(\bm v)\, x_1\big),
\qquad
\elabel{eq:explicit}
\end{eqnarray}
where $\lambda = 1$ or $2$, $t_1 = t$, and the primes on the
wave functions indicate that the velocity transformed
polarization vectors provide the basis vectors for $\lambda_1 =
1,2$.  For $\lambda_1 = 0$,
\begin{eqnarray}
&&\int\rd \bm x_1 \
\psi_{\bm{k}_1,0}^{\prime\,(+)\dagger}(\bm x_1)
\cV(\bm v)\,
\psi_{\bm{k},\lambda}^{(\kappa)}
\big(V^{-1}(\bm v)\, x_1\big)
\nonumber\\&& \qquad
= \frac{1}{\sqrt{2}(2\pi)^3}\int \rd \bm x_1
\left(\begin{array}{cc}
\bm{\hat \epsilon}_0^{\prime\dagger}(\bm{\hat k}_1) \,
& \0
\end{array}\right)
\cV(\bm v) 
\left(\begin{array}{c}\bm{\hat \epsilon}_\lambda(\bm{\hat k}) \\
       \bm \tau \cdot \bm{\hat k} \
       \bm{\hat \epsilon}_\lambda(\bm{\hat k})
       \end{array}\right)
\re^{-\ri \bm k_1\cdot \bm x_1}
\re^{-\kappa\ri k\cdot V^{-1}(\bm v) x_1}
\qquad
\nonumber\\&& \qquad
= \frac{1}{\sqrt{2}}\,\delta_{+ \kappa}
\, \xi \,
\bm{\hat \epsilon}_0^{\prime\dagger}(\bm{\hat k}^\prime) \,
\bm{\hat \epsilon}_\lambda^\prime(\bm{\hat k}^\prime) \,
\delta(\bm k_1 - \bm k^\prime) \,
\re^{-\ri k^{\prime0} c t}
= 0 ,
\elabel{eq:vme0p}
\\[10 pt]
&&\int\rd \bm x_1 \
\psi_{\bm{k}_1,0}^{\prime\,(-)\dagger}(\bm x_1)
\cV(\bm v)\,
\psi_{\bm{k},\lambda}^{(\kappa)}
\big(V^{-1}(\bm v)\, x_1\big)
\nonumber\\&& \qquad
= \frac{1}{\sqrt{2}(2\pi)^3}\int \rd \bm x_1
\left(\begin{array}{cc}
\0 &
\bm{\hat \epsilon}_0^{\prime\dagger}(\bm{\hat k}_1)
\end{array}\right)
\cV(\bm v) 
\left(\begin{array}{c}\bm{\hat \epsilon}_\lambda(\bm{\hat k}) \\
       \bm \tau \cdot \bm{\hat k} \
       \bm{\hat \epsilon}_\lambda(\bm{\hat k})
       \end{array}\right)
\re^{\ri \bm k_1\cdot \bm x_1}
\re^{-\kappa\ri k\cdot V^{-1}(\bm v) x_1}
\nonumber\\&& \qquad
= \frac{1}{\sqrt{2}}\,\delta_{- \kappa}
\, \xi \,
\bm{\hat \epsilon}_0^{\prime\dagger}(\bm{\hat k}^\prime) \,
\bm\tau\cdot\bm{\hat k}^\prime 
\bm{\hat \epsilon}_\lambda^\prime(\bm{\hat k}^\prime) \,
\delta(\bm k_1 - \bm k^\prime) \,
\re^{\ri k^{\prime0}c t}
= 0 .
\elabel{eq:vme0m}
\end{eqnarray}
In the exponent in Eqs.~(\ref{eq:vme0p}) and (\ref{eq:vme0m}),
$k\cdot V^{-1}(\bm v) x_1 = V(\bm v) k \cdot  x_1 = k^\prime
\cdot x_1$.  As for rotations, the factors $\delta_{+ \kappa}$
and $\delta_{- \kappa}$ result from the requirement that
$k^\prime \rightarrow k$ as the velocity $|\bm v| \rightarrow
0$, and the last equalities follow from Eqs.~(\ref{eq:key1}) and
(\ref{eq:key2}). For $\lambda_1 = 1$ or $2$, we have
\begin{eqnarray}
&&\int\rd \bm x_1 \
\psi_{\bm{k}_1,\lambda_1}^{\prime(\kappa_1)\dagger}(\bm x_1)
\cV(\bm v)\,
\psi_{\bm{k},\lambda}^{(\kappa)}
\big(V^{-1}(\bm v)\, x_1\big)
\nonumber\\&& \qquad\qquad 
= \frac{1}{2(2\pi)^3}\int \rd \bm x_1
\left(\begin{array}{cc}
\bm{\hat \epsilon}_{\lambda_1}^{\prime\dagger}(\bm{\hat k}_1) \,
& 
\bm{\hat \epsilon}_{\lambda_1}^{\prime\dagger}(\bm{\hat k_1}) \,
       \bm \tau \cdot \bm{\hat k}_1 \
\end{array}\right)
\cV(\bm v) 
\left(\begin{array}{c}\bm{\hat \epsilon}_\lambda(\bm{\hat k}) \\
       \bm \tau \cdot \bm{\hat k} \
       \bm{\hat \epsilon}_\lambda(\bm{\hat k})
       \end{array}\right)
       \qquad
       \nonumber\\&&\qquad\qquad\qquad\times
\re^{-\kappa_1\ri \bm k_1\cdot \bm x_1}
\re^{-\kappa\ri k\cdot V^{-1}(\bm v) x_1}
\nonumber\\&& \qquad\qquad 
= \delta_{\kappa_1 \kappa} \,
\delta_{\lambda_1 \lambda} \,
\xi \,
\delta(\bm k_1 - \bm k^\prime) \,
\re^{-\kappa\ri k^{\prime0}c t} .
\elabel{eq:vme12}
\end{eqnarray}
Substitution of Eq.~(\ref{eq:vme12}) into (\ref{eq:explicit})
yields
\begin{eqnarray}
\cV(\bm v)\,
\psi_{\bm{k},\lambda}^{(\kappa)}
\big(V^{-1}(\bm v)\, x\big) &=& \xi\,
\psi_{\bm{k}^\prime,\lambda}^{\,\prime\,(\kappa)}(x), 
\end{eqnarray}
where
\begin{eqnarray}
\xi\,\bm{\hat \epsilon}_\lambda^\prime(\bm{\hat k}^\prime) &=&
\left\{ I +(\bm \tau\cdot\bm{\hat v})^2 \, (\cosh{\zeta}-1)
+\bm \tau\cdot\bm{\hat v}\,\bm \tau \cdot \bm{\hat k} \,
\sinh{\zeta}
\right\} \bm{\hat \epsilon}_\lambda(\bm{\hat k}) ,
\end{eqnarray}
in accord with Eq.~(\ref{eq:trtr}).

\subsubsection{Longitudinal wave functions}
\elabel{sssec:lwf}

For a longitudinal solution, the four-vector $k$ in the
invariant phase factor $k\cdot x$, given by 
\begin{eqnarray} k
&=& \left(\begin{array}{c} 0 \\ \bm k\rc \end{array}\right)
\end{eqnarray} 
from Eqs.~(\ref{eq:longfreq})-(\ref{eq:kvector}), has a zero
time component.  However, the transformed phase, given by
$k\cdot V^{-1}({\bm v})\,x = V({\bm v})k\cdot x = k^\prime\cdot
x$, where
\begin{eqnarray}
k^\prime &=& V(\bm v) \, k
=
\left(\begin{array}{c} 
\bm{\hat v}\cdot\bm{ k}\sinh{\zeta}
\\
\bm k\rc + \bm{\hat v}\rc  \,
\bm{\hat v}\cdot\bm{k}(\cosh{\zeta}-1)
\end{array}\right), \qquad
\end{eqnarray}
does have time dependence.  
Thus the wave function is expected to transform according to
\begin{eqnarray}
\sum_{\kappa^\prime\rightarrow\pm}
\sum_{\lambda^\prime=0}^3\xi_{\lambda^\prime\,0}^{\kappa^\prime\kappa}(t)\,
\psi_{\bm{k}^\prime\!,\,\lambda^\prime}^{\,\prime\,(\kappa^\prime)}(x) &=&
\cV(\bm v)\,
\psi_{\bm{k},0}^{(\kappa)}
\big(V^{-1}(\bm v)\, x\big),
\label{eq:longtr}
\end{eqnarray}
where the coefficients include the extra time dependence
introduced by the velocity transformation.  The sum over
polarization states is necessary, because unlike the case of the
transverse wave function, the transformed longitudinal wave
function is mixture of longitudinal and transverse components.
This is expected on physical grounds, since moving charges cause
radiative atomic transitions.  On the other hand, the
transformed space-like wave vector does not the match the wave
vector of either the longitudinal or transverse basis functions.
Since the solutions are classified according to their
three-wave-vector, there is a residual time dependence in the
expansion that is included in the transformation coefficients.  

To be explicit, for $\lambda \ne 0$, we consider specific
polarization vectors.  Let $\bm{\hat
\epsilon}^\prime_1(\bm{\hat k}^\prime)$ be a linear
polarization vector in the plane of $\bm{\hat k}^\prime$ and
$\bm{\hat v}$ and $\bm{\hat \epsilon}^\prime_2(\bm{\hat
k}^\prime)$ be a linear polarization vector perpendicular to
$\bm{\hat k}^\prime$ and $\bm{\hat v}$.  These conditions,
together with $\bm k\rs^{\prime\,\dagger}\,\bm{\hat
\epsilon}^\prime_1(\bm{\hat k}^\prime) = 0$, yield (up to
phase factors)
\begin{eqnarray}
\bm{\hat \epsilon}_1^\prime(\bm{\hat k}^\prime) &=&
\frac{  \bm{\hat v}\cdot\bm{\hat k}^\prime\,
\bm{\hat k}\rs^\prime - \bm{\hat v}\rs}
{\sqrt{1-(\bm{\hat v}\cdot\bm{\hat k}^\prime)^2}} ,
\label{eq:leps1}
\\
\bm{\hat \epsilon}_2^\prime(\bm{\hat k}^\prime) &=&
\frac{\bm\tau\cdot\bm{\hat v}\,\bm{\hat k}\rs^\prime}
{\sqrt{1-(\bm{\hat v}\cdot\bm{\hat k}^\prime)^2}} ,
\label{eq:leps2}
\end{eqnarray}
where
\begin{eqnarray}
\bm{\hat \epsilon}^\prime_2(\bm{\hat k}^\prime) &=&
\bm\tau\cdot\bm{\hat k}^\prime \,
\bm{\hat \epsilon}^\prime_1(\bm{\hat k}^\prime) ,
\\
\bm{\hat \epsilon}^\prime_1(\bm{\hat k}^\prime) &=&
\bm\tau\cdot\bm{\hat k}^\prime \,
\bm{\hat \epsilon}^\prime_2(\bm{\hat k}^\prime) .
\end{eqnarray}
The transformed upper- and lower-component longitudinal wave
functions follow from the expressions (for $t=0$ and $\bm x =
0$)
\begin{eqnarray}
\cV(\bm v) 
\left(\begin{array}{c}\bm{\hat \epsilon}_0(\bm{\hat k}) \\
\0
\end{array}\right) &=&
\left(\begin{array}{c}
\left[ \bm I +(\bm \tau\cdot\bm{\hat v})^2 \, (\cosh{\zeta}-1)
\right] \bm{\hat \epsilon}_0(\bm{\hat k})
\\
\bm \tau\cdot\bm{\hat v} \sinh{\zeta} \,
\bm{\hat \epsilon}_0(\bm{\hat k})
\end{array}\right)
= \xi_{00}^{++}(0)
\left(\begin{array}{c}\bm{\hat \epsilon}^\prime_0(\bm{\hat k}^\prime) \\
       \0
       \end{array}\right)
       \nonumber\\[10 pt]&&
    +  \frac{\xi_{10}^{++}(0)}{\sqrt{2}}
\left(\begin{array}{c}\bm{\hat \epsilon}^\prime_1(\bm{\hat k}^\prime) \\
       \bm \tau\cdot\bm{\hat k}^\prime \,
       \bm{\hat \epsilon}^\prime_1(\bm{\hat k}^\prime)
       \end{array}\right)
    + \frac{\xi_{10}^{-+}(0)}{\sqrt{2}}
\left(\begin{array}{c}\bm{\hat \epsilon}^\prime_1(\bm{\hat k}^\prime) \\
       -\bm \tau\cdot\bm{\hat k}^\prime \,
       \bm{\hat \epsilon}^\prime_1(\bm{\hat k}^\prime)
       \end{array}\right),
\qquad
\elabel{eq:vmepp}
\end{eqnarray}
and
\begin{eqnarray}
\cV(\bm v) 
\left(\begin{array}{c}
\0 \\ \bm{\hat \epsilon}_0(\bm{\hat k})
\end{array}\right) &=&
\left(\begin{array}{c}
\bm \tau\cdot\bm{\hat v} \sinh{\zeta} \,
\bm{\hat \epsilon}_0(\bm{\hat k})
\\
\left[ \bm I +(\bm \tau\cdot\bm{\hat v})^2 \, (\cosh{\zeta}-1)
\right] \bm{\hat \epsilon}_0(\bm{\hat k})
\end{array}\right)
= \xi_{00}^{--}(0)
\left(\begin{array}{c}
\0 \\
\bm{\hat \epsilon}^\prime_0(\bm{\hat k}^\prime)
\end{array}\right)
\nonumber\\[10 pt]&&
    + \frac{\xi_{20}^{--}(0)}{\sqrt{2}}
\left(\begin{array}{c}\bm{\hat \epsilon}^\prime_2(\bm{\hat k}^\prime) \\
       \bm \tau\cdot\bm{\hat k}^\prime \,
       \bm{\hat \epsilon}^\prime_2(\bm{\hat k}^\prime)
       \end{array}\right)
    + \frac{\xi_{20}^{+-}(0)}{\sqrt{2}}
\left(\begin{array}{c}-\bm{\hat \epsilon}^\prime_2(\bm{\hat k}^\prime) \\
       \bm \tau\cdot\bm{\hat k}^\prime \,
       \bm{\hat \epsilon}^\prime_2(\bm{\hat k}^\prime)
       \end{array}\right),
\qquad
\elabel{eq:vmemm}
\end{eqnarray}
based on the relations
\begin{eqnarray}
\bm{\hat k} &=& \frac{\bm{\hat k}^\prime
- \bm{\hat v}\,\bm{\hat v}\cdot\bm{\hat
  k}^\prime\left(1-{\rm sech}{\,\zeta}\right)}
  {\sqrt{1 - (\bm{\hat v}\cdot\bm{\hat
  k}^\prime)^2\tanh^2{\zeta}}} ,
  \end{eqnarray}
  \begin{eqnarray}
\bm{\hat k}\rs^{\prime}\bm{\hat k}\rs^{\prime\dagger} 
\left[ \bm I +(\bm \tau\cdot\bm{\hat v})^2 \,
(\cosh{\zeta}-1)
\right] \bm{\hat k}\rs 
&=&
\cosh{\zeta}\,\sqrt{1-(\bm{\hat v}\cdot\bm{\hat
k}^\prime)^2\tanh^2{\zeta}} \  \bm{\hat k}\rs^{\prime},
\label{eq:xi0pu}
\\[8 pt]
\bm{\hat k}\rs^{\prime}\bm{\hat k}\rs^{\prime\dagger}
 \bm \tau\cdot\bm{\hat v} \sinh{\zeta} 
\,\bm{\hat k}\rs &=& 0,
\\[8 pt]
(\bm I - \bm{\hat k}\rs^{\prime}\bm{\hat
k}\rs^{\prime\dagger}) 
\left[ \bm I +(\bm \tau\cdot\bm{\hat v})^2 \,
(\cosh{\zeta}-1)
\right] \bm{\hat k}\rs &=&
\frac{\sinh{\zeta}\tanh{\zeta}\,
\bm{\hat v}\cdot\bm{\hat k}^\prime
(\bm{\hat v}\cdot\bm{\hat k}^\prime\bm{\hat
k}^\prime\rs-\bm{\hat v}\rs)}
{ \sqrt{1-(\bm{\hat v}\cdot\bm{\hat
k}^\prime)^2\tanh^2{\zeta}}}, \qquad
\\[8 pt]
(\bm I - \bm{\hat k}\rs^{\prime}\bm{\hat
k}\rs^{\prime\dagger})  \,
 \bm \tau\cdot\bm{\hat v} \sinh{\zeta}
\, \bm{\hat k}\rs &=&
\frac{\sinh{\zeta}\,
\bm\tau\cdot\bm{\hat k}^\prime(\bm{\hat v}\cdot\bm{\hat
k}^\prime\bm{\hat
k}^\prime\rs-\bm{\hat v}\rs)}
{ \sqrt{1-(\bm{\hat v}\cdot\bm{\hat
k}^\prime)^2\tanh^2{\zeta}}},
\label{eq:xi12pl}
\end{eqnarray}
with coefficients given by
\begin{eqnarray}
\xi_{00}^{++}(0) = \xi_{00}^{--}(0)
&=& \cosh{\zeta}\,\sqrt{1-(\bm{\hat
v}\cdot\bm{\hat k}^\prime)^2\tanh^2{\zeta}} \ ,
\\
\xi_{10}^{++}(0) = \xi_{20}^{--}(0) 
&=& \frac{\sinh{\zeta}}{\sqrt{2}}
\sqrt{1-(\bm{\hat v}\cdot\bm{\hat k}^\prime)^2} \,
\sqrt{\frac{1+\bm{\hat v}\cdot\bm{\hat k}^\prime \tanh{\zeta}}
{1-\bm{\hat v}\cdot\bm{\hat k}^\prime \tanh{\zeta}}} \ ,
\\
\xi_{10}^{-+}(0) = \xi_{20}^{+-}(0) 
&=& - \frac{\sinh{\zeta}}{\sqrt{2}}
\sqrt{1-(\bm{\hat v}\cdot\bm{\hat k}^\prime)^2} \,
\sqrt{\frac{1-\bm{\hat v}\cdot\bm{\hat k}^\prime \tanh{\zeta}}
{1+\bm{\hat v}\cdot\bm{\hat k}^\prime \tanh{\zeta}}} \ .
\end{eqnarray}
Terms that do not appear in Eq.~(\ref{eq:vmepp}) or
(\ref{eq:vmemm}) make no contribution
\begin{eqnarray}
\xi_{00}^{+-}(0) = \xi_{00}^{-+}(0) = 
\xi_{10}^{--}(0) = \xi_{10}^{+-}(0) = 
\xi_{20}^{++}(0) = \xi_{20}^{-+}(0) = 0.
\end{eqnarray}
An explicit calculation of Eq.~(\ref{eq:longtr}) is made by
evaluating
\begin{eqnarray}
&&\cV(\bm v)\,
\psi_{\bm{k},0}^{(\kappa)}
\big(V^{-1}(\bm v)\, x\big) =
\int \rd \bm x_1 \, \delta(\bm x - \bm x_1) \,
\cV(\bm v)\,
\psi_{\bm{k},0}^{(\kappa)}
\big(V^{-1}(\bm v)\, x_1\big)
\nonumber\\ &&\qquad\qquad=
\sum_{\kappa_1\rightarrow\pm}\sum_{\lambda_1=0}^2 \int \rd \bm k_1
\int\rd \bm x_1 \
\psi_{\bm{k}_1,\lambda_1}^{\prime\,(\kappa_1)}(\bm x)
\psi_{\bm{k}_1,\lambda_1}^{\prime\,(\kappa_1)\dagger}(\bm x_1)
\cV(\bm v)\,
\psi_{\bm{k},0}^{(\kappa)} 
\big(V^{-1}(\bm v)\, x_1\big).
\qquad
\elabel{eq:vtlong}
\end{eqnarray}
For $\lambda_1 = 0$,
\begin{eqnarray}
&&\int\rd \bm x_1 \,
\psi_{\bm{k}_1,0}^{\prime\,(+)\dagger}(\bm x_1)
\cV(\bm v)\,
\psi_{\bm{k},0}^{(+)}
\big(V^{-1}(\bm v)\, x_1\big)
\nonumber\\&&\qquad\qquad=
 \frac{1}{(2\pi)^3}\int \rd \bm x_1
\left(\begin{array}{cc}
\bm{\hat \epsilon}_0^{\prime\dagger}(\bm{\hat k}_1) \,
& \0
\end{array}\right)
\cV(\bm v) 
\left(\begin{array}{c}\bm{\hat \epsilon}_0(\bm{\hat k}) \\
       \0
       \end{array}\right)
\re^{-\ri \bm k_1\cdot \bm x_1}
\re^{-\ri k\cdot V^{-1}(\bm v) x_1}
\qquad
\nonumber\\&&\qquad\qquad=
\xi_{00}^{++}(0) \,
\delta(\bm k_1 - \bm k^\prime) \,
\re^{-\ri k^{\prime0} c t},
\elabel{eq:vme00pp}
\end{eqnarray}
\begin{eqnarray}
&&\int\rd \bm x_1 \,
\psi_{\bm{k}_1,0}^{\prime\,(-)\dagger}(\bm x_1)
\cV(\bm v)\,
\psi_{\bm{k},0}^{(-)}
\big(V^{-1}(\bm v)\, x_1\big)
\nonumber\\&&\qquad\qquad
= \frac{1}{(2\pi)^3}\int \rd \bm x_1
\left(\begin{array}{cc}
\0
& \bm{\hat \epsilon}_0^{\prime\dagger}(\bm{\hat k}_1)
\end{array}\right)
\cV(\bm v) 
\left(\begin{array}{c}
       \0 \\
       \bm{\hat \epsilon}_0(\bm{\hat k})
       \end{array}\right)
\re^{\ri \bm k_1\cdot \bm x_1}
\re^{\ri k\cdot V^{-1}(\bm v) x_1}
\qquad
\nonumber\\&&\qquad\qquad=
\xi_{00}^{--}(0) \,
\delta(\bm k_1 - \bm k^\prime) \,
\re^{\ri k^{\prime0} ct},
\elabel{eq:vme00mm}
\end{eqnarray}
where
\begin{eqnarray}
\bm{\hat k}^\prime &=& \frac{\bm{\hat k} + \bm{\hat v} \,
\bm{\hat v}\cdot\bm{\hat k}(\cosh{\zeta}-1)}
{\sqrt{1+(\bm{\hat v}\cdot\bm{\hat k})^2\sinh^2{\zeta}~}} ,
\end{eqnarray}
and $k^{\prime0}$ is the time component associated with
the transformed space-like vector $k$
\begin{eqnarray}
k^{\prime0} &=& \bm{\hat v}\cdot\bm k \, \sinh{\zeta}
= \bm{\hat v}\cdot\bm k^\prime \, \tanh{\zeta}.
\end{eqnarray}
For $\lambda_1 = 1$ or $2$,
\begin{eqnarray}
&&\int\rd \bm x_1 \,
\psi_{\bm{k}_1,\lambda_1}^{\prime\,(\kappa_1)\dagger}(\bm x_1)
\cV(\bm v)\,
\psi_{\bm{k},0}^{(+)}
\big(V^{-1}(\bm v)\, x_1\big)
\nonumber\\&& \qquad\qquad 
= \frac{1}{\sqrt{2}(2\pi)^3}\int \rd \bm x_1
\left(\begin{array}{cc}
\bm{\hat \epsilon}_{\lambda_1}^{\prime\dagger}(\bm{\hat k}_1) \,
&  \quad
\bm{\hat \epsilon}_{\lambda_1}^{\prime\dagger}(\bm{\hat k}_1) \,
       \bm \tau \cdot \bm{\hat k}_1 \
\end{array}\right)
\cV(\bm v) 
\left(\begin{array}{c}\bm{\hat \epsilon}_0(\bm{\hat k}) \\
       \0
       \end{array}\right)
\nonumber\\&&\qquad\qquad\qquad\times
\re^{-\kappa_1\ri \bm k_1\cdot \bm x_1}
\re^{-\ri k\cdot V^{-1}(\bm v) x_1}
\nonumber\\&& \qquad\qquad 
= \delta_{\kappa_1+}\delta_{\lambda_11} \,
\xi_{10}^{++}(0)\,\delta(\bm k_1 - \bm k^\prime) \,
\re^{-\ri k^{\prime0} c  t}
+ \delta_{\kappa_1-}\delta_{\lambda_11} \,
\xi_{10}^{-+}(0)\,\delta(\bm k_1 + \bm k^\prime) \,
\re^{-\ri k^{\prime0} c t} ,
\qquad
\end{eqnarray}
\begin{eqnarray}
&&\int\rd \bm x_1 \,
\psi_{\bm{k}_1,\lambda_1}^{\prime\,(\kappa_1)\dagger}(\bm x_1)
\cV(\bm v)\,
\psi_{\bm{k},0}^{(-)}
\big(V^{-1}(\bm v)\, x_1\big)
\nonumber\\&& \qquad\qquad 
= \frac{1}{\sqrt{2}(2\pi)^3}\int \rd \bm x_1
\left(\begin{array}{cc}
\bm{\hat \epsilon}_{\lambda_1}^{\prime\,\dagger}(\bm{\hat k}_1) \,
&  \quad
\bm{\hat \epsilon}_{\lambda_1}^{\prime\,\dagger}(\bm{\hat k}_1) \,
       \bm \tau \cdot \bm{\hat k}_1 \
\end{array}\right)
\cV(\bm v) 
\left(\begin{array}{c}
\0 \\
\bm{\hat \epsilon}_0(\bm{\hat k})
       \end{array}\right)
\nonumber\\&&\qquad\qquad\qquad\times
\re^{-\kappa_1\ri \bm k_1\cdot \bm x_1}
\re^{\ri k\cdot V^{-1}(\bm v) x_1}
\nonumber\\[6 pt]&& \qquad\qquad 
= \delta_{\kappa_1-}\delta_{\lambda_12} \,
\xi_{20}^{--}(0)\,\delta(\bm k_1 - \bm k^\prime) \,
\re^{\ri k^{\prime0} c t}
+ \delta_{\kappa_1+}\delta_{\lambda_12} \,
\xi_{20}^{+-}(0)\,\delta(\bm k_1 + \bm k^\prime) \,
\re^{\ri k^{\prime0} c t} .
\qquad
\end{eqnarray}
Here both signs of $\kappa_1$ are included, because there is no
continuity condition on the transverse solutions, which are
absent in the limit of small velocity transformations.  These
results yield Eq.~(\ref{eq:longtr}) with the non-zero
coefficients given by
\begin{eqnarray}
\xi_{00}^{++}(t) &=& \xi_{00}^{++}(0)\,
\re^{-\ri \bm{\hat v}\cdot\bm k^\prime
\tanh{\zeta}\,ct} \,, \qquad
\\
\xi_{00}^{--}(t) &=& \xi_{00}^{--}(0)\,
\re^{\ri \bm{\hat v}\cdot\bm k^\prime
\tanh{\zeta}\,ct} \,, \qquad
\\
\xi_{10}^{++}(t) &=& \xi_{10}^{++}(0)\,
\re^{\ri (|\bm k^\prime| - \bm{\hat v}\cdot\bm k^\prime
\tanh{\zeta})\,ct } \,, \qquad
\\
\xi_{10}^{-+}(t) &=& \xi_{10}^{-+}(0)\,
\re^{-\ri (|\bm k^\prime| + \bm{\hat v}\cdot\bm k^\prime
\tanh{\zeta})\,ct } \,, \qquad
\\
\xi_{20}^{+-}(t) &=& \xi_{20}^{+-}(0)\,
\re^{\ri (|\bm k^\prime| + \bm{\hat v}\cdot\bm k^\prime
\tanh{\zeta})\,ct } \,, \qquad
\\
\xi_{20}^{--}(t) &=& \xi_{20}^{--}(0)\,
\re^{-\ri (|\bm k^\prime| - \bm{\hat v}\cdot\bm k^\prime
\tanh{\zeta})\,ct } \,. \qquad
\end{eqnarray}
These transformed longitudinal solutions can be used, for
example, to describe the fields of a moving charge by means of
the expansion in Eq.~(\ref{eq:coultr}).

\subsection{Standing-wave parity eigenfunctions}
\elabel{ssec:ptwf}

The parity operator $\fP$ changes the sign of the coordinates
and includes multiplication by the matrix $\cP = -\gamma^0$, so
that the transformed wave function is also a solution of the
Maxwell equations, as discussed in Sec.~\ref{ssec:ptpsi}.  We
thus have
\begin{eqnarray}
\fP \psi_{\bm{k},\lambda}^{\,(\kappa)}(\bm x) &=& 
  -\gamma^0 \psi_{\bm{k},\lambda}^{\,(\kappa)}(-\bm x). 
\elabel{eq:parityform}
\end{eqnarray}
With this definition, the parity operator commutes with the
Hamiltonian in Eq.~(\ref{eq:phham})
\begin{eqnarray}
\fP \cH &=& \cH \fP,
\end{eqnarray}
so we may identify eigenstates of both parity and energy.
Since
\begin{eqnarray}
\fP^2 \psi_{\bm{k},\lambda}^{\,(\kappa)}(\bm x) &=&
 \psi_{\bm{k},\lambda}^{\,(\kappa)}(\bm x),
\end{eqnarray}
the parity eigenvalues are $\pm 1$.

Transverse parity and energy eigenstates are
\begin{eqnarray}
\psi_{\bm{k},\lambda}^{\,(\kappa,+)}(\bm x) &=& 
\frac{1}{\sqrt{(2\pi)^3}}
\left(\begin{array}{c} \kappa\ri\,
\bm{\hat \epsilon}_\lambda(\bm{\hat k})\sin{\bm k\cdot\bm x} \\
\bm\tau\cdot\bm{\hat k} \,
\bm{\hat \epsilon}_\lambda(\bm{\hat k})\cos{\bm k\cdot\bm x}
		      \end{array}\right),
		      \elabel{eq:ptpe}
\\
\psi_{\bm{k},\lambda}^{\,(\kappa,-)}(\bm x) &=& 
\frac{1}{\sqrt{(2\pi)^3}}
\left(\begin{array}{c}
\bm{\hat \epsilon}_\lambda(\bm{\hat k})\cos{\bm k\cdot\bm x} \\
\kappa\ri\,\bm \tau\cdot\bm{\hat k} \,
\bm{\hat \epsilon}_\lambda(\bm{\hat k})\sin{\bm k\cdot\bm x}
		      \end{array}\right),
		      \quad
		      \elabel{eq:ntpe}
\end{eqnarray}
where
\begin{eqnarray}
\fP \psi_{\bm{k},\lambda}^{\,(\kappa,\pm)}(\bm x) &=& 
\pm\psi_{\bm{k},\lambda}^{\,(\kappa,\pm)}(\bm x),
\label{eq:parityeqn}
\\ \cH \psi_{\bm{k},\lambda}^{\,(\kappa,\pm)}(\bm x) &=& 
\kappa \hbar c | \bm k|
\psi_{\bm{k},\lambda}^{\,(\kappa,\pm)}(\bm x),
\end{eqnarray}
and $\lambda = 1,~2$.  These states are linear combinations of
plane-wave states that form standing plane waves.  Orthogonality
relations are  calculated with the aid of the integrals
\begin{eqnarray}
\int\rd\bm x\,\cos{\bm k_2\cdot\bm x} \,\cos{\bm k_1\cdot\bm x}
&=&
 4\pi^3\left[\delta(\bm k_2-\bm k_1)+\delta(\bm k_2+\bm k_1)
\right],
\\
\int\rd\bm x\,\sin{\bm k_2\cdot\bm x} \,\sin{\bm k_1\cdot\bm x}
&=&
 4\pi^3\left[\delta(\bm k_2-\bm k_1)-\delta(\bm k_2+\bm k_1)
\right],\qquad
\\
\int\rd\bm x\,\cos{\bm k_2\cdot\bm x} \,\sin{\bm k_1\cdot\bm x}
&=& 0,
\end{eqnarray}
which show that the states that differ only by the sign of $\bm
k$ are not orthogonal.  One of the overlapping states may be
removed by including only states with $\bm k$ such that $\bm
k\cdot\bm k_0 > 0$, where $\bm k_0$ is a fixed direction in
space.  With this restriction on $\bm k_2$ and $\bm k_1$, the
orthonormality of the states is given by
\begin{eqnarray}
\int \rd \bm x \, \psi_{\bm{k}_2,\lambda_2}^{\,(\kappa_2,\pi_2)\dagger}(\bm x)
 \,\psi_{\bm{k}_1,\lambda_1}^{\,(\kappa_1,\pi_1)}(\bm x)
= \delta_{\kappa_2 \kappa_1}  \delta_{\pi_2 \pi_1}
\delta_{\lambda_2 \lambda_1}
\delta(\bm k_2 - \bm k_1).
\qquad
\end{eqnarray}
Completeness of these eigenfunctions, including the
restriction on $\bm k$ provided by a factor $\theta(\bm
k\cdot\bm k_0)$, where the theta function is defined as
\begin{eqnarray}
\theta(x) &=& \left\{\begin{array}{cc} 1 & \mbox{for}~x>0 \\
\frac{1}{2} & \mbox{for}~x=0 \rule{0pt}{10 pt} \\
0 & \mbox{for}~x<0 \rule{0pt}{10 pt} \\
\end{array}\right.,
\elabel{eq:thetadef}
\end{eqnarray}
follows from
\begin{eqnarray}
&&\sum_{\kappa,\pi\rightarrow\pm}
\sum_{\lambda=1}^2\int\rd\bm k\,
\theta(\bm k\cdot\bm k_0)\,\psi_{\bm{k},\lambda}
^{\,(\kappa,\pi)}(\bm x_2)
\,\psi_{\bm{k},\lambda} ^{\,(\kappa,\pi)\dagger}(\bm x_1)
\nonumber\\&&\qquad=
\frac{2}{(2\pi)^3}
\int\rd\bm k\,
\theta(\bm k\cdot\bm k_0)
\iPi\rT(\bm{\hat k})
(
\cos{\bm k\cdot\bm x_2}\,\cos{\bm k\cdot\bm x_1}
+\sin{\bm k\cdot\bm x_2}\,\sin{\bm k\cdot\bm x_1}
) 
\qquad
\nonumber\\&&\qquad=
\iPi\rT(\bm\nabla_2)\,\delta(\bm x_2 - \bm x_1) .
\end{eqnarray}

Longitudinal parity and energy eigenstates are given by
\begin{eqnarray}
\psi_{\bm{k},0}^{\,(+,+)}(\bm x) &=& 
\frac{1}{\sqrt{4\pi^3}}
\left(\begin{array}{c}
\bm{\hat \epsilon}_0(\bm{\hat k})\sin{\bm k\cdot\bm x} \\
\0
\end{array}\right),
\elabel{eq:lpe1}
\\
\psi_{\bm{k},0}^{\,(+,-)}(\bm x) &=& 
\frac{1}{\sqrt{4\pi^3}}
\left(\begin{array}{c}
\bm{\hat \epsilon}_0(\bm{\hat k})\cos{\bm k\cdot\bm x} \\
\0
\end{array}\right),
\elabel{eq:lpe2}
\\ 
\psi_{\bm{k},0}^{\,(-,+)}(\bm x) &=& 
\frac{1}{\sqrt{4\pi^3}}
\left(\begin{array}{c} 
\0 \\
\bm{\hat \epsilon}_0(\bm{\hat k})\cos{\bm k\cdot\bm x}
\end{array}\right),
\elabel{eq:lpe3}
\\ 
\psi_{\bm{k},0}^{\,(-,-)}(\bm x) &=& 
\frac{1}{\sqrt{4\pi^3}}
\left(\begin{array}{c} 
\0 \\
\bm{\hat \epsilon}_0(\bm{\hat k})\sin{\bm k\cdot\bm x}
\end{array}\right),
\elabel{eq:lpe4}
\end{eqnarray}
where
\begin{eqnarray}
\fP \psi_{\bm{k},0}^{\,(\kappa,\pm)}(\bm x) &=& 
\pm\psi_{\bm{k},0}^{\,(\kappa,\pm)}(\bm x),
\\ \cH \psi_{\bm{k},0}^{\,(\kappa,\pm)}(\bm x) &=& 
0.
\end{eqnarray}
As for the transverse eigenfunctions, there is overlap
between states with opposite signs of $\bm k$, so the same
condition on $\bm k$ may be applied here, that is $\theta(\bm
k\cdot\bm k_0) > 0$, to eliminate the redundancy.  With this
condition on $\bm k_2$ and $\bm k_1$, the orthonormality
relation is
\begin{eqnarray}
\int\rd\bm x \, \psi_{\bm{k}_2,0}^{\,(\kappa_2,\pi_2)\dagger}(\bm x)
 \, \psi_{\bm{k}_1,0}^{\,(\kappa_1,\pi_1)}(\bm x)
 &=& 
 \delta_{\kappa_2\kappa_1}
 \delta_{\pi_2\pi_1}
 \delta(\bm k_2 - \bm k_1) .
 \nonumber\\
\end{eqnarray}
In addition, the longitudinal parity eigenfunctions are
orthogonal to the transverse parity eigenfunctions.
Completeness of the longitudinal parity eigenfunctions is given
by
\begin{eqnarray}
&&\sum_{\kappa,\pi\rightarrow\pm}
\int\rd\bm k\,
\theta(\bm k\cdot\bm k_0)\,\psi_{\bm{k},0}
^{\,(\kappa,\pi)}(\bm x_2)
\,\psi_{\bm{k},0} ^{\,(\kappa,\pi)\dagger}(\bm x_1)
\nonumber\\&&\qquad=
\frac{2}{(2\pi)^3}
\int\rd\bm k\,
\theta(\bm k\cdot\bm k_0)
\iPi\rL(\bm{\hat k})
(
\cos{\bm k\cdot\bm x_2}\,\cos{\bm k\cdot\bm x_1}
+\sin{\bm k\cdot\bm x_2}\,\sin{\bm k\cdot\bm x_1}
)
\qquad
\nonumber\\&&\qquad=
\iPi\rL(\bm\nabla_2)\,\delta(\bm x_2 - \bm x_1) .
\end{eqnarray}
The combined completeness relation is thus
\begin{eqnarray}
\sum_{\kappa,\pi\rightarrow\pm}
\sum_{\lambda=0}^2\int\rd\bm k\,
\theta(\bm k\cdot\bm k_0)\,\psi_{\bm{k},\lambda}
^{\,(\kappa,\pi)}(\bm x_2)
\,\psi_{\bm{k},\lambda} ^{\,(\kappa,\pi)\dagger}(\bm x_1)
=\cI\delta(\bm x_2 - \bm x_1) .
\end{eqnarray}

\subsection{Wave packets}
\elabel{ssec:wp}

The plane wave solutions considered in this section are not
normalizable as ordinary functions.  Rather, integrals over
products of solutions should be interpreted in the sense of
distributions or generalized functions, like the delta function
\citep{1964013}.  That is, they provide a well-defined value for
an integral when they are included in the integrand together
with a suitable weight, or test function.  However, the plane
waves can provide the basis for an expansion of a normalizable
wave packet as a sum and integral over a complete set of
solutions of the Maxwell equation.  If $f_\lambda^{(\kappa)}(\bm
k)$ is a suitable function, we write
\begin{eqnarray} 
\iP_f(x) &=& \sum_{\kappa\,\lambda}
\int \rd \bm k \, f_\lambda^{(\kappa)}(\bm k) \,
\psi_{\bm{k},\lambda}^{(\kappa)}(x), 
\elabel{eq:wpex}
\end{eqnarray}
and $\iP_f$ is a solution of the Maxwell equation
\begin{eqnarray}
\gamma^\mu\partial_\mu\iP_f(x) &=&
\sum_{\kappa\,\lambda}
\int \rd \bm k \, f_\lambda^{(\kappa)}(\bm k) \,
\gamma^\mu\partial_\mu\,
\psi_{\bm{k},\lambda}^{(\kappa)}(x) 
= 0.
\end{eqnarray}
Further, $\iP_f$ will be normalized if $f_\lambda^{(\kappa)}$
is, because
\begin{eqnarray}
\int\rd \bm x\,\iP_f^\dagger(x)\iP_f(x) &=&
\sum_{\kappa\,\lambda}
\sum_{\kappa^\prime\,\lambda^\prime}
\int \rd \bm k 
\int \rd \bm k ^\prime
\, f_\lambda^{(\kappa)*}(\bm k)
\int\rd\bm x\,
\psi_{\bm{k},\lambda}^{(\kappa)\dagger}(x) \,
\psi_{\bm{k^\prime},\lambda^\prime}^{(\kappa^\prime)}(x)
 f_{\lambda^\prime}^{(\kappa^\prime)}(\bm k^\prime)
\nonumber\\&=&
 \sum_{\kappa\,\lambda}
 \int \rd \bm k 
 \, \left|f_\lambda^{(\kappa)}(\bm k)\right|^2  = 1 .
\end{eqnarray}
In view of the orthonormality (in the generalized sense) of the
plane-wave solutions, Eq.~(\ref{eq:wpex}) may be inverted to
give
\begin{eqnarray}
f_\lambda^{(\kappa)}(\bm k) &=&
\int \rd \bm x \, \psi_{\bm{k},\lambda}^{(\kappa)\dagger}(\bm x)
 \iP_f(\bm x),
 \elabel{eq:invert}
\end{eqnarray}
where we have specified $\iP_f(\bm x) = \iP_f(x)\big|_{t=0}$,
and the time dependence of the wave function is given by
Eq.~(\ref{eq:wpex}).  

An example is a normalized photon wave packet which at $t=0$ has
(approximately) a wave vector $\bm k_0$, a transverse
polarization vector $\bm{\hat \epsilon}_1(\bm{\hat k}_0)$, and a
Gaussian envelope of length $a$ and width $b$ centered at $\bm x
= 0$:
\begin{eqnarray}
\iP_f(\bm x) &=& 
\frac{1}{a^\frac{1}{2}b}
\left({\frac{2}{\pi^3}}\right)^{\!\frac{1}{4}}
\left(\begin{array}{c}\bm{\hat \epsilon}_1(\bm{\hat k}_0) \\
                      \bm \tau \cdot \bm{\hat k}_0 \
                      \bm{\hat \epsilon}_1(\bm{\hat k}_0)
		      \end{array}\right)
\re^{\ri\bm k_0 \cdot \bm x}
\,\re^{-\left(\bm x_\parallel^2/a^2+\bm x_\perp^2/b^2\right)} ,
\quad
\elabel{eq:gaussian}
\end{eqnarray}
where $\bm x_\parallel = \bm x \cdot \bm{\hat k}_0 \,\bm{\hat
k}_0$ and $\bm x_\perp = \bm x - \bm x_\parallel$.  For $a$ and
$b$ large compared to $|\bm k_0|^{-1}$, the packet has a
functional form that resembles a positive-energy transverse
plane wave.  From Eq.~(\ref{eq:invert}), we have
\begin{eqnarray}
f_\lambda^{(\kappa)}(\bm k) &=&
\frac{a^\frac{1}{2}b}{2}
\left({\frac{2}{\pi^3}}\right)^{\!\frac{1}{4}}
F_\lambda^{(\kappa)}(\bm{\hat k})\,
\, \re^{-\left[\left(\bm k_0 - \kappa \bm
k_\parallel\right)^2a^2/4
+ \bm k_\perp^2 b^2/4 \right]} ,
\elabel{eq:ftrans}
\end{eqnarray}
where $\bm k_\parallel = \bm k \cdot \bm{\hat k}_0 \,\bm{\hat
k}_0$, $\bm k_\perp = \bm k - \bm k_\parallel$, for
$\lambda = 1,2$,
\begin{eqnarray}
F_\lambda^{(\kappa)}(\bm{\hat k}) &=&
\frac{1}{2} \,
\bm{\hat \epsilon}_\lambda^\dagger(\bm{\hat k})\left(
\bm I + \bm \tau\cdot\bm{\hat k}\,\bm \tau \cdot \bm{\hat
k}_0\right) \bm{\hat \epsilon}_1(\bm{\hat k}_0),
\qquad
\elabel{eq:ffact}
\end{eqnarray}
and
\begin{eqnarray}
F_0^{(+)}(\bm{\hat k}) &=&
\frac{1}{\sqrt{2}}\,
\bm{\hat \epsilon}_0^\dagger(\bm{\hat k}) \,
\bm{\hat \epsilon}_1(\bm{\hat k}_0) ,
\\[10 pt]
F_0^{(-)}(\bm{\hat k}) &=&
\frac{1}{\sqrt{2}}\,
\bm{\hat \epsilon}_0^\dagger(\bm{\hat k})
\,\bm \tau \cdot \bm{\hat
k}_0 \bm{\hat \epsilon}_1(\bm{\hat k}_0) .
\end{eqnarray}
The wave packet has small longitudinal components, because
$F_0^{(\kappa)}(\bm{\hat k})$ is not necessarily zero unless
$\bm{\hat k} = \pm\bm{\hat k}_0$.  It has negative-energy
components, but they are also suppressed, particularly as $a,b
\rightarrow \infty$, because for $\kappa\rightarrow -$, the
exponential factors in Eq.~(\ref{eq:ftrans}) favor $\bm k = -\bm
k_0$, and for $\lambda = 1,2$, $F_\lambda^{(\kappa)}(\bm{-\hat
k}_0) = 0$ as compared to $F_\lambda^{(\kappa)}(\bm{\hat k}_0) =
\delta_{\lambda \, 1}$.  Thus a cancellation between the
upper-three and lower-three components of the wave function
suppresses the contribution of negative-energy eigenstates to
the wave packet.

The expectation value of the Hamiltonian $\cH$,
Eq.~(\ref{eq:phham}), is 
\begin{eqnarray} 
\left<\iP_f
\left| \, \cH \, \right| \iP_f \right> &=& 
 -\ri\,\hbar  c 
\int \rd \bm x \iP_f^\dagger(\bm x) \,
\bm \alpha \cdot \bm \nabla 
 \iP_f(\bm x)
=
\hbar c |\bm k_0| = \hbar \omega_0,
\elabel{eq:exham} 
\end{eqnarray} 
and the expectation value of the momentum $\bm \cP $,
Eq.~(\ref{eq:phmom}), is 
\begin{eqnarray} \left<\iP_f \left|\,
\bm \cP \, \right| \iP_f \right> &=& \hbar \bm k_0.
\elabel{eq:exmom} 
\end{eqnarray} 
The initial probability density $Q(\bm x)$ is
\begin{eqnarray} Q(\bm x) &=& \iP_f^\dagger(\bm x) \iP_f(\bm x)
= \frac{2}{ab^2}
\left({\frac{2}{\pi^3}}\right)^{\!\frac{1}{2}}
\,\re^{-2\left(\bm x_\parallel^2/a^2+\bm x_\perp^2/b^2\right)} ,
\qquad 
\elabel{eq:initprob}
\end{eqnarray} 
with 
\begin{eqnarray} \int \rd \bm x \,
Q(\bm x) &=& 1 , \qquad 
\end{eqnarray} 
and the initial energy density $E(\bm x)$ is 
\begin{eqnarray} E(\bm x) &=&
\iP_f^\dagger(\bm x)\, \cH \, \iP_f(\bm x) = \hbar
\omega_0 Q(\bm x) 
- \frac{\ri}{2}\,\hbar c \bm{\hat k}_0 \cdot \bm \nabla Q(\bm
  x).  \qquad 
\end{eqnarray}
The real part of the energy density is proportional to the
probability density for the photon, and the imaginary term, which
vanishes upon integration to arrive at the expectation value in
Eq.~(\ref{eq:exham}), reflects the change in the initial
probability density at the point $\bm x$ due to the motion of
the wave packet.  At a fixed point in the path of the wave
packet, the probability density increases as the packet
approaches and decreases after the maximum of the wave packet
has passed by.  The time-dependent probability density is
\begin{eqnarray} 
Q(x) &=&
\iP_f^\dagger(x) \iP_f(x) = \left|\re^{-\ri \, \cH \, t/\hbar
} \iP_f(\bm x)\right|^2 , 
\elabel{eq:probden}
\end{eqnarray} 
and the change at $t=0$ is 
\begin{eqnarray} 
\frac{\partial Q(x)}{\partial
t}\,\bigg|_{t=0} &=&  - \frac{\ri}{\hbar} \iP_f^\dagger(\bm x)
\left( \cH - \overleftarrow{\cH}^\dagger \right)\!\iP_f(\bm x) 
\nonumber\\ &=&
\frac{2}{\hbar}\,{\rm Im} \iP_f^\dagger(\bm x)  \cH \iP_f(\bm
x)  =
 - c \bm{\hat k}_0 \cdot \bm \nabla Q(\bm x) .  
 \elabel{eq:wpm}
\end{eqnarray} 
The gradient produces a vector that points toward the maximum of
the wave packet, so that on the forward side of the packet
$\bm{\hat k}_0\cdot \bm \nabla Q(\bm x)$ is negative and the
probability density is increasing, as expected.
Eq.~(\ref{eq:wpm}) also shows that the wave packet is initially
moving with velocity c in the direction of $\bm{\hat k}_0$.

The time dependence of the wave packet, Eq.~(\ref{eq:gaussian})
at $t=0$, is given exactly by Eq.~(\ref{eq:wpex}).  However,
approximations may be made in order to obtain a more transparent
expression.  In view of the exponential factors in
Eq.~(\ref{eq:ftrans}), the assumption that $a,b \gg |\bm
k_0|^{-1}$ implies $\bm k \approx \kappa \bm k_0$, and
\begin{eqnarray}
F_\lambda^{(\kappa)}(\bm{\hat k}) &\approx& 
F_\lambda^{(\kappa)}(\kappa\bm{\hat k}_0) 
= \delta_{\lambda  1} \delta_{\kappa+}  ,
\elabel{eq:approx1}
\end{eqnarray}
so that from Eq.~(\ref{eq:wpex}),
$\iP_f\rightarrow\iP_f^\prime$, where
\begin{eqnarray}
\iP_f^\prime(x) &=& 
\frac{a^\frac{1}{2}b}{8\pi^\frac{3}{2}}
\left({\frac{2}{\pi^3}}\right)^{\!\frac{1}{4}}
\int\rd\bm k
\left(\begin{array}{c}\bm{\hat \epsilon}_1(\bm{\hat k}) \\
                      \bm \tau \cdot \bm{\hat k} \
                      \bm{\hat \epsilon}_1(\bm{\hat k})
		      \end{array}\right)
\re^{-\ri\left(|\bm k|ct - \bm k \cdot \bm x\right)}
\,\re^{-\left[\left(\bm k_0 - \bm k_\parallel\right)^2a^2/4
+\bm k_\perp^2b^2/4\right]}. 
\qquad
\end{eqnarray}
This is a normalized positive energy wave function with
polarization $\bm {\hat \epsilon}_1$ that is an exact solution
of the Maxwell equation $\gamma^\mu\partial_\mu \iP_f^\prime(x)
= 0$.  Further simplifications are the replacements $\bm{\hat k}
\rightarrow \bm{\hat k}_0$ in the polarization vector matrix and
$|\bm k| \rightarrow \bm k \cdot \bm{\hat k}_0$ in the exponent,
which yield $\iP_f^\prime\rightarrow\iP_f^{\prime\prime}$, with
\begin{eqnarray}
\iP_f^{\prime\prime}(x) &=& 
\frac{a^\frac{1}{2}b}{8\pi^\frac{3}{2}}
\left({\frac{2}{\pi^3}}\right)^{\!\frac{1}{4}}
\left(\begin{array}{c}\bm{\hat \epsilon}_1(\bm{\hat k}_0) \\
                      \bm \tau \cdot \bm{\hat k}_0 \
                      \bm{\hat \epsilon}_1(\bm{\hat k}_0)
		      \end{array}\right)
\int\rd\bm k \,
\re^{-\ri\left(\bm k \cdot \bm{\hat k}_0 ct - 
\bm k \cdot \bm x\right)}
\,\re^{-\left[\left(\bm k_0 - \bm k_\parallel\right)^2a^2/4
+\bm k_\perp^2b^2/4\right]}
\qquad
\nonumber\\ &=& 
\frac{1}{a^\frac{1}{2}b}
\left({\frac{2}{\pi^3}}\right)^{\!\frac{1}{4}}
\left(\begin{array}{c}\bm{\hat \epsilon}_1(\bm{\hat k}_0) \\
                      \bm \tau \cdot \bm{\hat k}_0 \
                      \bm{\hat \epsilon}_1(\bm{\hat k}_0)
		      \end{array}\right)
\re^{-\ri\left(\omega_0t - \bm k_0 \cdot \bm x\right)}
\,\re^{-\left[\left(ct - \bm {\hat k}_0 \cdot \bm x\right)^2/a^2
+\bm x_\perp^2/b^2\right]} ,
\elabel{eq:tdgaussian}
\end{eqnarray}
which is an approximate wave function with a normalized Gaussian
probability distribution
\begin{eqnarray}
Q^{\prime\prime}(x) &=& 
\frac{2}{ab^2}
\left({\frac{2}{\pi^3}}\right)^{\!\frac{1}{2}}
\,\re^{-2\left[\left(ct - \bm {\hat k}_0\cdot \bm x\right)^2/a^2
+\bm x_\perp^2/b^2\right]} , \qquad
\elabel{eq:tdgaussianp}
\end{eqnarray}
that moves with velocity $c$ in the $\bm{\hat k}_0$ direction.

\subsection{Conservation of probability}
\elabel{ssec:cop}

The formulation of the Poynting theorem in Sec.~\ref{sec:mmaxeq}
can be reinterpreted here to demonstrate conservation of
probability.  We define the probability density four-vector to be
\begin{eqnarray}
q^\mu(x) &=& \overline{\iP}(x){\gamma^\mu}\iP(x),
\elabel{eq:probdenv}
\end{eqnarray}
where in the previous section $Q(x) = q^0(x)$.  For the
source-free case, $\iX(x)=0$, Eq.~(\ref{eq:derivpsis}) is
\begin{eqnarray}
\partial_\mu \overline{\iP}(x){\gamma^\mu}\iP(x) &=& 0,
\end{eqnarray}
which is the statement of conservation of probability
\begin{eqnarray}
\frac{\partial}{\partial t}\,q^0(x) + c \, \bm\nabla\cdot
\bm q(x) &=& 0.
\elabel{eq:consprob}
\end{eqnarray}

Applying this relation to plane-wave states gives consistent,
although trivial, results.  We have
\begin{eqnarray}
q^0(x) &=& 
\psi_{\bm{k},\lambda}^{(\kappa)\dagger}(x)
\,\psi_{\bm{k},\lambda}^{(\kappa)}(x)
= (2\pi)^{-3} ,
\end{eqnarray}
which reflects the fact that the probability distribution for a
plane wave is uniform throughout space and not normalizable.
For transverse plane waves, $\lambda = 1,2$, the probability
flux vector is
\begin{eqnarray}
\bm q(x) &=& 
\psi_{\bm{k},\lambda}^{(\kappa)\dagger}(x)
\,\bm\alpha\,\psi_{\bm{k},\lambda}^{(\kappa)}(x)
= (2\pi)^{-3} \bm {\hat k},
\end{eqnarray}
and for longitudinal plane waves $\bm q(x) = 0$.

For the wave packet in Eq.~(\ref{eq:gaussian}), at $t=0$
\begin{eqnarray}
\bm q(x)
&=& \bm{\hat k}_0 \, q^0(x),
\end{eqnarray}
so Eq.~(\ref{eq:wpm}) is essentially the conservation of
probability equation evaluated at $t=0$.  Conservation of
probability is valid for any solution of the homogeneous Maxwell
equation from the definition of the probability density
operator.  However, it also happens to be valid for the wave
packet represented by $\iP_f^{\prime\prime}$, which is not an
exact solution of the wave equation.  The wave equation is not
satisfied by $\iP_f^{\prime\prime}$, because there is a small
extra term resulting from the gradient operator acting on the
perpendicular coordinate $\bm x_\perp$.  On the other hand, $\bm
q(x)$ for $\iP_f^{\prime\prime}$ is proportional to $\bm{\hat
k}_0$ and $\bm{\hat k}_0\cdot\bm\nabla\,\bm x_\perp = 0$, so
there is no corresponding extra term in $\bm\nabla\cdot\bm
q(x)$.

\section{Angular momentum eigenfunctions} \elabel{sec:pswf}

Radiation emitted in atomic transitions is characterized by its
angular momentum and parity.  In this section, wave functions
that are eigenfunctions of energy, angular momentum, and parity
are given; they are also classified according to whether they
are transverse or longitudinal.  The three-component
angular-momentum matrices given here are to some extent
parallels of the three-vector functions of \cite{1940003}.

\subsection{Angular momentum}
\elabel{ssec:am}

The spatial angular-momentum operator is given by 
\begin{eqnarray}
\bm L &=& \bm x \times \bm p 
= -\ri\,\hbar\,\bm x\times \bm\nabla ,
\end{eqnarray}
and following the example of the Dirac equation, the total
angular momentum is given as a $3\times3$ matrix by
\cite{1931004}
\begin{eqnarray}
\bm J &=& \bm L + \hbar \bm\tau ,
\elabel{eq:j}
\end{eqnarray}
where it is understood that the first term on the right side is
a $3\times3$ matrix with $\bm L$ for diagonal elements and zeros
for the rest.\footnote{In some works, where electric and
magnetic fields or the vector potential are three-vector valued
fields, the spin operator is represented by a cross product.
For example, Corben and Schwinger \cite{1940003} write $J_z\bm
\Phi = L_z \bm\Phi + \ri \, \bm e_z\times\bm\Phi$, where $\bm
\Phi$ is a vector potential, and in Edmonds \cite{1960006}, spin
is represented symbolically as $\ri\,\bm{\hat e} \bm\times$.}
(In order to adhere to convention, we denote both the current
three-vector and the angular-momentum matrix by $\bm J$.  In
either case, the meaning should be clear from the context.) The
extension to a $6\times6$ matrix is
\begin{eqnarray}
\bm\cJ &=& \left(\begin{array}{cc}
\bm J & \0 \\ \0 & \bm J
\end{array}\right)
 =
\bm x \times \bm \cP + \hbar \bm \cS
\elabel{eq:cj}
\end{eqnarray}
and we have
\begin{eqnarray}
\left[\cH,\bm\cJ\right] &=& 0,
\elabel{eq:hjcom}
\end{eqnarray}
so eigenfunctions of both energy and angular momentum may be
constructed.  The vanishing of the commutator follows from the
relations
\begin{eqnarray}
\left[\bm\tau\cdot\bm \nabla,\bm L\right] &=&
-\ri\hbar\bm\tau\times\bm \nabla,
\\
\left[\bm\tau\cdot\bm \nabla,\bm \tau\right] &=& 
\ri\bm\tau\times\bm \nabla,
\\
\left[\bm\tau\cdot\bm \nabla,\bm J\right] &=& 0.
\elabel{eq:tddcj}
\end{eqnarray}
It is of interest to see that $\bm \cJ$ commutes with $\cH$ only for
the (relative) combination of $\bm L$ and $\bm\tau$ given in
Eq.~(\ref{eq:j}).
To obtain eigenstates of the square of the total angular
momentum $\bm\cJ^2$ and the third component of angular momentum
$\cJ^3$, given by
\begin{eqnarray}
\bm\cJ^2 &=& 
\left(\begin{array}{cc}
\bm J^2 & \0 \\ \0 & \bm J^2
\end{array}\right)
\end{eqnarray}
and
\begin{eqnarray}
\cJ^{3} &=& 
\left(\begin{array}{cc}
J^3 & \0 \\ \0 & J^3
\end{array}\right),
\end{eqnarray}
we construct matrix spherical harmonics that are analogous to
conventional vector spherical harmonics and are three-component
extensions of the Dirac two-component spin-angular-momentum
eigenfunctions.  Orthonormal basis matrices are given by
\begin{eqnarray}
\bm{\hat \epsilon}^{(1)} = \left(\begin{array}{c}
1 \\ 0 \\ 0 \end{array}\right)
,\qquad
\bm{\hat \epsilon}^{(0)} = \left(\begin{array}{c}
0 \\ 1 \\ 0 \end{array}\right)
,\qquad
\bm{\hat \epsilon}^{(-1)} = \left(\begin{array}{c}
0 \\ 0 \\ 1 \end{array}\right)
,
\end{eqnarray}
and they satisfy the eigenvalue equations
\begin{eqnarray}
\bm\tau^2\,\bm{\hat \epsilon}^{(\nu)}
&=& 2\,\bm{\hat \epsilon}^{(\nu)},
\\
\tau^3\,\bm{\hat \epsilon}^{(\nu)}
&=& \nu\,\bm{\hat \epsilon}^{(\nu)},
\end{eqnarray}
where the 2 may be regarded as $s(s+1)$, with $s=1$ as the spin
eigenvalue.  The matrix spherical harmonics are 
\begin{eqnarray}
\bm Y_{jl}^m(\bm{\hat x})
&=& \sum_{\nu}\,
(l \ m-\nu \ 1 \ \nu|l \ 1 \ j \ m)
Y_l^{m-\nu}(\bm{\hat x})\,\bm{\hat \epsilon}^{(\nu)},
\end{eqnarray}
with vector addition coefficients and spherical harmonics in the
notation of \cite{1960006}.  The spherical harmonics satisfy
the eigenvalue equations
\begin{eqnarray}
\bm L^2 Y_l^m(\bm{\hat x}) &=& \hbar^2 l(l+1)\,Y_l^m(\bm{\hat x}),
\\
L^3 Y_l^m(\bm{\hat x}) &=& \hbar \, m\,Y_l^m(\bm{\hat x}),
\end{eqnarray}
and by their construction, the matrix spherical harmonics
are eigenfunctions of the total angular momentum:
\begin{eqnarray}
\bm J^2 \bm Y_{jl}^m(\bm{\hat x}) &=& \hbar^2 j(j+1)\,
\bm Y_{jl}^m(\bm{\hat x}),
\\
J^3 \bm Y_{jl}^m(\bm{\hat x}) &=& \hbar\,m\,\bm Y_{jl}^m(\bm{\hat x}).
\end{eqnarray}
Explicit expressions in terms of spherical harmonics are
\begin{eqnarray}
\bm Y_{jj}^m(\bm{\hat x}) &=& \left(\begin{array}{c}
-\sqrt{\frac{(j+m)(j+1-m)}{2j(j+1)}}
\,Y_j^{m-1}(\bm{\hat x})
\\[0 pt]
\frac{m}{\sqrt{j(j+1)}}
\,Y_j^{m}(\bm{\hat x})
\\[6 pt]
\sqrt{\frac{(j-m)(j+1+m)}{2j(j+1)}}
\,Y_j^{m+1}(\bm{\hat x})
\end{array}\right), \qquad
\elabel{eq:msh1}
\end{eqnarray}
\begin{eqnarray}
\bm Y_{jj+1}^m(\bm{\hat x}) &=& \left(\begin{array}{c}
\sqrt{\frac{(j+1-m)(j+2-m)}{(2j+2)(2j+3)}}
\,Y_{j+1}^{m-1}(\bm{\hat x})
\\[6 pt]
-\sqrt{\frac{(j+1-m)(j+1+m)}{(j+1)(2j+3)}}
\,Y_{j+1}^{m}(\bm{\hat x})
\\[6 pt]
\sqrt{\frac{(j+2+m)(j+1+m)}{(2j+2)(2j+3)}}
\,Y_{j+1}^{m+1}(\bm{\hat x})
\end{array}\right), \qquad
\elabel{eq:msh2}
\end{eqnarray}
\begin{eqnarray}
\bm Y_{jj-1}^m(\bm{\hat x}) &=& \left(\begin{array}{c}
\sqrt{\frac{(j-1+m)(j+m)}{(2j-1)2j}}
\,Y_{j-1}^{m-1}(\bm{\hat x})
\\[6 pt]
\sqrt{\frac{(j-m)(j+m)}{(2j-1)j}}
\,Y_{j-1}^{m}(\bm{\hat x})
\\[6 pt]
\sqrt{\frac{(j-1-m)(j-m)}{(2j-1)2j}}
\,Y_{j-1}^{m+1}(\bm{\hat x})
\end{array}\right). \qquad
\elabel{eq:msh3}
\end{eqnarray}
These functions are orthonormal
\begin{eqnarray}
\int\rd\bm\iO \,\bm Y_{j_2l_2}^{m_2\dagger}(\bm{\hat x})
\,\bm Y_{j_1l_1}^{m_1}(\bm{\hat x})
&=&
\delta_{j_2j_1} \delta_{l_2l_1}\delta_{m_2m_1}, \qquad
\end{eqnarray}
which follows from the relations
\begin{eqnarray}
\int\rd\bm\iO \, Y_{l_2}^{m_2*}(\bm{\hat x})
\, Y_{l_1}^{m_1}(\bm{\hat x})
=  \delta_{l_2l_1}\delta_{m_2m_1}, 
\elabel{eq:sho}
\end{eqnarray}
\vspace{-20pt}
\begin{eqnarray}
\bm{\hat \epsilon}^{(\nu_2)\dagger}\,
\bm{\hat \epsilon}^{(\nu_1)} = \delta_{\nu_2\nu_1},
\end{eqnarray}
\begin{eqnarray}
&&\sum_\nu(l \ 1 \ j_2 \ m | l \ m-\nu \ 1 \ \nu)
(l \ m-\nu \ 1 \ \nu|l \ 1 \ j_1 \ m) = \delta_{j_2j_1}, \qquad
\end{eqnarray}
and they are complete
\begin{eqnarray}
&&\sum_{jlm} \bm Y_{jl}^m(\bm{\hat x}_2)\bm Y_{jl}^{m\dagger}(\bm{\hat
x}_1)
=
 \bm I \,
 \delta(\cos{\theta_2}-\cos{\theta_1})\,\delta(\phi_2-\phi_1),
 \qquad
\end{eqnarray}
based on the relations
\begin{eqnarray}
&&\sum_{j}(l \ m-\nu_2 \ 1 \ \nu_2|l \ 1 \ j \ m)
(l \ 1 \ j \ m|l \ m-\nu_1\ 1 \ \nu_1) 
= \delta_{\nu_2\nu_1}, \qquad
\end{eqnarray}
\begin{eqnarray}
\sum_\nu \bm{\hat \epsilon}^{(\nu)}\,\bm{\hat
\epsilon}^{(\nu)\dagger} = \bm I,
\end{eqnarray}
\vspace{-15pt}
\begin{eqnarray}
&&\sum_{lm}  Y_{l}^m(\bm{\hat x}_2)
Y_{l}^{m*}(\bm{\hat
x}_1)
=
  \delta(\cos{\theta_2}-\cos{\theta_1})\,\delta(\phi_2-\phi_1),
   \qquad
\end{eqnarray}
where $\theta_i,\phi_i$ are the spherical coordinates of $\bm{\hat
x}_i$.

An alternative set of matrix angular-momentum eigenfunctions is
\begin{eqnarray}
\bm X^{jm}_{1} (\bm{\hat x}) &=&
\frac{1}{\hbar\sqrt{j(j+1)}}\,\bm L\rs Y_j^m(\bm{\hat x}),
\elabel{eq:amf1}
\\
\bm X^{jm}_{2} (\bm{\hat x}) &=& 
\frac{1}{\hbar\sqrt{j(j+1)}}\,\bm\tau\cdot\bm{\hat x}\bm L\rs
Y_j^m(\bm{\hat x}),
\elabel{eq:amf2}
\\
\bm X^{jm}_{3} (\bm{\hat x}) &=& 
\bm{\hat x}\rs Y_j^m(\bm{\hat x}).
\elabel{eq:amf3}
\end{eqnarray}
For $j=0$, $\bm X^{00}_1 (\bm{\hat x}) = \bm X^{00}_2 (\bm{\hat
x}) = 0$.  
From a comparison of Eqs.~(\ref{eq:msh1})-(\ref{eq:msh3})
to Eqs.~(\ref{eq:amf1})-(\ref{eq:amf3}), one has
\begin{eqnarray}
\bm X^{jm}_1(\bm{\hat x}) &=& 
\bm Y_{jj}^m(\bm{\hat x}),
\elabel{eq:xyeq1}
\\
\bm X^{jm}_2(\bm{\hat x}) &=& 
-\sqrt{\frac{j}{2j+1}}\,\bm Y_{jj+1}^m(\bm{\hat x})
-\sqrt{\frac{j+1}{2j+1}}\,\bm Y_{jj-1}^m(\bm{\hat x}),
\elabel{eq:xyeq2}
\\
\bm X^{jm}_3(\bm{\hat x}) &=& 
-\sqrt{\frac{j+1}{2j+1}}\,\bm Y_{jj+1}^m(\bm{\hat x})
+\sqrt{\frac{j}{2j+1}}\,\bm Y_{jj-1}^m(\bm{\hat x}).
\elabel{eq:xyeq3}
\end{eqnarray}
In view of the relations in
Eqs.~(\ref{eq:xyeq1})-(\ref{eq:xyeq3}), the functions $\bm
X_i^{jm}(\bm{\hat x})$ are eigenfunctions of $\bm
J^2$ and $J^3$ with 
\begin{eqnarray}
\bm J^2 \bm X_{i}^{jm}(\bm{\hat x}) &=& \hbar^2 j(j+1)\,
\bm X_{i}^{jm}(\bm{\hat x}),
\\
J^3 \bm X_{i}^{jm}(\bm{\hat x}) &=& \hbar\,m\,\bm X_{i}^{jm}(\bm{\hat x}).
\end{eqnarray}
This can be confirmed directly from
the definitions in Eqs.~(\ref{eq:amf1})-(\ref{eq:amf3}) with the
aid of the commutation relations
\begin{eqnarray}
\left[L^i,L^j\right]  &=& \ri\hbar\epsilon_{ijk}L^k,
\\
\left[L^i,x^j\right]  &=& \ri\hbar\epsilon_{ijk}x^k
\end{eqnarray}
and the tau matrix identities in Sec.~\ref{sec:spinmat},
which provide the operator identities
\begin{eqnarray}
J^i\bm L\rs &=& \bm L\rs L^i,
\elabel{eq:jlcr}
\\
\left[J^i,\bm \tau\cdot\bm{\hat x}\right] &=& 0,
\\
J^i  \bm{\hat x}\rs &=& \bm{\hat x}\rs L^i.
\elabel{eq:jxcr}
\end{eqnarray}
These functions are orthonormal
\begin{eqnarray}
\int\rd\bm\iO\,\bm X^{j_2m_2\dagger}_{i_2}(\bm{\hat x})
\, \bm X^{j_1m_1}_{i_1}(\bm{\hat x}) 
&=&\delta_{i_2i_1} \delta_{j_2j_1}\delta_{m_2m_1}, \qquad
\end{eqnarray}
and they are complete
\begin{eqnarray}
&&\sum_{ijm}\bm X_i^{jm}(\bm{\hat x}_2)\,
\bm X_i^{jm\dagger}(\bm{\hat x}_1)
 =
 \bm I \,
 \delta(\cos{\theta_2}-\cos{\theta_1})\,\delta(\phi_2-\phi_1),
 \qquad
\end{eqnarray}
where the latter fact may be seen from the completeness of the
matrix spherical harmonics and the relation
\begin{eqnarray}
\sum_{i}\bm X_i^{jm}(\bm{\hat x}_2)\,
\bm X_i^{jm\dagger}(\bm{\hat x}_1) &=&
\sum_{l} \bm Y_{jl}^m(\bm{\hat x}_2)\bm Y_{jl}^{m\dagger}(\bm{\hat
x}_1). \qquad
\end{eqnarray}
The parity of the eigenfunctions is given by
\begin{eqnarray}
\bm X^{jm}_1(-\bm{\hat x}) &=&
(-1)^j\bm X^{jm}_1(\bm{\hat x}),
\\
\bm X^{jm}_2(-\bm{\hat x}) &=&
(-1)^{j+1}\bm X^{jm}_2(\bm{\hat x}),
\\
\bm X^{jm}_3(-\bm{\hat x}) &=&
(-1)^{j+1}\bm X^{jm}_3(\bm{\hat x}),
\end{eqnarray}
which follows from $Y_j^m(-\bm{\hat x}) = 
(-1)^j Y_j^m(\bm{\hat x})$.

\subsection{Helicity eigenstates}
\elabel{ssec:he}

Special cases of the transverse plane-wave solutions in
Eq.~(\ref{eq:trwf}) are circularly polarized states with the
polarization vectors in Eq.~(\ref{eq:circ}).  They can be
grouped with the longitudinal plane-wave states in
Eqs.~(\ref{eq:plwf}) and (\ref{eq:nlwf}) with the polarization
vector in Eq.~(\ref{eq:lpv}).  These polarization vectors are
summarized here as
\begin{eqnarray}
\bm{\hat \epsilon}_1(\bm{\hat e}^3) \!=\! 
\left(\begin{array}{c}
1 \\ 
0 \\ 
0 \end{array}\right) ;\quad
\bm{\hat \epsilon}_0(\bm{\hat e}^3) \!=\! 
\left(\begin{array}{c}
0 \\ 
1 \\ 
0 \end{array}\right) ;\quad
\bm{\hat \epsilon}_{-1}(\bm{\hat e}^3) \!=\! 
\left(\begin{array}{c}
0 \\ 
0 \\ 
1 \end{array}\right),
\nonumber
\end{eqnarray}
where we have changed the label for $\lambda$ from 2 to $-1$ for
this section.

The states have a well-defined helicity; they are eigenfunctions
of the operator for the projection of angular momentum in the
direction of the wave vector $\bm\cJ\cdot\bm{\hat k}$
\cite{1931004,1959008}.  In view of the relations
\begin{eqnarray}
\bm L \cdot \bm{\hat k} \, \re^{\pm\ri\bm k\cdot\bm x} &=& 0,
\\
\bm \tau\cdot\bm {\hat k} \, \bm{\hat\epsilon}_\lambda(\bm {\hat k})
&=& \lambda\,\bm{\hat\epsilon}_\lambda(\bm {\hat k}) 
\end{eqnarray}
for the polarizations considered here, we have 
\begin{eqnarray}
\bm\cJ\cdot\bm{\hat k} \, \psi_{\bm k,\lambda}^{(\pm)}(\bm x)
&=& \lambda \hbar \, \psi_{\bm k,\lambda}^{(\pm)}(\bm x)
\end{eqnarray}
for these states.

\subsection{Transverse spherical photons}
\elabel{ssec:stp}

Transverse spherical wave functions are given by
\begin{eqnarray}
\psi_{\omega,jm}^{{\rm T}(\kappa,+)}(\bm x) &=&
\frac{1}{\sqrt{2}}
\left(\begin{array}{c} 
f_{\omega,j}(r) \bm X_1^{jm}(\bm{\hat x})
\\[10 pt]
-\kappa\ri\frac{c}{\omega}\bm\tau\cdot\bm\nabla
f_{\omega,j}(r) \bm X_1^{jm}(\bm{\hat x})
\end{array}\right), \qquad
\elabel{eq:tswf1}
\\
\psi_{\omega,jm}^{{\rm T}(\kappa,-)}(\bm x) &=&
\frac{1}{\sqrt{2}}
\left(\begin{array}{c} 
\frac{c}{\omega}\bm\tau\cdot\bm\nabla
f_{\omega,j}(r) \bm X_1^{jm}(\bm{\hat x})
\\[10 pt]
\kappa\ri f_{\omega,j}(r) \bm X_1^{jm}(\bm{\hat x})
\end{array}\right),
\elabel{eq:tswf2}
\end{eqnarray}
where $r = |\bm x|$ and $j\ge1$.  They are transverse because $
\bm\nabla\rs^\dagger \, \bm L\rs = \bm\nabla\rs^\dagger \, \bm
\tau\cdot\bm\nabla = 0$, so that
\begin{eqnarray}
\iPi\rT(\bm\nabla)\,\psi_{\omega,jm}^{{\rm T}(\kappa,\pi)}(\bm x) &=&
\psi_{\omega,jm}^{{\rm T}(\kappa,\pi)}(\bm x),
\\
\iPi\rL(\bm\nabla)\,\psi_{\omega,jm}^{{\rm T}(\kappa,\pi)}(\bm x) &=&
0.
\end{eqnarray}
The wave functions are eigenfunctions of angular momentum with
eigenvalues given by [see Eq.~(\ref{eq:tddcj})]
\begin{eqnarray}
\bm\cJ^2 \,\psi_{\omega,jm}^{{\rm T}(\kappa,\pi)}(\bm x) &=&
\hbar^2j(j+1)\,\psi_{\omega,jm}^{{\rm T}(\kappa,\pi)}(\bm x),
\\
\cJ^{3} \,\psi_{\omega,jm}^{{\rm T}(\kappa,\pi)}(\bm x) &=&
\hbar\,m\,\psi_{\omega,jm}^{{\rm T}(\kappa,\pi)}(\bm x),
\end{eqnarray}
and they are eigenfunctions of $\cH$, with eigenvalue
$\kappa\hbar\omega$
\begin{eqnarray}
\cH \,\psi_{\omega,jm}^{{\rm T}(\kappa,\pi)}(\bm x)
&=& -\ri\,\hbar c \,\bm\alpha\cdot\bm \nabla
\psi_{\omega,jm}^{{\rm T}(\kappa,\pi)}(\bm x)
= \kappa\hbar\omega\,
\psi_{\omega,jm}^{{\rm T}(\kappa,\pi)}(\bm x),
\end{eqnarray}
provided
\begin{eqnarray}
\left(\bm\nabla^2 + \frac{\omega^2}{c^2}\right)
f_{\omega,j}(r) \bm X_1^{jm}(\bm{\hat x})
= 0 ,
\end{eqnarray}
which is true if
\begin{eqnarray}
\left(\frac{1}{r}\,\frac{\partial^2}{\partial r^2}\,r
-\frac{j(j+1)}{r^2} + \frac{\omega^2}{c^2}\right)f_{\omega,j}(r)
&=& 0.
\elabel{eq:radeq}
\end{eqnarray}
Solutions of Eq.~(\ref{eq:radeq}) are spherical Bessel functions
given by \cite{1965020}
\begin{eqnarray}
f_{\omega,j}(r) &\propto& \left\{\begin{array}{c}
j_j(\omega r/c) \\[10 pt] h_j^{(1)}(\omega r/c)
\end{array}\right. .
\end{eqnarray}
We employ the normalized solution
\begin{eqnarray}
f_{\omega,j}(r) &=& \frac{\omega}{c}\,\sqrt{\frac{2}{\pi c}}
 \ j_j(\omega r/c) 
\end{eqnarray}
for the wave functions; any other linear combination of
spherical Bessel functions (with $j\ge1$) is not integrable as
$r \rightarrow 0$.  The parity of the wave functions is
\begin{eqnarray}
\fP \, \psi_{\omega,jm}^{{\rm T}(\kappa,+)}(\bm x) &=&
(-1)^{j+1} \, \psi_{\omega,jm}^{{\rm T}(\kappa,+)}(\bm x),
\\
\fP \, \psi_{\omega,jm}^{{\rm T}(\kappa,-)}(\bm x) &=&
(-1)^{j} \, \psi_{\omega,jm}^{{\rm T}(\kappa,-)}(\bm x).
\end{eqnarray}
This provides the conventional parity and angular-momentum
attributes for electric and magnetic multipole radiation.
Namely, $\psi_{\omega,jm}^{{\rm T}(\kappa,+)}(\bm x)$ is
magnetic $2j$-pole or M$j$ radiation and $\psi_{\omega,jm}^{{\rm
T}(\kappa,-)}(\bm x)$ is electric $2j$-pole or E$j$ radiation.

Alternative forms for the lower three components in
Eq.~(\ref{eq:tswf1}) or the upper three components in
Eq.~(\ref{eq:tswf2}) are obtained by writing (see
\ref{app:stlgo})
\begin{eqnarray}
\bm\tau\cdot\bm\nabla &=&
\frac{1}{r}\,\frac{\partial}{\partial r}\,r \,
\bm\tau\cdot\,\bm{\hat x}\,
+\frac{1}{\hbar r}\left(\bm L\rs\,\bm{\hat x}\rs^\dagger
+\bm{\hat x}\rs\,\bm L\rs^\dagger\right),
\qquad
\end{eqnarray}
which yields
\begin{eqnarray}
\bm\tau\cdot\bm\nabla \,
f_{\omega,j}(r) \bm X_1^{jm}(\bm{\hat x}) =
\frac{1}{r}\,\frac{\partial}{\partial r}\,r \,
f_{\omega,j}(r) \bm X_2^{jm}(\bm{\hat x})
-
\frac{\sqrt{j(j+1)}}{r}\,
f_{\omega,j}(r) \bm X_3^{jm}(\bm{\hat x}).
\end{eqnarray}
Relations among spherical Bessel functions provide
\begin{eqnarray}
\frac{1}{r}\,\frac{\partial}{\partial r}\,r \, f_{\omega,j}(r) 
&=&  \frac{\omega}{c}\,
\frac{1}{2j+1}\big[(j+1)f_{\omega,j-1}(r) -j\, 
f_{\omega,j+1}(r)\big], \qquad
\\
\frac{1}{r}\,f_{\omega,j}(r) 
&=& \frac{\omega}{c}\,
\frac{1}{2j+1}\big[f_{\omega,j-1}(r)+f_{\omega,j+1}(r)\big],
\end{eqnarray}
which together with Eqs.~(\ref{eq:xyeq2}) and (\ref{eq:xyeq3})
yield the second alternative form 
\begin{eqnarray}
\frac{c}{\omega} \bm\tau\cdot\bm\nabla \,
f_{\omega,j}(r) \bm X_1^{jm}(\bm{\hat x}) 
=
-\sqrt{\frac{j+1}{2j+1}}f_{\omega,j-1}(r)\,\bm Y_{jj-1}^m
(\bm{\hat x})
+\sqrt{\frac{j}{2j+1}}f_{\omega,j+1}(r)\,\bm Y_{jj+1}^m 
(\bm{\hat x}). \quad
\end{eqnarray}
This latter form is useful in calculating the wave function
orthonormality and completeness relations.

An analogous longitudinal function is obtained by writing
[Eq.~(\ref{eq:slg})]
\begin{eqnarray}
  \bm\nabla\rs &=&
   \bm{\hat x}\rs\,\frac{\partial}{\partial r}
   -\frac{1}{\hbar r}\,\bm \tau\cdot\bm{\hat x}\,\bm L\rs 
   \elabel{eq:snab}
\end{eqnarray}
and
\begin{eqnarray}
\frac{\omega}{c}\,\bm F_\omega^{jm}(\bm x) &=&
\bm\nabla\rs \,f_{\omega,j}(r)\,Y_j^m(\bm{\hat x})
\nonumber\\&=&
 \frac{\partial}{\partial r}\,f_{\omega,j}(r)
\bm X^{jm}_3(\bm{\hat x})
-\frac{\sqrt{j(j+1)}}{r}\,f_{\omega,j}(r)\bm X^{jm}_2(\bm{\hat x}),
\qquad
\end{eqnarray}
so that
\begin{eqnarray}
\bm\tau\cdot\bm\nabla \, \bm F_\omega^{jm}(\bm x) &=& 0.
\end{eqnarray}
From the additional relation
\begin{eqnarray}
\frac{\partial}{\partial r} \, f_{\omega,j}(r) 
&=&  \frac{\omega}{c}\,
\frac{1}{2j+1}\big[j\,f_{\omega,j-1}(r) -(j+1)
f_{\omega,j+1}(r)\big],
\end{eqnarray}
together with Eqs.~(\ref{eq:xyeq2}) and (\ref{eq:xyeq3}),
one has
\begin{eqnarray}
&& \bm F_\omega^{jm}(\bm x) =
\sqrt{\frac{j}{2j+1}}\ f_{\omega,j-1}(r)\,\bm Y_{jj-1}^m
(\bm{\hat x})
+\sqrt{\frac{j+1}{2j+1}}\ f_{\omega,j+1}(r)\,\bm Y_{jj+1}^m
(\bm{\hat x}) . \qquad
\elabel{eq:lonex}
\end{eqnarray}

The orthonormality of the transverse wave functions is given by
\begin{eqnarray}
\int\rd \bm x \,
\psi_{\omega_2,j_2m_2}^{{\rm T}(\kappa_2,\pi_2)\dagger}(\bm x)
\psi_{\omega_1,j_1m_1}^{{\rm T}(\kappa_1,\pi_1)}(\bm x)
&=& \delta_{\kappa_2\kappa_1}
\delta_{\pi_2\pi_1}\delta_{j_2j_1}\delta_{m_2m_1}
\delta(\omega_2-\omega_1),
\qquad
\elabel{eq:sphortht}
\end{eqnarray}
which takes into account the integral
\begin{eqnarray}
\int_0^\infty \rd r\,r^2\,f_{\omega_2,j}(r)
\,f_{\omega_1,j}(r) &=& \delta(\omega_2-\omega_1).
\qquad
\end{eqnarray}
The completeness relation for the transverse wave functions is
\begin{eqnarray}
\int_0^\infty\rd\omega\sum_{\kappa\pi j m}
\psi_{\omega,jm}^{{\rm T}(\kappa,\pi)}(\bm x_2)
\psi_{\omega,jm}^{{\rm T}(\kappa,\pi)\dagger}(\bm x_1)
&=& \iPi\rT(\bm\nabla_2)\,\delta(\bm x_2 - \bm x_1),
\elabel{eq:tcrel}
\end{eqnarray}
which is shown in some detail by writing
\begin{eqnarray}
\sum_{\kappa\pi}
\psi_{\omega,jm}^{{\rm T}(\kappa,\pi)}(\bm x_2)
\psi_{\omega,jm}^{{\rm T}(\kappa,\pi)\dagger}(\bm x_1)
&=&
\left(\begin{array}{cc}
\bm S_\omega^{jm}(\bm x_2,\bm x_1) & \0 \\
\0 & \bm S_\omega^{jm}(\bm x_2,\bm x_1)
\end{array}\right) ,
\end{eqnarray}
where
\begin{eqnarray}
\bm S_\omega^{jm}(\bm x_2,\bm x_1) &=&
f_{\omega,j}(r_2) \bm X_1^{jm}(\bm{\hat x}_2)
f_{\omega,j}(r_1) \bm X_1^{jm\dagger}(\bm{\hat x}_1)
\nonumber\\&&+
\frac{c^2}{\omega^2}\left[\bm\tau\cdot\bm\nabla_2
f_{\omega,j}(r_2) \bm X_1^{jm}(\bm{\hat x}_2)\right]
\left[\bm\tau\cdot\bm\nabla_1
f_{\omega,j}(r_1) \bm X_1^{jm}(\bm{\hat x}_1)\right]^\dagger,
\qquad
\end{eqnarray}
and
\begin{eqnarray}
\bm S_\omega^{jm}(\bm x_2,\bm x_1) &=&
\bm{\iPi}\rs\rT(\bm\nabla_2)\Big[
\bm S_\omega^{jm}(\bm x_2,\bm x_1) + \bm F_\omega^{jm}(\bm x_2)
\bm F_\omega^{jm\dagger}(\bm x_1) \Big] 
\nonumber\\&=&
 \bm{\iPi}\rs\rT(\bm\nabla_2) \Big[
f_{\omega,j}(r_2) \bm Y_{jj}^m(\bm{\hat x}_2)
f_{\omega,j}(r_1) \bm Y_{jj}^{m\dagger}(\bm{\hat x}_1)
\nonumber\\[5 pt]&&
+f_{\omega,j-1}(r_2)\,\bm Y_{jj-1}^m(\bm{\hat x}_2)
f_{\omega,j-1}(r_1)\,\bm
Y_{jj-1}^{m\dagger}(\bm{\hat x}_1)
\nonumber\\[5 pt]&& +
f_{\omega,j+1}(r_2)\,\bm Y_{jj+1}^m(\bm{\hat x}_2)
f_{\omega,j+1}(r_1)\,\bm Y_{jj+1}^{m\dagger}(\bm{\hat x}_1)
\Big] ,
\end{eqnarray}
which gives
\begin{eqnarray}
&&\int_0^\infty\rd\omega\sum_{j m} \bm S_\omega^{jm}(\bm x_2,\bm x_1) =
\bm{\iPi}\rs\rT(\bm\nabla_2)
\frac{1}{r_2r_1}\,\delta(r_2-r_1)
\nonumber\\&&\qquad\qquad\times
\sum_{j m} \Big[
\bm Y_{jj}^m(\bm{\hat x}_2)
\bm Y_{jj}^{m\dagger}(\bm{\hat x}_1) +
\bm Y_{jj-1}^m(\bm{\hat x}_2)
\bm Y_{jj-1}^{m\dagger}(\bm{\hat x}_1) +
\bm Y_{jj+1}^m(\bm{\hat x}_2)
\bm Y_{jj+1}^{m\dagger}(\bm{\hat x}_1)
\Big]
\qquad
\nonumber\\&&\qquad=
\bm{\iPi}\rs\rT(\bm\nabla_2) \,
\delta(\bm x_2-\bm x_1) ,
\end{eqnarray}
based on
\begin{eqnarray}
\int_0^\infty\rd\omega \,f_{\omega,j}(r_2)\,f_{\omega,j}(r_1) &=&
\frac{1}{r_2r_1}\,\delta(r_2-r_1) . \qquad
\end{eqnarray}

\subsection{Longitudinal spherical photons}
\elabel{ssec:lpswf}

Longitudinal spherical wave functions are 
\begin{eqnarray}
\psi_{k,jm}^{{\rm L}(+)}(\bm x) &=& \frac{1}{k}
\left(\begin{array}{c} 
\0
\\[10 pt]
\bm\nabla\rs \, g_{k,j}(r) Y_j^m(\bm{\hat x})
\end{array}\right), \qquad
\elabel{eq:lswf1}
\\
\psi_{k,jm}^{{\rm L}(-)}(\bm x) &=& \frac{1}{k}
\left(\begin{array}{c} 
\bm\nabla\rs \, g_{k,j}(r) Y_j^m(\bm{\hat x})
\\[10 pt]
\0
\end{array}\right).
\elabel{eq:lswf2}
\end{eqnarray}
It follows from the identity $ \bm \tau\cdot\bm\nabla
\,\bm\nabla\rs = 0$ that they are longitudinal
\begin{eqnarray}
\iPi\rL(\bm\nabla)\,\psi_{k,jm}^{{\rm L}(\pi)}(\bm x) &=&
\psi_{k,jm}^{{\rm L}(\pi)}(\bm x),
\\
\iPi\rT(\bm\nabla)\,\psi_{k,jm}^{{\rm L}(\pi)}(\bm x) &=&
0
\end{eqnarray}
and that they are eigenfunctions of $\cH$, with eigenvalue $0$
\begin{eqnarray}
\cH \,\psi_{k,jm}^{{\rm L}(\pi)}(\bm x)
&=& 0 ,
\end{eqnarray}
with no condition on $g$.  However, in order to have a complete
set of longitudinal wave functions, a set of functions, indexed
by the parameter $k$ is specified here.  The form of the
plane-wave longitudinal solutions in Eqs.~(\ref{eq:plwf}) and
(\ref{eq:nlwf}) and of the expansion of a plane wave in
spherical waves suggest the choice
\begin{eqnarray}
g_{k,j}(r) &=& k\,\sqrt{\frac{2}{\pi}}
 \ j_j(k r) 
\end{eqnarray}
for the radial wave function, where $k$ is a free parameter.  
This set of functions provides an infinite orthonormal set of
degenerate ($\omega = 0$) basis functions for each $j$.

From the form of $\bm\nabla\rs$ in Eq.~(\ref{eq:snab}) and the
fact that $\bm X_2^{jm}(\bm{\hat x})$ and $\bm X_3^{jm}(\bm{\hat
x})$ are eigenfunctions of angular momentum, it follows that the
longitudinal spherical wave functions are also eigenfunctions of
angular momentum
\begin{eqnarray}
\bm\cJ^2 \,\psi_{k,jm}^{{\rm L}(\pi)}(\bm x) &=&
\hbar^2j(j+1)\,\psi_{k,jm}^{{\rm L}(\pi)}(\bm x),
\\
\cJ^{3} \,\psi_{k,jm}^{{\rm L}(\pi)}(\bm x) &=&
\hbar\,m\,\psi_{k,jm}^{{\rm L}(\pi)}(\bm x).
\end{eqnarray}
They have parity given by
\begin{eqnarray}
\fP \, \psi_{k,jm}^{{\rm L}(+)}(\bm x) &=&
(-1)^{j+1} \, \psi_{k,jm}^{{\rm L}(+)}(\bm x),
\\
\fP \, \psi_{k,jm}^{{\rm L}(-)}(\bm x) &=&
(-1)^{j} \, \psi_{k,jm}^{{\rm L}(-)}(\bm x).
\end{eqnarray}

With spherical Bessel functions for the radial
wave functions, we have, following Eqs.~(\ref{eq:snab}) to
(\ref{eq:lonex}),
\begin{eqnarray}
\frac{\bm\nabla\rs}{k} \, g_{k,j}(r) Y_j^m(\bm{\hat x})
&=&
\sqrt{\frac{j}{2j+1}}\ g_{k,j-1}(r)\,\bm Y_{jj-1}^m
(\bm{\hat x})
+\sqrt{\frac{j+1}{2j+1}}\ g_{k,j+1}(r)\,\bm Y_{jj+1}^m
(\bm{\hat x}) . 
\qquad
\end{eqnarray}
This identity facilitates calculation of the orthonormality relation for
the longitudinal wave functions, which is 
\begin{eqnarray}
\int\rd \bm x \,
\psi_{k_2,j_2m_2}^{{\rm L}(\pi_2)\dagger}(\bm x)
\psi_{k_1,j_1m_1}^{{\rm L}(\pi_1)}(\bm x)
&=& \delta_{\pi_2\pi_1}
\delta_{j_2j_1}\delta_{m_2m_1}
\delta(k_2-k_1),
\qquad
\elabel{eq:sphorthl}
\end{eqnarray}
with
\begin{eqnarray}
\int_0^\infty \rd r\,r^2\,g_{k_2,j}(r)
\,g_{k_1,j}(r) &=& \delta(k_2-k_1).
\qquad
\end{eqnarray}
The completeness relation for the longitudinal wave functions is
\begin{eqnarray}
\int_0^\infty\rd k\sum_{\pi j m}
\psi_{k,jm}^{{\rm L}(\pi)}(\bm x_2)
\psi_{k,jm}^{{\rm L}(\pi)\dagger}(\bm x_1)
&=& \iPi\rL(\bm\nabla_2)\,\delta(\bm x_2 - \bm x_1), \qquad
\elabel{eq:lcrel}
\end{eqnarray}
which follows from
\begin{eqnarray}
\sum_{\pi }
\psi_{k,jm}^{{\rm L}(\pi)}(\bm x_2)
\psi_{k,jm}^{{\rm L}(\pi)\dagger}(\bm x_1)
&=&
\left(\begin{array}{cc}
\bm T_k^{jm}(\bm x_2,\bm x_1) & \0 \\
\0 & \bm T_k^{jm}(\bm x_2,\bm x_1)
\end{array}\right) ,
\end{eqnarray}
where
\begin{eqnarray}
\bm T_k^{jm}(\bm x_2,\bm x_1) &=& -
\frac{\bm\nabla_{\!2\,\rm s}\bm\nabla_{\!1\,\rm s}^\dagger}
{\bm\nabla_2^2}
\, g_{k,j}(r_2) g_{k,j}(r_1)
Y_j^m(\bm{\hat x}_2)Y_j^{m*}(\bm{\hat x}_1),
\end{eqnarray}
and
\begin{eqnarray}
\bm\nabla^2 g_{k,j}(r) Y_j^m(\bm{\hat x})
&=& - k^2 g_{k,j}(r) Y_j^m(\bm{\hat x}).
\end{eqnarray}
Thus from
\begin{eqnarray}
\int_0^\infty \rd k\,g_{k,j}(r_2) g_{k,j}(r_1)
&=& \frac{1}{r_2r_1}\,\delta(r_2-r_1),
\end{eqnarray}
we have
\begin{eqnarray}
\int_0^\infty\rd k\sum_{j m}
\bm T_k^{jm}(\bm x_2,\bm x_1) &=&
\bm\iPi\rL\rs(\bm\nabla_2)\delta(\bm x_2 - \bm x_1),
\qquad
\end{eqnarray}
which provides Eq.~(\ref{eq:lcrel}).  That result, together with
Eq.~(\ref{eq:tcrel}) gives the full completeness relation.

As an illustration of a role of the spherical functions, we
revisit the example of a point charge at the origin, for which
\begin{eqnarray}
\iP\rp(\bm x) &=& -\frac{q}{4 \pi\epsilon_0}
\left(\begin{array}{c} \bm \nabla\rs \,\frac{\textstyle
1}{\textstyle r} \\ [8 pt]
\0 \end{array}\right).
\end{eqnarray}
In view of the integral
\begin{eqnarray}
\int_0^\infty \rd k \,\frac{1}{k} \, g_{k,0}(r) &=& 
\sqrt{\frac{\pi}{2}} \,\frac{1}{r} ,
\end{eqnarray}
one has
\begin{eqnarray}
\iP\rp(\bm x) &=&-\frac{q}{\sqrt{2}\, \pi\epsilon_0}
\int_0^\infty \rd k \, 
\psi_{k,00}^{{\rm L}(-)}(\bm x),
\end{eqnarray}
which is the analog, for spherical solutions, of
Eq.~(\ref{eq:coultr}) for plane wave solutions.

\section{Maxwell Green function}
\elabel{sec:prop}

A solution of the Maxwell equation for the electric and
magnetic fields $\iP(x)$ given a specified current source
$\iX(x)$, as they are related in Eq.~(\ref{eq:dms})
\begin{eqnarray*}
\gamma_\mu\partial^\mu \iP(x) &=& \iX(x),
\end{eqnarray*}
can be found with the aid of the $6\times6$ matrix Maxwell
Green function $\cD\rM(x_2 - x_1)$, given by
\begin{eqnarray}
\cD\rM(x_2 - x_1)
&=&
\sum_{\lambda=0}^2 \int \rd \bm k \ 
\psi_{\bm{k},\lambda}^{(+)}(x_2)
\overline{\psi}_{\bm{k},\lambda}^{(+)}(x_1)
\,\theta(t_2-t_1)
\nonumber\\
&-&
\sum_{\lambda=0}^2 \int \rd \bm k \ 
\psi_{\bm{k},\lambda}^{(-)}(x_2)
\overline{\psi}_{\bm{k},\lambda}^{(-)}(x_1)
\,\theta(t_1-t_2),
\nonumber\\
\elabel{eq:propdef}
\end{eqnarray}
where $\psi_{\bm{k},\lambda}^{(\pm)}(x)$ is given by
Eq.~(\ref{eq:trwf}), (\ref{eq:plwf}) or (\ref{eq:nlwf}), and
Eq.~(\ref{eq:tdwf}).
In view of the relations
\begin{eqnarray}
\gamma_\mu\partial_2^\mu\,
\psi_{\bm{k},\lambda}^{(\pm)}(x_2) &=& 0,
\\
\gamma_\mu\partial_2^\mu \, \theta(t_2-t_1) &=& 
\gamma_0\,\delta(ct_2-ct_1),
\\
\gamma_\mu\partial_2^\mu \, \theta(t_1-t_2) &=& 
-\gamma_0\,\delta(ct_2-ct_1)
\end{eqnarray}
and the completeness of the wave functions, we have
\begin{eqnarray}
\gamma_\mu\partial_2^\mu\,
\cD\rM(x_2 - x_1) &=& \cI \, \delta(x_2-x_1)
\elabel{eq:mpeq}
\end{eqnarray}
and
\begin{eqnarray}
\cD\rM(x_2 - x_1)\,\gamma_\mu\overleftarrow{\partial}_1^\mu 
&=& -\cI \, \delta(x_2-x_1),
\end{eqnarray}
where
\begin{eqnarray}
\delta(x_2-x_1) &=& \delta(ct_2-ct_1)\,\delta(\bm x_2 - \bm x_1).
\end{eqnarray}
In terms of the Green function, a solution for the
electric and magnetic fields is
\begin{eqnarray}
\iP(x_2) &=& \int \rd^4 x_1 \,
\cD\rM(x_2 - x_1)\,\iX(x_1),
\elabel{eq:solsource}
\end{eqnarray}
as is confirmed by the application of $\gamma_\mu\partial_2^\mu$
to both sides.  In Eq.~(\ref{eq:solsource}) $\rd^4 x_1 = c\,\rd
t_1 \rd\bm x_1$.  A separation into transverse or longitudinal
solutions may be made by restricting the sum over polarizations
to $\lambda = 1,2$ for a transverse solution or $\lambda = 0$
for a longitudinal solution.

The Maxwell Green function also may be written as an integral
over the four-vector $k$ of the plane-wave solutions.  For this
it is useful to make the separation into transverse and
longitudinal components.  For the transverse part we have
\begin{eqnarray}
\cD\rM\rT(x_2 - x_1)
&=&
\sum_{\lambda=1}^2 \int \rd \bm k \ 
\psi_{\bm{k},\lambda}^{(+)}(\bm x_2)
\overline{\psi}_{\bm{k},\lambda}^{(+)}(\bm x_1)
\,\re^{-\ri\omega (t_2-t_1)}\,\theta(t_2-t_1)
\nonumber\\
&&-
\sum_{\lambda=1}^2 \int \rd \bm k \ 
\psi_{\bm{k},\lambda}^{(-)}(\bm x_2)
\overline{\psi}_{\bm{k},\lambda}^{(-)}(\bm x_1)
\,\re^{\ri\omega (t_2-t_1)}\,\theta(t_1-t_2),
\end{eqnarray}
and we employ the identities
\begin{eqnarray}
\re^{-\ri\omega (t_2-t_1)}\,\theta(t_2-t_1) &=&
\frac{\ri}{2\pi}\int_{-\infty}^\infty \rd k_0\,\frac{\re^{-\ri
k_0 (ct_2-ct_1)}}
{k_0(1+\ri\delta) - \omega/c},
\\
-\re^{\ri\omega (t_2-t_1)}\,\theta(t_1-t_2) &=&
\frac{\ri}{2\pi}\int_{-\infty}^\infty \rd k_0\,\frac{\re^{-\ri
k_0 (ct_2-ct_1)}}
{k_0(1+\ri\delta) + \omega/c},
\end{eqnarray}
where the limit $\delta\rightarrow0^+$ for the integral is
understood.  We also have
\begin{eqnarray}
\bm\alpha\cdot\bm k\,\psi_{\bm{k},\lambda}^{(\pm)}(\bm x_2) &=&
|\bm k|\,\psi_{\bm{k},\lambda}^{(\pm)}(\bm x_2)
=\frac{\omega}{c}\,\psi_{\bm{k},\lambda}^{(\pm)}(\bm x_2).
\qquad
\end{eqnarray}
Together these relations yield [see Eq.~(\ref{eq:trcomp})]
\begin{eqnarray}
\cD\rM\rT(x_2 - x_1)
&=& \frac{\ri}{2\pi}\sum_{\lambda=1}^2
\int \rd^4 k 
\Bigg[
\frac{\re^{-\ri k_0 (ct_2-ct_1)}}{k_0(1+\ri\delta) - \bm\alpha\cdot\bm k}
\,\psi_{\bm{k},\lambda}^{(+)}(\bm x_2)
\overline{\psi}_{\bm{k},\lambda}^{(+)}(\bm x_1)
\nonumber\\&&\qquad
+\frac{\re^{-\ri k_0 (ct_2-ct_1)}}{k_0(1+\ri\delta) + \bm\alpha\cdot\bm k}
\,\psi_{\bm{k},\lambda}^{(-)}(\bm x_2)
\overline{\psi}_{\bm{k},\lambda}^{(-)}(\bm x_1)
\Bigg]
\nonumber\\ &=& \frac{\ri}{(2\pi)^4}
\int \rd^4 k \ \re^{-\ri k_0 (ct_2-ct_1)}
\,
\frac{\re^{\ri\bm k\cdot(\bm x_2-\bm
x_1)}}{\gamma^0k_0(1+\ri\delta) - \bm\gamma\cdot\bm k}
\left(\begin{array}{cc}(\bm\tau\cdot\bm{\hat k})^2 & 
\0 \\ \0 & (\bm\tau\cdot\bm{\hat k})^2 \end{array}\right)
\nonumber\\ &=& \frac{\ri}{(2\pi)^4} \, \iPi\rT(\bm\nabla_2)
\int_{\rm C_F} \rd^4 k \ \frac{\re^{-\ri k\cdot (x_2-x_1)}}
{\gamma^\mu k_\mu} ,
\elabel{eq:transprop}
\end{eqnarray}
where $\rd^4k = \rd k_0\,\rd\bm k$, and ${\rm C_F}$ indicates
that the contour of integration over $k_0$ is the Feynman
contour, which passes from $-\infty$ below the negative real
axis, through $0$, and above the positive real axis to
$+\infty$; this is equivalent to including the factor
$(1+\ri\delta)$ multiplying $k_0$ in the denominator and
integrating along the real axis.

For applications, it is useful to consider an alternative form
for the transverse Green function.  Taking into account
the relation
\begin{eqnarray}
\iPi\rT(\bm{\hat k})
\frac{1}{\gamma^0 k_0(1+\ri\delta) - \bm\gamma\cdot\bm k}
&=&
\iPi\rT(\bm{\hat k}) \,
\frac{\gamma^0 k_0 - \bm\gamma\cdot\bm k}
{k_0^2 - \bm k^2 + \ri \delta}
\nonumber\\&=&
\left(\begin{array}{cc}k_0\,(\bm\tau\cdot\bm{\hat k})^2 & 
-\bm\tau\cdot\bm{k} \\ \bm\tau\cdot\bm{k} & 
-k_0\,(\bm\tau\cdot\bm{\hat k})^2 \end{array}\right)
\frac{1}{k^2 + \ri \delta} ,
\elabel{eq:trgfid}
\end{eqnarray}
we have, from Eq.~(\ref{eq:transprop}), 
\begin{eqnarray}
\cD\rM\rT(x_2 - x_1)
&=&
\frac{\ri}{(2\pi)^4} 
\int_{-\infty}^\infty \rd k_0 \
\re^{-\ri k_0 (ct_2-ct_1)} \,
\left(\begin{array}{cc}k_0\,\bm \iPi\rs\rT(\bm\nabla) & 
\ri\,\bm\tau\cdot\bm\nabla \\
-\ri\,\bm\tau\cdot\bm\nabla & 
-k_0\,\bm \iPi\rs\rT(\bm\nabla) \end{array}\right)
\int \rd \bm k \,
\frac{\re^{\ri\bm k\cdot\bm r}}{k^2 + \ri \delta}
\nonumber\\&=&
\frac{1}{8\pi^2\ri} 
\int_{-\infty}^\infty \rd k_0 \
\re^{-\ri k_0 (ct_2-ct_1)} \,
\left(\begin{array}{cc}k_0\,\bm \iPi\rs\rT(\bm\nabla) & 
\ri\,\bm\tau\cdot\bm\nabla \\
-\ri\,\bm\tau\cdot\bm\nabla & 
-k_0\,\bm \iPi\rs\rT(\bm\nabla) \end{array}\right)
\frac{\re^{\ri(k_0^2+\ri\delta)^{1/2}| \bm r|}}
{|\bm r|}
\nonumber\\&\rightarrow&
\frac{1}{8\pi^2\ri} 
\int_{-\infty}^\infty \rd k_0 \
\re^{-\ri k_0 (ct_2-ct_1)}
\nonumber\\&&\qquad\times
\left(\begin{array}{cc}k_0(\bm{\bm \tau\cdot\hat r})^2 & 
-(k_0^2+\ri\delta)^{1/2}\,\bm\tau\cdot\bm{\hat r} \\
(k_0^2+\ri\delta)^{1/2}\,\bm\tau\cdot\bm{\hat r} & 
-k_0(\bm{\bm \tau\cdot\hat r})^2 \end{array}\right)
\frac{\re^{\ri(k_0^2+\ri\delta)^{1/2}| \bm r|}}
{|\bm r|},
\elabel{eq:trprop}
\end{eqnarray}
where $\bm r = \bm x_2 - \bm x_1$, the gradient $\bm \nabla$ is
with respect to $\bm r$, and the branch of the square root in
the exponent is determined by the condition ${\rm
Im}(k_0^2+\ri\delta)^{1/2} > 0$, which specifies that
$(k_0^2+\ri\delta)^{1/2} \rightarrow |k_0|$ for real values of
$k_0$.  In the last line of Eq.~(\ref{eq:trprop}), higher-order
terms in $(k_0\,|\bm r|)^{-1}$ are not included,  but the exact
expression follows from the formulas in \ref{app:extr}.

For the longitudinal Green function, we write
\begin{eqnarray}
\cD\rM\rL(x_2 - x_1)
&=&
\int \rd \bm k \ 
\psi_{\bm{k},0}^{(+)}(\bm x_2)
\overline{\psi}_{\bm{k},0}^{(+)}(\bm x_1)
\,\re^{-\epsilon (ct_2-ct_1)}\,\theta(t_2-t_1)
\nonumber\\
&&-
\int \rd \bm k \ 
\psi_{\bm{k},0}^{(-)}(\bm x_2)
\overline{\psi}_{\bm{k},0}^{(-)}(\bm x_1)
\,\re^{\epsilon(ct_2-ct_1)}\,\theta(t_1-t_2),
\elabel{eq:lprop}
\end{eqnarray}
where damping factors with $\epsilon>0$ are added so that the
Green function falls off for large time differences.  In
addition, we assume that the longitudinal wave functions are
solutions of the Maxwell equation with an infinitesimal mass
$m_\epsilon$ included, as given in Eq.~(\ref{eq:dmeq}), in order
to be able to use the Feynman contour to specify the path of
integration over $k_0$ in relation to the poles of the
integrand.  We employ the identities
\begin{eqnarray}
\re^{-\ri m_\epsilon c^2 (t_2-t_1)/\hbar}\,
\re^{-\epsilon(ct_2-ct_1)} \,
\theta(t_2-t_1)
&=&
\frac{\ri}{2\pi}\int_{-\infty}^\infty \rd k_0\,\frac{\re^{-\ri
k_0 (ct_2-ct_1)}}
{k_0 + \ri\epsilon - m_\epsilon c/\hbar},
\qquad
\\
-\re^{\ri m_\epsilon c^2 (t_2-t_1)/\hbar}\,
\re^{\epsilon (ct_2-ct_1)}\,\theta(t_1-t_2)
&=&
\frac{\ri}{2\pi}\int_{-\infty}^\infty \rd k_0\,\frac{\re^{-\ri
k_0 (ct_2-ct_1)}}
{k_0 - \ri\epsilon + m_\epsilon c/\hbar}
\end{eqnarray}
and
\begin{eqnarray}
\gamma^0\,\psi_{\bm{k},0}^{(\kappa)}(\bm x_2) &=&
\kappa\,\psi_{\bm{k},0}^{(\kappa)}(\bm x_2),
\\
\bm\alpha\cdot\bm k\,\psi_{\bm{k},0}^{(\kappa)}(\bm x_2) &=& 0
\end{eqnarray}
to obtain
\begin{eqnarray}
\cD\rM\rL(x_2 - x_1) &=& 
\frac{\ri}{2\pi} \int \rd^4 k 
\Bigg[
\frac{\re^{-\ri k_0 (ct_2-ct_1)}}{k_0 + \ri\epsilon -
m_\epsilon c/\hbar}
\,\psi_{\bm{k},0}^{(+)}(\bm x_2)
\overline{\psi}_{\bm{k},0}^{(+)}(\bm x_1)
\nonumber\\&&\qquad\qquad
+\frac{\re^{-\ri k_0 (ct_2-ct_1)}}{k_0 - \ri\epsilon + m_\epsilon
c/\hbar}
\,\psi_{\bm{k},0}^{(-)}(\bm x_2)
\overline{\psi}_{\bm{k},0}^{(-)}(\bm x_1)
\Bigg]
\nonumber\\&=&
\frac{\ri}{2\pi} \int_{\rm C_F} \rd^4 k 
\frac{\re^{-\ri k_0 (ct_2-ct_1)}}{k_0 - \gamma^0m_\epsilon
c/\hbar}
\sum_{\kappa \rightarrow \pm}
\psi_{\bm{k},0}^{(\kappa)}(\bm x_2)
\overline{\psi}_{\bm{k},0}^{(\kappa)}(\bm x_1)
\nonumber\\ &=& \frac{\ri}{(2\pi)^4} \,
\iPi\rL(\bm\nabla_2)
\int_{\rm C_F} \rd^4 k \ \frac{\re^{-\ri k\cdot (x_2-x_1)}}
{\gamma^\mu k_\mu - m_\epsilon c/\hbar} \, .
\elabel{eq:longprop}
\end{eqnarray}
Here the limit $\epsilon\rightarrow0$ would be undefined without
the mass term.  A concise alternative expression for the
longitudinal Green function is obtained by the substitution of
the partial completeness relations that follow from
Eqs.~(\ref{eq:plongcomp}) and (\ref{eq:nlongcomp}) into
Eq.~(\ref{eq:lprop}):
\begin{eqnarray}
\cD\rM\rL(x_2 - x_1) &=& 
\iPi\rL(\bm\nabla_2) \,
\delta(\bm x_2 - \bm x_1)
\left(\begin{array}{cc}
\bm I \, \theta(t_2-t_1) & \0 \\
\0 & \bm I \, \theta(t_1-t_2)
\end{array}\right) .
\end{eqnarray}

The transverse and longitudinal Green functions in
Eqs.~(\ref{eq:transprop}) and (\ref{eq:longprop}) differ only by
the type of projection operator and the infinitesimal mass term
in Eq.~(\ref{eq:longprop}).  However, such a mass term in the
last line of Eqs.~(\ref{eq:transprop}) would not change the
relation of the path of integration over $k_0$ to the location
of the poles of the integrand, so it could also be included in
that expression.  In particular, the poles in
Eq.~(\ref{eq:trgfid}) at $k_0 = \pm \left(\bm k^2 - \ri
\delta\right)^{1/2}$ would move to $k_0 = \pm\left[\bm k^2
+(m_\epsilon c/\hbar)^2 - \ri \delta\right]^{1/2}$.  These poles
lie on curves in the second and fourth quadrants of the complex
$k_0$ plane, whereas the Feynman contour passes through the
first and third quadrants.  Thus, we may write $\cD\rM\rT(x_2 -
x_1) + \cD\rM\rL(x_2 - x_1) = \cD\rM(x_2 - x_1)$, with
\begin{eqnarray} \cD\rM(x_2 - x_1) 
&=& 
\frac{\ri}{(2\pi)^4} \, \int_{C_F} \rd^4 k \
\frac{\re^{-\ri k\cdot (x_2-x_1)}} {\gamma^\mu k_\mu -
m_\epsilon c/\hbar}.
\elabel{eq:mprop}
\end{eqnarray}
This result is a covariant Green function for the Maxwell
equation which is of the same form as the well-known Green
function for the Dirac equation.  A formal
coordinate-representation is
\begin{eqnarray}
\cD\rM(x_2 - x_1) 
&=&
\frac{1}{\gamma_\mu\partial_2^\mu} \, \delta(x_2-x_1) .
\end{eqnarray}

The fields given by Eq.~(\ref{eq:solsource}) represent a
particular solution of the Maxwell equation.  Any solution of
Eq.~(\ref{eq:solsource}) for $\iX(x_1)=0$, such as the field of
a static charge distribution, may be added to the particular
solution, and the sum will be a solution with the same source
function.  In fact, even if the three-vector current density
vanishes in the distant past and future, there could be a static
charge distribution that persists, with or without a net total
charge, for which the fields would be non-zero indefinitely.  To
deal with this case, we obtain an expression that takes into
account the possible fields in the past and future by writing
the time derivative 
\begin{eqnarray}
\frac{\partial}{\partial(ct_1)}
\int{\rm d}\bm x_1\,
\cD\rM(x_2 - x_1)\gamma_0\iP(x_1)
&=& \int{\rm d}\bm x_1 \, \cD\rM(x_2 - x_1)
\gamma_0
\overleftarrow{\partial}_1^{0}
\iP(x_1)
\nonumber\\&&
+ \int{\rm d}\bm x_1 \, \cD\rM(x_2 - x_1)
\gamma_0
\partial_1^{\,0}
\iP(x_1),  \quad
\label{eq:mpr1}
\end{eqnarray}
and for fields that vanish for large space-like distances, we
write
\begin{eqnarray}
\int{\rm d}\bm x_1 \, \bm\nabla_1\cdot \, \cD\rM(x_2 - x_1) \bm\gamma \iP(x_1)
&=& 
\int{\rm d}\bm x_1 \, \cD\rM(x_2 - x_1) \bm\gamma \cdot
\overleftarrow{\bm \nabla}_1
\iP(x_1)
\nonumber\\&&
+
\int{\rm d}\bm x_1 \, \cD\rM(x_2 - x_1) \bm \gamma\cdot\bm\nabla_1  \iP(x_1)
= 0, \qquad
\label{eq:mpr2}
\end{eqnarray}
where the result is zero, because it may be written as an
integral over the bounding surface, by the Gauss-Ostrogradsky
theorem.  The sum of Eqs.~(\ref{eq:mpr1}) and (\ref{eq:mpr2}) is
\begin{eqnarray}
&&
\frac{\partial}{\partial(ct_1)}
\int{\rm d}\bm x_1\,
\cD\rM(x_2 - x_1)\gamma_0\iP(x_1)
\nonumber\\&&\qquad\qquad
= \int{\rm d}\bm x_1 \, \cD\rM(x_2 - x_1)
\gamma_\mu
\overleftarrow{\partial}_1^{\,\mu}
\iP(x_1)
 + \int{\rm d}\bm x_1 \, \cD\rM(x_2 - x_1)
\gamma_\mu
\partial_1^{\,\mu}
\iP(x_1) \qquad
\nonumber\\&&\qquad\qquad
= - \int{\rm d}\bm x_1 \, \delta(x_2 - x_1) \iP(x_1)
 + \int{\rm d}\bm x_1 \, \cD\rM(x_2 - x_1)
\,\iX(x_1) .
\label{eq:mpr3}
\end{eqnarray}
Integration of Eq.~(\ref{eq:mpr3}) over $t_1$ from $t_\ri$ to $t_\rf$, where
$t_\ri < t_2 < t_\rf$, yields
\begin{eqnarray}
\iP(x_2) &=& \int{\rm d}\bm x_1\,
\cD\rM(x_2 - x_1)\gamma_0\iP(x_1)\Big|_{t_1 = t_\ri}
-\int{\rm d}\bm x_1\,
\cD\rM(x_2 - x_1)\gamma_0\iP(x_1)\Big|_{t_1 = t_\rf}
\nonumber\\&&
+ c\int_{t_\ri}^{t_\rf}{\rm d}t_1\int{\rm d}\bm x_1 \, \cD\rM(x_2 - x_1)
\,\iX(x_1)
\end{eqnarray}
or
\begin{eqnarray}
\iP(x_2) &=& \int{\rm d}\bm x_1\,
\sum_{\lambda=0}^2 \int \rd \bm k  
\left[
\psi_{\bm{k},\lambda}^{(+)}(x_2)
\psi_{\bm{k},\lambda}^{(+)\dagger}(x_1)
\iP(x_1)\Big|_{t_1 = t_\ri}
+
\psi_{\bm{k},\lambda}^{(-)}(x_2)
\psi_{\bm{k},\lambda}^{(-)\dagger}(x_1)
\iP(x_1)\Big|_{t_1 = t_\rf}
\right]
\nonumber\\ &&+
 c\int_{t_\ri}^{t_\rf}{\rm d}t_1\int{\rm d}\bm x_1 \, \cD\rM(x_2 - x_1)
\,\iX(x_1).
\elabel{eq:propbdy}
\end{eqnarray}

As a consistency check of this expression, we note that it
properly reduces to the expected result for the field of a
constant charge distribution.  In this case, the current source
term vanishes, the initial and final fields are the same, and
they are purely longitudinal.  As a result, only longitudinal
functions with no time dependence will contribute to the sum
over states, which is just the longitudinal completeness
relation, and Eq.~(\ref{eq:propbdy}) reduces to the proper
identity.

\section{Applications of the Maxwell Green function}
\elabel{sec:appmpf}

The Maxwell Green function is used here to calculate the
radiation fields of a point dipole source as an example of an
application.  Only the large distance transverse fields are
considered, and they are given by
\begin{eqnarray}
\iP_\rd(x_2) &=& \int \rd^4 x_1 \,
\cD\rM\rT(x_2 - x_1)\,\iX_\rd(x_1),
\end{eqnarray}
with the source term
\begin{eqnarray}
\iX_\rd(x) &=& \left(\begin{array}{c} -\mu_0c\,
\bm j\rs(\bm x) \,
\re^{-\ri\omega_\rd t} \\ \0 \end{array} \right).
\end{eqnarray}
The classical source for dipole radiation is a
charge $q$ with position 
\begin{eqnarray}
\bm x_{\rm d}(t) &=& \bm x_0 \cos{\omega_{\rm d} t}
\end{eqnarray}
which produces a current density
\begin{eqnarray}
\bm j_{\rm cl}(x) &=& q \,\delta(\bm x - \bm x_{\rm d}(t))
\, \dot{\bm x}_{\rm d}(t)
\nonumber\\&\approx&
- \omega_{\rm d} \, \bm d \,\delta(\bm x) \sin{\omega_{\rm d} t},
\end{eqnarray}
where $\bm d = q\,\bm x_0$.  This is the real part of
\begin{eqnarray}
\bm j(x) &=&
- \ri \, \omega_{\rm d} \, \bm d \,\delta(\bm x) 
\re^{-\ri\omega_{\rm d} t},
\end{eqnarray}
which is the source current for the radiation \cite{1998165}.
For the transverse Maxwell Green function, we use the
expression on the last line of Eq.~(\ref{eq:trprop}).
Integration over $t_1$ yields a factor
$2\pi\delta(k_0c-\omega_\rd)$, and evaluation of the
integration over $k_0$ follows.  The result is
\begin{eqnarray}
&&c\int\rd t_1 \, \cD\rM\rT(x_2 - x_1)\,\iX_\rd(x_1)
\nonumber\\&&\qquad\qquad=
-\frac{\mu_0c\,k}{4\pi\ri} \,
\left(\begin{array}{cc}(\bm \tau\cdot\bm{\hat r})^2 & 
-\bm\tau\cdot\bm{\hat r} \\
\bm\tau\cdot\bm{\hat r} & 
-(\bm \tau\cdot\bm{\hat r})^2 \end{array}\right)
\left(\begin{array}{c} \bm j\rs(\bm x_1)
\\ \0 \end{array} \right)
\frac{\re^{\ri k |\bm r|}}
{|\bm r|}
\, \re^{-\ri \omega_\rd t_2} + \dots \ ,
\qquad
\end{eqnarray}
where $k = \omega_\rd/c$.  Since the source is point-like at the
origin $\bm x_1 = 0$, $\bm r = \bm x_2$, and
\begin{eqnarray}
\iP_\rd(x) &=& 
\frac{k^2}{4\pi\epsilon_0} \,
\left(\begin{array}{c} (\bm \tau\cdot\bm{\hat x})^2 \,\bm d\rs
\\ \bm \tau\cdot\bm{\hat x} \,\bm d\rs \end{array} \right)
\frac{\re^{\ri k |\bm x|}}
{|\bm x|}
\, \re^{-\ri \omega_\rd t} + \dots \ .
\end{eqnarray}
The time-average differential radiated power, based on
Eqs.~(\ref{eq:emdo}) to (\ref{eq:srelg}) with a factor $1/2$
from the time averaging \cite{1998165} is
\begin{eqnarray}
\frac{\rd I_{\rm d}}{\rd\iO} &=& 
\frac{1}{2} \, \bm x^2\,\bm{\hat x}\cdot \bm S(x)
=
\frac{c\epsilon_0}{4} \, \bm x^2 \,
\overline{\iP}_{\rm d}(x) \bm \gamma \cdot \bm{\hat x}
\iP_{\rm d}(x) 
\nonumber\\&=&
\frac{c k^4}{32\,\pi^2\epsilon_0} \,
\bm d\rs^\dagger (\bm \tau\cdot\bm{\hat x})^2 \, \bm d\rs
=
\frac{c k^4}{32\,\pi^2\epsilon_0}
\left[\, \bm d^2 - (\bm{\hat x}\cdot\bm d)^2\right],
\end{eqnarray}
which is the well-known result.

A more realistic example is the radiation produced by a Dirac
transition current, which is given by 
\begin{eqnarray} \iX_{\rm
D}(x) &=& \left(\begin{array}{c} -\mu_0c\, \bm j\rs^{if}(x) \,
\\ \0 \end{array} \right) =
\left(\begin{array}{c} 
\frac{\textstyle 2e}{\textstyle\epsilon_0}\,
\phi_f^\dagger(\bm x) \, \bm \alpha\rs\,\phi_i(\bm x) \,
\re^{-\ri\omega_{if} t} \\ \0 \end{array} \right),
\end{eqnarray} 
where $\phi_i$ and $\phi_f$ are the initial and final hydrogen
atom Dirac wave functions, here $\bm \alpha$ is the $4\times4$
Dirac matrix, and
\begin{eqnarray}
\omega_{if} &=& \frac{E_i - E_f}{\hbar}
\\\nonumber
\end{eqnarray}
is the frequency corresponding to the energy difference of the
transition.  
The factor of 2 multiplying the matrix element accounts for the
difference between a classical dipole moment and the quantum
mechanical dipole moment operator in Eq.~(\ref{eq:diprate}).
(See the footnote on p. 407 of \cite{1998165}.)  ~We have
\begin{eqnarray}
&&c\int\rd t_1 \, \cD\rM\rT(x_2 - x_1)\,\iX_{\rm D}(x_1)
\nonumber\\&&\qquad\qquad=
\frac{\ri k}{4\pi\epsilon_0c} \,
\left(\begin{array}{cc}(\bm \tau\cdot\bm{\hat r})^2 & 
-\bm\tau\cdot\bm{\hat r} \\
\bm\tau\cdot\bm{\hat r} & 
-(\bm \tau\cdot\bm{\hat r})^2 \end{array}\right)
\left(\begin{array}{c} \bm j\rs^{if}(\bm x_1)
\\ \0 \end{array} \right)
\frac{\re^{\ri k |\bm r|}}
{|\bm r|}
\, \re^{-\ri \omega_{if} t_2} + \dots \ ,
\qquad
\end{eqnarray}
where $k = \omega_{if}/c$.
For distances far from the source atom, $|\bm
x_2| \gg |\bm x_1|$, $\bm {\hat k} \approx \bm {\hat r} \approx
\bm{\hat x}_2$, and in the exponent
$k|\bm r| = k|\bm x_2| - \bm k\cdot\bm x_1 + \dots \ $,
which yields
\begin{eqnarray}
\iP_{\rm D}(x_2) &=& 
\int \rd^4 x_1 \,
\cD\rM\rT(x_2 - x_1)\,\iX_{\rm D}(x_1)
\nonumber\\&=&
\frac{\ri k}{4\pi\epsilon_0c} \,
\int \rd \bm x_1 
\left(\begin{array}{c} (\bm \tau\cdot\bm{\hat k})^2 \,\bm
j\rs^{if}(\bm x_1)
\\ \bm \tau\cdot\bm{\hat k} \,\bm j\rs^{if}(\bm x_1)
\end{array} \right)
\re^{-\ri \bm k\cdot\bm x_1} \,
\frac{\re^{\ri k |\bm x_2|}}
{|\bm x_2|}
\, \re^{-\ri \omega_{if} t_2} + \dots \ .
\end{eqnarray}
The average radiated power is
\begin{eqnarray}
\frac{\rd I_{\rm D}}{\rd\iO} &=& 
\frac{1}{2} \, \bm x_2^2\,\bm{\hat x_2}\cdot \bm S(x_2)
=
\frac{c\epsilon_0}{4} \, \bm x_2^2 \,
\overline{\iP}_{\rm D}(x_2) \bm \gamma \cdot \bm{\hat k}
\iP_{\rm D}(x_2) 
\nonumber\\[8 pt] &=&
\frac{k^2}{32\,\pi^2\epsilon_0c} \,
\int\rd\bm x_1\,\bm j\rs^{if\dagger}
(\bm x_1)\,\re^{\ri\bm k\cdot\bm x_1} \,
(\bm \tau\cdot\bm{\hat k})^2 \, 
\int\rd\bm x_1^\prime\,\bm j\rs^{if}(\bm
x_1^\prime)\,\re^{-\ri\bm k\cdot\bm x_1^\prime}
\nonumber\\&=&
\hbar\omega_{if} \,\frac{\alpha k c}{2\pi}
\sum_{\lambda=1}^2
\int\rd\bm x\, \phi_i^\dagger(\bm x) \, 
\bm{\hat \epsilon}_\lambda(\bm{\hat k})
\cdot \bm \alpha \, \re^{\ri\bm k\cdot\bm x}
\phi_f(\bm x)
\int\rd\bm x^\prime\, \phi_f^\dagger(\bm x^\prime) \, 
\bm{\hat \epsilon}_\lambda(\bm{\hat k})
\cdot \bm \alpha \, \re^{-\ri\bm k\cdot\bm x^\prime}
\phi_i(\bm x^\prime) ,
\nonumber\\
\elabel{eq:opdr}
\end{eqnarray}
where $\alpha = e^2/4\pi\epsilon_0\hbar c$ is the fine-structure
constant.  The radiated power integrated over directions of the
vector $\bm{\hat k}$ may be interpreted as
$\hbar\omega_{if}A_{if}$, where $A_{if}$ is the radiative
transition rate for $i\rightarrow f$, that is, the probability
that the atom providing the source current makes a transition
from state $i$ to state $f$ in one second.  This gives
\begin{eqnarray}
A_{if} &=& 
\frac{\alpha k c}{2\pi}
\int\rd\iO_{\bm k}
\sum_{\lambda=1}^2
\left< i \left|
\bm{\hat \epsilon}_\lambda(\bm{\hat k})
\cdot \bm \alpha \, \re^{\ri\bm k\cdot\bm x}
\right|f\right>
\left< f \left|
\bm{\hat \epsilon}_\lambda(\bm{\hat k})
\cdot \bm \alpha \, \re^{-\ri\bm k\cdot\bm x}
\right|i\right>,
\elabel{eq:transrate}
\end{eqnarray}
which is the same as the relativistic radiative transition rate
given by QED (see \ref{app:transrate}).  In the dipole
approximation $\re^{\ri \bm k \cdot \bm x} \rightarrow 1$,
$\left<i\left| \bm \alpha \right|f\right> = \ri\,k
\left<i\left|\bm x \right|f\right>$, which follows from the
identity $[H,\bm x] = [c\bm \alpha\cdot \bm p,\bm x] = -\ri\hbar
c \bm \alpha$, where $H$ is the Dirac Hamiltonian, and
integration over $\bm{\hat k}$ yields the familiar result
\begin{eqnarray}
A_{if} &\rightarrow& \frac{4\alpha\omega_{if}^3}{3c^2}
\left|\left< f \left|
\bm x
\right|i\right> \right|^2.
\elabel{eq:diprate}
\end{eqnarray}

\section{Summary}
\elabel{sec:summary}

In Eq.~(\ref{eq:dms}), two of the Maxwell equations,
Eqs.~(\ref{eq:maxeq2}) and (\ref{eq:maxeq3}), are written in the
form of the Dirac equation without a mass, but with the addition
of a source term $\iX(x)$:
\begin{eqnarray*}
\gamma^\mu\partial_\mu  \iP(x) &=& \iX(x),
\end{eqnarray*}
where the gamma matrices are $6\times6$ versions of the
Dirac gamma matrices in Eq.~(\ref{eq:ggammas}), and
\begin{eqnarray*}
\iP(x) = \left(\begin{array}{c}
\bm E\rs(x) \\ \ri \, c \bm B\rs(x) \msp \end{array}\right),
&&
\iX(x) = \left(\begin{array}{c}
-\mu_0c\bm J\rs(x) \\ \0 \msp \end{array}\right)
\end{eqnarray*}
from Eqs.~(\ref{eq:dwf}) and (\ref{eq:des}).
The source-free version of this equation, with $\iX(x)=0$, can be
written as a Schr\"odinger-like equation from
Eq.~(\ref{eq:block6}) or (\ref{eq:freqeq})
\begin{eqnarray}
\ri\hbar\,\frac{\partial}{\partial t} \iP(x) = 
\cH\iP(x),
\elabel{eq:seq}
\end{eqnarray}
where the Hamiltonian, Eq.~(\ref{eq:phham}),
\begin{eqnarray*}
\cH = -\ri \,\hbar c \,\bm \alpha\cdot\bm\nabla 
\end{eqnarray*}
is the analog of the Dirac Hamiltonian for the electron.  The
factors of $\hbar$ are not essential here, but they are
introduced to provide the conventional units of frequency and
energy.  As with the Dirac wave functions, where all four
components are necessary to describe an electron bound in an
atom relativistically, all six of the components of the photon
wave function apparently are necessary to properly account for
the space-time properties of electromagnetic fields.  There are
three polarization degrees of freedom, two for radiation and one
for electrostatic interactions, and relativistic covariance
requires twice that many components.  Alternatively stated, six
complex functions are necessary to describe the six components
of the electric and magnetic fields, and they are coupled by the
Maxwell equation and Lorentz transformations.

According to Eq.~(\ref{eq:seq}), as in Eq.~(\ref{eq:tdep}), the
time dependence of the solution is given by
\begin{eqnarray}
\iP(x) = 
\re^{-\ri \cH t/\hbar}
\iP(\bm x) .
\end{eqnarray}
The time-independent solutions may be expanded in eigenfunctions
of the Hamiltonian with eigenvalues $E_n$ given by
\begin{eqnarray}
\cH\iP_n(\bm x) &=& E_n \iP_n(\bm x),
\elabel{eq:eveq}
\end{eqnarray}
where $n$ is a set of parameters that characterize the state
represented by the wave function.  For each eigenfunction, one
has
\begin{eqnarray}
\iP_n(x) = 
\re^{-\ri E_n t/\hbar}
\iP_n(\bm x) .
\end{eqnarray}
The eigenfunctions are orthonormal
\begin{eqnarray}
\int \rd \bm x\,\iP_{n_2}^\dagger(\bm x)\iP_{n_1}(\bm x)
= \delta_{n_2n_1},
\elabel{eq:conorth}
\end{eqnarray}
and they are complete
\begin{eqnarray}
\sum_n  \iP_n(\bm x_2) \iP_n^\dagger(\bm x_1) = \delta(\bm
x_2-\bm x_1).
\elabel{eq:concomp}
\end{eqnarray}
The state index $n$ includes continuous variables, so
Eq.~(\ref{eq:conorth}) has delta functions in those variables
on the right-hand side, and the summation symbol in
Eq.~(\ref{eq:concomp}) includes integration over those
variables.

States considered in detail in this paper are propagating plane
waves in Secs.~\ref{ssec:tph} and \ref{ssec:lph}, standing plane
waves in Sec.~\ref{ssec:ptwf}, and angular-momentum eigenstates
in Secs.~\ref{ssec:stp} and \ref{ssec:lpswf}.  The propagating
plane-wave states are eigenfunctions of the momentum operator in
Eq.~(\ref{eq:phmom})
\begin{eqnarray*}
\bm \cP = -\ri \, \hbar \, \cI \, \bm \nabla.
\end{eqnarray*}
This operator commutes with the Hamiltonian,
$\left[\cH,\bm\cP\right] = 0$, and eigenstates of both energy
and momentum are given in Eqs.~(\ref{eq:trwf}), (\ref{eq:plwf}),
and (\ref{eq:nlwf}).  These plane-wave states are further
characterized by polarization vectors given in
Eq.~(\ref{eq:epst}) or (\ref{eq:epsl}).  Linear combinations of
the traveling plane waves, combined to give standing-wave parity
eigenfunctions, are in Eqs.~(\ref{eq:ptpe}), (\ref{eq:ntpe}),
and (\ref{eq:lpe1}) to (\ref{eq:lpe4}).  The angular-momentum
operator, Eq.~(\ref{eq:cj}), is
\begin{eqnarray*}
\bm\cJ &=& 
\bm x \times \bm \cP + \hbar \, \bm \cS,
\end{eqnarray*}
where the spin matrix, Eq.~(\ref{eq:spinop}), is
\begin{eqnarray*}
\bm \cS &=& \left(\begin{array}{cc} \bm \tau & \0 \\
\0 & \bm \tau \end{array}\right),
\end{eqnarray*}
and $\bm \tau$ is given by Eqs.~(\ref{eq:t3}) to (\ref{eq:t2}).
One has $\left[\cH,\bm\cJ\right] = 0$ in Eq.~(\ref{eq:hjcom}),
and simultaneous eigenfunctions of energy, angular momentum
squared $\bm\cJ^2$, third component of angular momentum
$\cJ^{3}$, and parity are given in Eq~(\ref{eq:tswf1}),
(\ref{eq:tswf2}), (\ref{eq:lswf1}), and (\ref{eq:lswf2}).  All
three sets of eigenfunctions listed above are shown to be
orthogonal and complete, as in Eqs.~(\ref{eq:conorth}) and
(\ref{eq:concomp}).

The eigenfunctions considered here are not normalizable wave
functions.  However, they provide basis functions for the
expansion of a normalizable wave packet, as discussed in
Sec.~\ref{ssec:wp}.  For the sum
\begin{eqnarray}
\iP_f(\bm x) = \sum_n f_n \iP_n(\bm x),
\end{eqnarray}
from the orthonormality of the eigenfunctions one has
\begin{eqnarray}
f_n = \int \rd \bm x \iP_n^\dagger(\bm x)
\iP_f(\bm x)
\end{eqnarray}
and
\begin{eqnarray}
\int\rd\bm x \, \iP_f^\dagger(\bm x)\iP_f(\bm x)
= \sum_n |f_n|^2 = 1
\end{eqnarray}
for suitably chosen $f_n$.
For the example of a Gaussian wave packet in
Eq.~(\ref{eq:gaussian}), the expectation value of
the Hamiltonian, in Eq.~(\ref{eq:exham}), is
\begin{eqnarray*} 
\left<\iP_f
\left| \, \cH \, \right| \iP_f \right> &=& 
\hbar \omega_0,
\end{eqnarray*}
where $\omega_0 = c|\bm k_0|$ is the frequency corresponding to
the wave vector $\bm k_0$ of the wave packet.  It is clear that
Eq.~(\ref{eq:exham}) applies in more generality than just to the
wave packet in Eq.~(\ref{eq:gaussian}).  If the Gaussian shape
function were replaced by any normalized real function, the
expectation value of the Hamiltonian would still be exactly
$\hbar\omega_0$.  For the wave packet in
Eq.~(\ref{eq:gaussian}), the expectation expectation value of
the momentum operator, Eq.~(\ref{eq:exmom}), is
\begin{eqnarray*} \left<\iP_f \left|\,
\bm \cP \, \right| \iP_f \right> &=& \hbar \bm k_0 ,
\end{eqnarray*} 
and the expectation value of the projection of the angular
momentum in the direction of the wave vector is
\begin{eqnarray} \left<\iP_f \left|\,
\bm \cJ \cdot \bm{\hat k}_0 \, \right| \iP_f \right> &=& 
\hbar \,
\bm{\hat \epsilon}_1^\dagger(\bm{\hat k}_0)
\,\bm\tau\cdot\bm{\hat k}_0 \,
\bm{\hat \epsilon}_1(\bm{\hat k}_0),
\end{eqnarray}
where the result depends on the polarization state represented
by $\bm{\hat \epsilon}_1(\bm{\hat k}_0)$.  For circular
polarization, Eq.~(\ref{eq:circ}), the expectation value is
$\pm\hbar$, while for linear polarization, Eq.~(\ref{eq:lin}),
it is 0.  The real part of the energy density is $\hbar
\omega_0$ times the probability density
\begin{eqnarray}
{\rm Re} \,
\iP_f^\dagger (x) \cH \iP_f(x) &=& 
\hbar \omega_0
\iP_f^\dagger (x) \iP_f(x)
\end{eqnarray}
for the wave packet.

For any wave packet, represented by $\iP_f$, the photon
probability density four-vector defined in
Eq.~(\ref{eq:probdenv}) is
\begin{eqnarray}
q_f^\mu(x) &=& \overline{\iP}_f(x){\gamma^\mu}\iP_f(x),
\end{eqnarray}
and the differential conservation of probability is given by
Eq.~(\ref{eq:consprob})
\begin{eqnarray}
\frac{\partial}{\partial t}\,q_f^0(x) + c \, \bm\nabla\cdot
\bm q_f(x) &=& 0.
\end{eqnarray}
This is valid for any solution of the homogeneous Maxwell
equation.  By integrating over a closed volume and converting
the divergence to an integral of the normal component of the
vector over the surface, one obtains a statement of conservation
of probability for the volume.  If the surface of the volume is
taken to infinity in all directions, where the wave function
vanishes, this expression shows the time independence of the
normalization of the wave function
\begin{eqnarray}
\frac{\partial}{\partial t}
\int\rd\bm x\, \iP_f^\dagger(x)\iP_f(x) &=& 0 .
\end{eqnarray}
Of course, this does not give meaningful results for a plane
wave, because in this case, the probability density is constant
over space and is not normalizable.

For electromagnetic fields and photons, Lorentz invariance is a
necessary consideration.  In Secs.~\ref{ssec:psirots} and
\ref{ssec:vtpsi} it is shown that 
\begin{eqnarray*}
\gamma^\mu\partial_\mu  \iP^\prime(x) &=& \iX^\prime(x),
\end{eqnarray*}
where the primes indicate that the field and source have been
transformed by either a rotation or a velocity boost.  For
rotations represented by the vector $\bm u=\theta\bm{\hat u}$,
the transformed field, in Eq.~(\ref{eq:solrot}), is
\begin{eqnarray*}
\iP^\prime(x) &=& \cR(\bm u)\iP\!\big(R^{-1}(\bm u)\,x\big),
\end{eqnarray*}
where $R(\bm u)$ is the coordinate rotation operator in
Eq.~(\ref{eq:4rotmat}) and 
\begin{eqnarray*}
\cR(\bm u) &=& \re^{-\ri\bm \cS\cdot\bm u}
\end{eqnarray*}
in Eq.~(\ref{eq:srotop}).  The source term transforms in the
same way.  For velocity transformations, corresponding to the
velocity $\bm v = c\tanh{\zeta}\,\bm{\hat v}$,
Eq.~(\ref{eq:solvel}) is
\begin{eqnarray*}
\iP^\prime(x) &=& \cV(\bm v)\iP\!\big(V^{-1}(\bm v)\,x\big),
\end{eqnarray*}
where $V(\bm v)$ is the coordinate velocity transformation
operator in Eq.~(\ref{eq:cvt}) and
\begin{eqnarray*}
\cV(\bm v) &=& \re^{\zeta\bm \cK \cdot \bm{\hat v}}
\end{eqnarray*}
in Eq.~(\ref{eq:vdef}), where
\begin{eqnarray*}
\bm \cK &=& \left(\begin{array}{cc} \0 & \bm \tau \\ \bm \tau & \0
\end{array}\right)
\end{eqnarray*}
in Eq.~(\ref{eq:kdef}).  The transformation of the source term
under a velocity boost is noteworthy.  In the presence of a
non-zero source, the Maxwell equation is invariant, but the
left- and right-hand sides do not transform separately.  As shown
in Sec.~\ref{ssec:vtpsi}, the derivatives acting on the fields
produce terms that combine with the original source term in such
a way as to produce the velocity transformed source term, even
though it is the three-vector current.  In the absence of
sources, the transformation reduces to a more conventional form.

In the case of photon wave functions, sources are taken to be
absent and the wave functions are solutions of the homogeneous
Maxwell equation.  The Lorentz transformations of the plane-wave
functions are explicitly shown in Secs.~\ref{ssec:rtwf} and
\ref{ssec:vtwf}.  As with the Dirac equation for an electron,
the eigenvalues in Eq.~(\ref{eq:eveq}) may be either positive or
negative, and here they also may be zero, for both plane-wave
and spherical-wave eigenfunctions.  The negative eigenvalues,
which are associated with relativistic invariance, are necessary
in order to have a complete set of solutions satisfying
Eq.~(\ref{eq:concomp}).  It is relevant to note that for the
six-component wave packet in Eq.~(\ref{eq:gaussian}), there is
an interaction between the upper-three components and
lower-three components, evident in Eq.~(\ref{eq:ffact}), that
suppresses the role of the negative energy states.

In Sec.~\ref{sec:tlfields}, Eqs.~(\ref{eq:ptop}) and
(\ref{eq:plop}), orthogonal transverse and longitudinal
projection operators are defined:
\begin{eqnarray*}
\iPi\rT(\bm \nabla) = \left(\begin{array}{ccc}
\bm{\iPi}\rs\rT(\bm\nabla) && \0 \\
\0 && \bm{\iPi}\rs\rT(\bm\nabla) \msp \end{array}\right),
&&
\iPi\rL(\bm \nabla) = \left(\begin{array}{ccc}
\bm{\iPi}\rs\rL(\bm\nabla) && \0 \\
\0 && \bm{\iPi}\rs\rL(\bm\nabla) \msp \end{array}\right),
\end{eqnarray*}
where from Eqs.~(\ref{eq:tpjop}) and (\ref{eq:lpjop})
\begin{eqnarray*}
\bm{\iPi}\rs\rT(\bm\nabla) = \frac{(\bm\tau\cdot\bm\nabla)^2}{\bm\nabla^2} \, ,
&&
\bm{\iPi}\rs\rL(\bm\nabla) =
\frac{\bm\nabla\rs\bm\nabla\rs^\dagger}{\bm\nabla^2} \, .
\end{eqnarray*}
These operators commute with the Hamiltonian, the momentum
operator, and the angular-momentum operator
\begin{eqnarray}
\left[\cH,\iPi\right] = 
\left[\bm \cP,\iPi\right] = 
\left[\bm \cJ,\iPi\right] = 0,
\end{eqnarray}
where $\iPi$ represents either projection operator, $\iPi\rT(\bm
\nabla)$ or $\iPi\rL(\bm \nabla)$, so all the eigenstates
considered in this paper are classified as being either
transverse or longitudinal, with
\begin{eqnarray}
\iPi\rT(\bm \nabla) \iP\rT_n(x) &=& \iP\rT_n(x),
\\
\iPi\rL(\bm \nabla) \iP\rL_n(x) &=& \iP\rL_n(x),
\end{eqnarray}
respectively.  The transverse states describe radiation and have
non-zero eigenvalues in Eq.~(\ref{eq:eveq}), while the
longitudinal states correspond to electrostatic interactions,
with an eigenvalue of zero.  An exception is that the
longitudinal states may have a non-zero eigenvalue if a
hypothetical mass term is considered, as discussed in
Sec.~\ref{ssec:tdwf}.  The projection operators commute with
rotations, but not, in general, with velocity boosts.  However,
as shown in Sec.~\ref{sssec:twf}, the velocity transformed
transverse plane-wave states are also transverse.  This
corresponds to the fact that radiation may be treated
relativistically independent of electrostatic interactions.  On
the other hand, as shown in Sec.~\ref{sssec:lwf}, the velocity
transformed longitudinal states have both longitudinal and
transverse components, corresponding to the fact that moving
charges may excite radiative transitions.  Both transverse and
longitudinal states are necessary in order to have a complete
set, as in Eq.~(\ref{eq:concomp}).

A solution of the inhomogeneous Maxwell equation may be obtained
with the Maxwell Green function, as discussed in
Sec.~\ref{sec:prop}.  The Green function satisfies the equation
\begin{eqnarray*}
\gamma_\mu\partial_2^\mu\,
\cD\rM(x_2 - x_1) &=& \cI \, \delta(x_2-x_1)
\end{eqnarray*}
in Eq.~(\ref{eq:mpeq}), and a solution of the Maxwell equation
is given by
\begin{eqnarray*}
\iP(x_2) &=& \int \rd^4 x_1 \,
\cD\rM(x_2 - x_1)\,\iX(x_1)
\end{eqnarray*}
in Eq.~(\ref{eq:solsource}).  It is shown, by summing over the
complete set of plane-wave solutions, that the Green function is
\begin{eqnarray*} 
\cD\rM(x_2 - x_1) 
&=& \frac{\ri}{(2\pi)^4} \, \int_{\rm C_F} \rd^4 k \
\frac{\re^{-\ri k\cdot (x_2-x_1)}} {\gamma^\mu k_\mu -
m_\epsilon c/\hbar}
\end{eqnarray*} 
in Eq.~(\ref{eq:mprop}), which is the same form as the Dirac
Green function, except that here it is a $6\times6$ matrix
instead of a $4\times4$ matrix.  In this equation, C$_{\rm F}$
is the Feynman contour and the infinitesimal mass is included to
resolve an ambiguity in the longitudinal contribution.

In Sec.~\ref{sec:appmpf}, applications of the Maxwell Green
function are made, including a calculation of radiation from a
Dirac-electron current source.  In this example the
six-component Maxwell formalism couples radiation to the Dirac
current relativistically with a result that is the same as the
result of a calculation that starts from Feynman-gauge QED.

\section{Conclusion}
\elabel{sec:conclusion}

We conclude that the criteria for properties of a single-photon
wave function proposed in the introduction are met by the
formalism described in the subsequent sections.  In particular,
the example of a photon wave packet provides a normalizable
solution of the wave equation whose properties can be verified
by explicit calculations.  It yields the unanticipated result
that for virtually any probability distribution, under rather
mild assumptions about the form of the wave packet, the
expectation value of the Hamiltonian is exactly
$\left<\,\cH\,\right> = \hbar\omega_0$, where $\omega_0$ is the
frequency associated with the wave vector of the packet.

\appendix

\section{Velocity transformation of electromagnetic fields}
\elabel{app:vtft}

The velocity transformation of electromagnetic fields is derived
here without invoking potentials for completeness.  With the
aid of the identity $\bm\nabla\rc^\top \bm{\tilde \tau}\cdot
c\bm B = (\bm\nabla\times c \bm B)\rc^\top$,
Eqs.~(\ref{eq:maxeq1}) and (\ref{eq:maxeq2}) may be written as
\begin{eqnarray}
\bm\nabla\rc^\top\bm E\rc &=& \mu_0 c^2 \rho,
\\
\frac{\partial\bm E\rc^\top}{\partial ct} - \bm\nabla\rc^\top
\bm{\tilde \tau}\cdot c\bm B &=& - \mu_0 c \bm J\rc^\top
\end{eqnarray}
or
\begin{eqnarray}
\partial\rc^\top g F &=& \mu_0 J^\top,
\elabel{eq:fseq}
\end{eqnarray}
where
\begin{eqnarray}
F &=& \frac{1}{c}\left(\begin{array}{ccc}
  0     && -\bm E^{\top}_{\rm c}  \\
  \bm E_{\rm c}    &&
  \bm{\tilde \tau}\cdot c \bm B
  \vbox to 14 pt {} 
  \end{array} \right)
\end{eqnarray}
is the field tensor [see Eq.~(\ref{eq:lrp})] and 
\begin{eqnarray}
J &=& \left(\begin{array}{c}
  c\rho \\ \bm J\rc \end{array}\right).
\end{eqnarray}
Since $V(\bm v) \, g \, V(\bm v) = g$,
Eq.~(\ref{eq:fseq}) is equivalent to
\begin{eqnarray}
\partial\rc^\top V(\bm v) \, g \, V(\bm v) \, F(x) V(\bm v) &=& 
\mu_0 J^\top(x)V(\bm v) \qquad
\end{eqnarray}
or
\begin{eqnarray}
\partial\rc^\top  g \, V(\bm v) \, F\left(V^{-1}(\bm v)x\right) V(\bm v) 
&=&
\mu_0 J^\top\!\!\left(V^{-1}(\bm v)x\right)V(\bm v). \qquad
\end{eqnarray}
Assuming the current transforms as a four-vector,
\begin{eqnarray}
J^\prime(x) &=& V(\bm v)\,J\!\left(V^{-1}(\bm v)x\right),
\end{eqnarray}
Eq.~(\ref{eq:fseq}) will be invariant if the field
tensor transforms according to
\begin{eqnarray}
F^\prime(x) &=& V(\bm v) \, F\!\left(V^{-1}(\bm v)x\right) V(\bm v).
\elabel{eq:fttr}
\end{eqnarray}
Direct calculation yields\footnote{The identity $\epsilon_{ijk}
= \epsilon_{ljk}\hat{v}^i\hat{v}^l + 
 \epsilon_{ilk}\hat{v}^j\hat{v}^l +
 \epsilon_{ijl}\hat{v}^k\hat{v}^l$ may be useful here.}
\begin{eqnarray}
F^\prime(x^\prime) &=& \frac{1}{c}\left(\begin{array}{ccc}
  0     && -\bm E^{\prime\top}_{\rm c}(x)  \\
  \bm E^\prime_{\rm c}(x)    &&
  \bm{\tilde \tau}\cdot c \bm B^\prime(x)
  \vbox to 14 pt {} 
  \end{array} \right),
\end{eqnarray}
where $x^\prime = V(\bm v)\,x$ and
\begin{eqnarray}
\bm E_{\rm c}^{\,\prime} &=&
\bm E_{\rm c}\cosh{\zeta}-\bm{\hat v}_{\rm c}\,\bm{\hat
v}\cdot\bm E
 \left(\cosh{\zeta}-1\right)
 -\bm{\tilde \tau}\cdot \bm{\hat v}
 \, c\bm B\rc\sinh{\zeta},
\elabel{eq:ecft}
  \\
c \bm B_{\rm c}^{\,\prime} &=& 
c \bm B_{\rm c}\cosh{\zeta}-\bm{\hat v}_{\rm c}\,\bm{\hat
v}\cdot c \bm B
 \left(\cosh{\zeta}-1\right)
 + \bm{\tilde \tau}\cdot \bm{\hat v}
 \,\bm E\rc\sinh{\zeta}.
\elabel{eq:bcft}
\end{eqnarray}
These relations are equivalent (up to the velocity sign
convention) to the electric and magnetic field transformations
in \cite{1998165}, and in the spherical basis they are
\begin{eqnarray}
\bm E\rs^\prime &=& \bm E\rs + (\bm \tau \cdot \bm{\hat v})^2 
\bm E\rs (\cosh{\zeta}-1) 
+ \ri \, \bm \tau \cdot \bm{\hat v} 
\, c \bm B\rs \sinh{\zeta},
\\ c\bm B\rs^\prime &=& c\bm B\rs + (\bm \tau \cdot \bm{\hat v})^2 
c\bm B\rs (\cosh{\zeta}-1) 
- \ri \, \bm \tau \cdot \bm{\hat v} \, \bm E\rs
\sinh{\zeta}.
\end{eqnarray}

The Cartesian transformation in Eqs.~(\ref{eq:ecft}) and
(\ref{eq:bcft}) can be written as
\begin{eqnarray}
\iP\rc^\prime(x^\prime) &=& \re^{-\zeta\bm\tilde{\bm \cK}\cdot
\bm{\hat v}}
\iP\rc(x) ,
\end{eqnarray}
where
\begin{eqnarray}
\bm\tilde{\bm \cK} &=& \left(\begin{array}{c@{\quad}c}
\0 & \bm{\tilde \tau} \\
-\bm{\tilde \tau} & \0
\end{array}\right) 
\elabel{eq:ktilde}
\end{eqnarray}
and
\begin{eqnarray}
\iP\rc(x) &=& \left(\begin{array}{c}
\bm E\rc(x) \\ c \bm B\rc(x)
\end{array}\right) .
\end{eqnarray}
Similarly, the Cartesian form of the rotation
transformation, corresponding to the operator in
Eq.~(\ref{eq:srotop}), is
\begin{eqnarray}
\iP\rc^\prime(x^\prime) &=& \re^{\bm \tilde{\bm\cS}\cdot \bm u}
\iP\rc(x) ,
\end{eqnarray}
where
\begin{eqnarray}
\bm\tilde{\bm \cS} &=& \left(\begin{array}{c@{\quad}c}
\bm{\tilde \tau} & \0 \\
\0 & \bm{\tilde \tau} 
\end{array}\right) .
\elabel{eq:stilde}
\end{eqnarray}
The components of the matrices in Eqs.~(\ref{eq:ktilde}) and
(\ref{eq:stilde}) have the commutation relations
\begin{eqnarray}
\left[\tilde \cS^i,\tilde \cS^j\right] &=& \epsilon_{ijk}\,\tilde \cS^k,
\\
\left[\tilde \cS^i,\tilde \cK^j\right] &=&
\epsilon_{ijk}\,\tilde \cK^k,
\\
\left[\tilde \cK^i,\tilde \cK^j\right] &=& -\epsilon_{ijk}\,\tilde \cS^k,
\end{eqnarray}
characteristic of the Lie algebra of Lorentz transformations.

\section{Inverse Laplacian}
\elabel{app:invlap}

For cases where the integral definition of the inverse Laplacian
converges poorly, we use a generalized definition that includes
a damping factor to resolve ambiguity in the intermediate steps
of the calculation.  From the equation
\begin{eqnarray}
\left(\bm\nabla^2 - \epsilon^2\right)
\frac{\re^{-\epsilon\,|\bm x - \bm x^\prime|}}
{|\bm x - \bm x^\prime|} 
&=& -4\pi\delta(\bm x - \bm x^\prime) ,
\end{eqnarray}
one has
\begin{eqnarray}
\frac{1}{\bm\nabla^2 - \epsilon^2} \, \delta(\bm x - \bm x^\prime)
&=&  - \frac{1}{4\pi} \,
\frac{\re^{-\epsilon\,|\bm x - \bm x^\prime|}}
{|\bm x - \bm x^\prime|} .
\end{eqnarray}
Multiplication by $f(\bm x^\prime)$ and integration over $\bm
x^\prime$ yields 
\begin{eqnarray}
\frac{1}{\bm\nabla^{2}-\epsilon^2}\,f(\bm x) &=&
-{1\over 4\pi} \int{\rd}\bm x^\prime
\,\frac{\re^{-\epsilon\,|\bm x - \bm x^\prime|}}
{|\bm x - \bm x^\prime|} \,f(\bm x^\prime).
\qquad
\end{eqnarray}
We thus have, for example,
\begin{eqnarray}
\frac{1}{\bm\nabla^{2}-\epsilon^2}\,
\re^{\ri\bm k\cdot \bm x}
&=& -\frac{1}{\bm k^2 + \epsilon^2} \,
\re^{\ri\bm k\cdot \bm x}
\rightarrow 
-\frac{1}{\bm k^2} \,
\re^{\ri\bm k\cdot \bm x}, \qquad
\end{eqnarray}
by direct calculation of the integral.

\section{Coulomb matrix element}
\elabel{app:coultr}

The calculation of the Coulomb field matrix element in
Eq.~(\ref{eq:coultr}) requires evaluation of the integral
\begin{eqnarray}
\int\rd\bm x \, \frac{\bm{\hat k}\cdot \bm x}{|\bm x|^3}\,
\re^{- \ri \bm k \cdot \bm x}
&=& \lim_{\epsilon\rightarrow 0}
\int\rd\bm x  \,
\re^{-\epsilon |\bm x|}
\, \frac{\bm{\hat k}\cdot \bm x}{|\bm x|^3}\,
\re^{- \ri \bm k \cdot \bm x} \qquad
\nonumber\\ &\rightarrow&
 -  \, \ri
\, \frac{4\pi}{|\bm k|},
\end{eqnarray}
which is defined with the convergence factor
$\re^{-\epsilon |\bm x|}$.
The inverse transformation requires the integral
\begin{eqnarray}
\int\rd\bm k \, \frac{\bm{\hat k}}{|\bm k|} \,
\re^{ \ri \bm k \cdot \bm x}
 &=&
\lim_{\epsilon\rightarrow 0}
\int\rd\bm k  \,
\re^{-\epsilon |\bm k|}
\, \frac{\bm{\hat k}}{|\bm k|}\,
\re^{ \ri \bm k \cdot \bm x} \qquad
\nonumber\\ &\rightarrow&  \,\ri
\, \frac{2\pi^2\bm x}{|\bm x|^3},
\end{eqnarray}
which confirms the result that 
\begin{eqnarray}
- \frac{\ri\,q}{\sqrt{(2\pi)^3}\,\epsilon_0}
\int \rd \bm k \,
\frac{1}{|\bm k|} \,
\psi_{\bm{k},0}^{(+)}(\bm x)
&=&
\frac{q}{4 \pi\epsilon_0 |\bm x|^3}
\left(\begin{array}{c} \bm x\rs \\ \0 \end{array}\right).
\nonumber\\
\end{eqnarray}

\section{Separation of the transverse and longitudinal gradient
operators}
\elabel{app:stlgo}

The transverse gradient operator $\bm\tau\cdot\bm\nabla$ is
separated into radial and angular parts by writing
\begin{eqnarray}
\bm\tau\cdot\bm\nabla &=& 
\bm\tau\cdot\bm{\hat x}\,\bm\tau\cdot\bm{\hat x} \,
\bm\tau\cdot\bm\nabla 
+\bm{\hat x}\rs\,\bm{\hat x}\rs^\dagger\,
\bm\tau\cdot\bm\nabla 
\nonumber\\&=&
\frac{\partial}{\partial r}\, \bm\tau\cdot\bm{\hat x}
-\bm\tau\cdot\bm{\hat x}\,\overline{\bm \nabla}\rs\,
\bm{\hat x}\rs^\dagger + \frac{1}{\hbar r}\,\bm{\hat x}\rs
\bm L\rs^\dagger \qquad
\end{eqnarray}
where the line over the gradient operator indicates that it does
not act on the unit vector directly to the right.  That term is
\begin{eqnarray}
\bm\tau\cdot\bm{\hat x}\,\overline{\bm \nabla}\rs\,
\bm{\hat x}\rs^\dagger &=&
\bm \tau\cdot\bm {\hat x}\left(\bm\nabla\rs \bm{\hat x}\rs^\dagger
+\frac{1}{r}\,\bm{\hat x}\rs\bm{\hat x}\rs^\dagger -
\frac{1}{r}\,\bm I \right)
\nonumber\\&=& - \frac{1}{\hbar r}\, \bm L\rs\bm{\hat
x}\rs^\dagger
-\frac{1}{r}\,\bm \tau\cdot\bm {\hat x} .
\end{eqnarray}
Thus
\begin{eqnarray}
\bm\tau\cdot\bm\nabla &=&
\frac{1}{r}\,\frac{\partial}{\partial r}\,r \,
\bm\tau\cdot\,\bm{\hat x}\,
+\frac{1}{\hbar r}\left(\bm L\rs\,\bm{\hat x}\rs^\dagger
+\bm{\hat x}\rs\,\bm L\rs^\dagger\right).
\qquad
\end{eqnarray}
Acting on $\bm L\rs$, the transverse gradient operator yields
\begin{eqnarray}
\bm\tau\cdot\bm\nabla \, \bm L\rs &=&
\frac{1}{r}\,\frac{\partial}{\partial r}\,r \,
\bm\tau\cdot\,\bm{\hat x}\,\bm L\rs
-\frac{1}{\hbar r}
\bm{\hat x}\rs\,\bm L^2 ,
\qquad
\end{eqnarray}
where $\bm L\rs^\dagger\bm L\rs = - \bm L^2$.

For the longitudinal gradient operator $\bm\nabla\rs$, 
the identity
\begin{eqnarray}
\bm\nabla
  &=& \bm{\hat x}\,\frac{\partial}{\partial r}
  -\frac{\ri}{\hbar r}\,\bm{\hat x}\times\bm L
  \end{eqnarray}
provides
\begin{eqnarray}
  \bm\nabla\rs &=&
   \bm{\hat x}\rs\,\frac{\partial}{\partial r}
   -\frac{1}{\hbar r}\,\bm \tau\cdot\bm{\hat x}\,\bm L\rs .
   \elabel{eq:slg}
\end{eqnarray}

\section{Exact transverse Green function}
\elabel{app:extr}

The exact transverse Maxwell Green function follows from
Eq.~(\ref{eq:trprop}) together with the relations

\begin{eqnarray}
\bm \tau\cdot\bm\nabla\,\frac{\re^{-w|\bm r|}}{|\bm r|}
&=& 
 - w \,\bm\tau\cdot\bm{\hat r}\,\frac{\re^{-w|\bm r|}}{|\bm
r|} \left(1 + \frac{1}{w|\bm r|}\right)
\end{eqnarray}
and
\begin{eqnarray}
\frac{1}{\nabla^2}\,
\frac{\re^{-w|\bm r|}}
{|\bm r|} 
&=& -\frac{1}{4\pi}
\int \rd \bm x \,
\frac{1}{|\bm r - \bm x|}
\frac{\re^{-w|\bm x|}}
{|\bm x|} 
= - \frac{1}{w^2\,|\bm r|}
\left(1 - \re^{-w|\bm r|}\right),
\end{eqnarray}
which yield
\begin{eqnarray}
\frac{\nabla^i\nabla^j}{\nabla^2}\,\frac{\re^{-w|\bm
r|}}{|\bm r|} &=&
\frac{r^ir^j}{|\bm r|^2}
\frac{\re^{-w|\bm r|}}{|\bm r|} +
\left(3\,\frac{r^ir^j}{|\bm r|^2} - \delta_{ij}\right)
\left(\frac{\re^{-w|\bm r|}}{w|\bm r|^2}
+\frac{\re^{-w|\bm r|}}{w^2|\bm r|^3} -
\frac{1}{w^2|\bm r|^3}
\right)
\end{eqnarray}
and
\begin{eqnarray}
\bm \iPi\rs\rT(\bm \nabla)\,
\frac{\re^{-w|\bm r|}}{|\bm r|}
&=&(\bm \tau\cdot\bm{\hat r})^2 \,
\frac{\re^{-w|\bm r|}}{|\bm r|}
+\left[(\bm \tau\cdot\bm{\hat r})^2 - 2 \, \bm{\hat
r}\rs\,\bm{\hat r}\rs^\dagger \right]
\left(\frac{\re^{-w|\bm r|}}{w|\bm r|^2}
+\frac{\re^{-w|\bm r|}}{w^2|\bm r|^3} -
\frac{1}{w^2|\bm r|^3}
\right).
\qquad
\end{eqnarray}

\section{Radiative decay in quantum electrodynamics}
\elabel{app:transrate}

In QED, the radiative decay rate of an excited state may be
obtained from the imaginary part of the radiative correction to
the energy level of that state
\begin{eqnarray}
\hbar\sum_f A_{if} &=& -2 \,{\rm Im} (\Delta E_i),
\end{eqnarray}
where the sum is over all states with a lower unperturbed
energy.  This gives a correction to the level that, roughly
speaking, results in an exponentially damped time dependence for
the population of the state:
\begin{eqnarray}
\big|\re^{-\ri\,\Delta \! E \, t/\hbar} \big|^2 =
\re^{-\sum_f A_{if} t} .
\end{eqnarray}
For one-photon decays, the rate is included in the second-order
self-energy correction to the level.  An expression derived from
Feynman-gauge QED that includes some of the real part and all of
the imaginary part of this level shift in hydrogen-like atoms is
given by \citep{1974003}
\begin{eqnarray}
\Delta E_i &=& -\frac{\alpha\hbar^2c^2}{4\pi^2}
\int_{\hbar c k < E_i}\rd \bm k\,\frac{1}{k}\left(\delta_{jl}-
\frac{k^jk^l}{\bm k^2}\right)
\left< \alpha^j \re^{\ri\bm k\cdot\bm x}
\frac{1}{H -E_i + \hbar c k - \ri \delta} \,
\alpha^l \re^{-\ri\bm k \cdot\bm x} \right>
\nonumber\\
&=& -\frac{\alpha\hbar^2c^2}{4\pi^2}
\int_{\hbar c k < E_i}\rd \bm k\,\frac{1}{k}
\sum_{\lambda=1}^2\sum_f
\left<i \left|\,\bm{\hat \epsilon}_\lambda(\bm{\hat k})\cdot\bm \alpha
\,\re^{\ri\bm k\cdot\bm x}\right|f\right>
\nonumber\\&&\times
\frac{1}{E_f -E_i + \hbar c k - \ri \delta} \,
\left<f\left|
\bm{\hat \epsilon}_\lambda(\bm{\hat k})
\cdot \bm \alpha \,\re^{-\ri\bm k \cdot\bm x} \right|i\right>,
\quad
\elabel{eq:ese}
\end{eqnarray}
where $H$ is the Dirac Hamiltonian.  The integrand is real,
except for the imaginary infinitesimal in the denominator, for
which
\begin{eqnarray}
{\rm Im}\,\frac{1}{E_f -E_i + \hbar c k - \ri \delta}  
&\rightarrow& 
\pi\delta(E_f -E_i + \hbar c k)
\end{eqnarray}
and hence
\begin{eqnarray}
\sum_f A_{if} 
&=& 
\sum_f
\frac{\alpha k c}{2\pi}
\int \rd \iO_{\bm k}
\sum_{\lambda=1}^2
\left<i \left|\,\bm{\hat \epsilon}_\lambda(\bm{\hat k})\cdot\bm \alpha
\,\re^{\ri\bm k\cdot\bm x}\right|f\right>
\left<f\left|
\bm{\hat \epsilon}_\lambda(\bm{\hat k})
\cdot \bm \alpha \,\re^{-\ri\bm k \cdot\bm x} \right|i\right>,
\end{eqnarray}
with the restriction $0 < E_f < E_i$ on the sum over $f$.  The
contribution to the decay rate from each final state $f$
coincides with Eq.~(\ref{eq:transrate}) for the transition rate
$A_{if}$.

\newpage

\bibliography{refs} 

\end{document}